%% file: arXiv-astro-camera-ready.tex
\documentclass{egpubl-noname}
\usepackage{eurovis2021-noname}

\usepackage[T1]{fontenc}
\usepackage{dfadobe}  

\usepackage{cite}  
\BibtexOrBiblatex
\electronicVersion
\PrintedOrElectronic
\ifpdf \usepackage[pdftex]{graphicx} \pdfcompresslevel=9
\else \usepackage[dvips]{graphicx} \fi

\usepackage{egweblnk} 
\usepackage{amsmath}
\usepackage{amssymb}

\usepackage{textcomp}

\usepackage{makecell}

\usepackage{xcolor,colortbl}
\definecolor{green}{rgb}{0.1,0.1,0.1}

\definecolor{amethyst}{rgb}{0.6, 0.4, 0.8}
\definecolor{aliceblue}{rgb}{0.94, 0.97, 1.0}
\definecolor{apricot}{rgb}{0.98, 0.81, 0.69}
\definecolor{aquamarine}{rgb}{0.5, 1.0, 0.83}	
\definecolor{ashgrey}{rgb}{0.7, 0.75, 0.71}
\definecolor{asparagus}{rgb}{0.53, 0.66, 0.42}
\definecolor{babyblue}{rgb}{0.54, 0.81, 0.94}
\definecolor{babypink}{rgb}{0.96, 0.76, 0.76}
\definecolor{burlywood}{rgb}{0.87, 0.72, 0.53}
\definecolor{brightlavender}{rgb}{0.75, 0.58, 0.89}

\usepackage{lipsum,multicol}

\graphicspath{{./figs/}}

\usepackage{enumitem}

\newcommand{\para}[1]        {\vspace{1mm}\noindent{\textbf{#1}}}

\newcommand{\etal}{\emph{et~al.}}
\newcommand{\etalnocite}{\emph{et~al. }}

\newcommand{\ie}{\emph{i.e.}}

\usepackage{amsfonts}
\usepackage{booktabs}
\usepackage{siunitx}

\usepackage{xcolor,colortbl}
\definecolor{green}{rgb}{0.1,0.1,0.1}

\usepackage{tikz}

\usepackage{tabularx}

\definecolor{tableaublue}{RGB}{31, 119, 180}
\definecolor{tableauorange}{RGB}{255, 127, 14}
\definecolor{tableaugreen}{RGB}{44, 160, 44}
\definecolor{tableaured}{RGB}{214, 39, 40}
\definecolor{tableaupurple}{RGB}{148, 103, 189}
\definecolor{tableaugrey}{RGB}{127, 127, 127}

\definecolor{amethyst}{rgb}{0.6, 0.4, 0.8}
\definecolor{aliceblue}{rgb}{0.94, 0.97, 1.0}
\definecolor{apricot}{rgb}{0.98, 0.81, 0.69}
\definecolor{aquamarine}{rgb}{0.5, 1.0, 0.83}	
\definecolor{ashgrey}{rgb}{0.7, 0.75, 0.71}
\definecolor{asparagus}{rgb}{0.53, 0.66, 0.42}
\definecolor{babyblue}{rgb}{0.54, 0.81, 0.94}
\definecolor{babypink}{rgb}{0.96, 0.76, 0.76}
\definecolor{burlywood}{rgb}{0.87, 0.72, 0.53}
\definecolor{coral}{rgb}{1, 0.5, 0.31}
\definecolor{dimgrey}{rgb}{0.41, 0.41, 0.41}
\definecolor{palevioletred}{rgb}{0.86, 0.44, 0.58}
\definecolor{olive}{rgb}{0.5, 0.5, 0}
\definecolor{dodgerblue}{rgb}{0.12, 0.56, 1}
\definecolor{saddlebrown}{rgb}{0.54, 0.27, 0.07}

\definecolor{dodgerblue}{rgb}{0.12, 0.56, 1}

\newcommand*{\img}[1]{%
    \raisebox{-.3\baselineskip}{%
        \includegraphics[
        height=\baselineskip,
        width=\baselineskip,
        keepaspectratio,
        ]{#1}%
    }%
}

\newcommand*{\imglarger}[1]{%
    \raisebox{-.3\baselineskip}{%
        \includegraphics[
        width=0.07\textwidth,
        keepaspectratio,
        ]{#1}%
    }%
}

\newcommand\mc[1]{\multicolumn{1}{l}{#1}} 

\title[Visualization in Astrophysics]
      {Visualization in Astrophysics: Developing New Methods, Discovering Our Universe, and Educating the Earth}

\author[Lan et al.]
{\parbox{\textwidth}{\centering 
        Fangfei Lan$^{1}$\orcid{0000-0002-8237-5919}, 
        Michael Young$^{1}$\orcid{0000-0001-8219-1215}, 
        Lauren Anderson$^{2}$\orcid{0000-0001-5725-9329}, 
      Anders Ynnerman$^{3}$\orcid{0000-0002-9466-9826}
        Alexander Bock$^{1,3}$\orcid{0000-0002-8700-7795},
        Michelle A. Borkin$^{4}$\orcid{0000-0002-8016-355X},\\
        Angus G. Forbes$^{5}$\orcid{0000-0002-8700-7795},
        Juna A. Kollmeier$^{2}$\orcid{0000-0001-9852-1610},       
        Bei Wang$^{1}$\orcid{0000-0002-9240-0700}
        }
        \\
{\parbox{\textwidth}{\centering 
$^1$Scientific Computing and Imaging Institute, University of Utah, USA\\
$^2$Carnegie Institution for Science, USA\\
$^3$Link\"oping University, Sweden\\
$^4$Northeastern University, USA\\
$^5$University of California, Santa Cruz, USA
}}}

\begin{document}

\maketitle
%-------------------------------------------------------------------------
\begin{abstract}
\input{sec-abstract}

\end{abstract}  
%-------------------------------------------------------------------------
\input{sec-introduction}

\input{sec-categories}

\input{sec-data-wrangling}

\input{sec-data-exploration}

\input{sec-feature-identification}

\input{sec-object-reconstruction}

\input{sec-data-accessibility}

\input{sec-challenges}

\input{sec-navigation-tool}

\input{sec-conclusion}

\section*{Acknowledgements} 
\label{sec:acknowledgements}
The authors would like to thank the constructive suggestions by the reviewers and the generous funding by the following entities:  
Carnegie Institution for Science Venture Grant no. BPO-700082; The Canadian Institute for Advanced Research Fellowship program; The Ahmanson Foundation; 
NASA under grant no. NNX16AB93A; the Swedish e-Science Research Centre; the Knut and Alice Wallenberg Foundation; and the Swedish Research Council DNR:2015-05462.  
The authors would also like to thank the speakers and participants of following two workshops, who helped us identify challenges and opportunities via great discussions: the Carnegie + SCI mini-workshop (April 2020), and the Visualization in Astrophysics workshop during IEEE VIS (October 2020), in particular, Chuck Hansen, Chris R. Johnson, Alyssa Goodman, Jackie Faherty, Matthew J. Turk, Ryan Wyatt, and Joseph N. Burchett.

%-------------------------------------------------------------------------
%\bibliographystyle{eg-alpha} 
%\bibliography{STAR-astro-refs.bib}    
\input{arXiv-astro-camera-ready.bbl}

%-------------------------------------------------------------------------
 
\input{sec-bio}

%-------------------------------------------------------------------------

\end{document}

%% file: sec-abstract.tex
We present a state-of-the-art report on visualization in astrophysics. 
We survey representative papers from both astrophysics and visualization and provide a taxonomy of existing approaches based on data analysis tasks. 
The approaches are classified based on five categories: data wrangling, data exploration, feature identification, object reconstruction, as well as  education and outreach. 
Our unique contribution is to combine the diverse viewpoints from both astronomers and visualization experts to identify challenges and opportunities for visualization in astrophysics. 
The main goal is to provide a reference point to bring modern data analysis and visualization techniques to the rich datasets in astrophysics.

%% file: sec-introduction.tex
\section{Introduction}
\label{sec:introduction}

Modern astronomers are recording an increasing amount of information for a larger number of astronomical objects and making more complex predictions about the nature of these objects and their evolution over cosmic time. Both successes are being driven by advances in experimental and computational infrastructure.  
As a result, the next generation of computations and surveys will put astronomers face to face with a ``digital tsunami'' of both simulated and observed data.  These data present opportunities to make enormous strides in discovering more about our universe and state-of-the-art visualization methodologies.

This state-of-the-art report serves as a starting point to bridge the knowledge gap between the astronomy and visualization communities and catalyze research opportunities. Astronomy has a long and rich history as a visual science. Images of the cosmos have been used to build theories of physical phenomena for millennia. This history makes astronomy a natural area for fruitful collaborations between visualization and astronomy. A substantial fraction of previous work at this scientific intersection has therefore focused on image reconstruction -- generating the most precise representation from a series of images of a patch of the sky -- typically using optimizations and signal processing techniques. Advances in image reconstruction have enabled great breakthroughs in astronomy, including the recent imaging of a black hole~\cite{EHTC19}.
However, in this report, we focus on modern visualization techniques, which include 3D rendering, interaction, uncertainty visualization, and new display platforms. 
This report, authored by experts in both astronomy and visualization, will help visualization experts better understand the needs and opportunities of astronomical visualization, and provide a mechanism for astronomers to learn more about cutting-edge methods and research in visualization as applied to astronomy. 

\para{Comparison with related surveys.}    
Several studies have focused on surveying visualization of astronomical data. 
Hassan \etal~\cite{HassanFluke2011} surveyed scientific visualization in astronomy from 1990 to 2010. 
They studied visualization approaches for N-body particle simulation data and spectral data cubes -- two areas they identified as the most active fields. 
They classified research papers in these areas based on how astronomical data are stored (i.e.,~as points, splats, isosurfaces, or volumes) and which visualization techniques are used. 
They also discussed visualization workflows and public outreach, and reviewed existing softwares for astronomical visualization. 

Lipsa \etal~\cite{LipsaLarameeCox2012}, on the other hand, took a broader view in surveying visualization for the physical sciences, which included astronomy and physics. 
For astronomy, the papers are classified based on the visualization challenges they tackle: multi-field visualization, feature detection, modeling and simulation, scalability, error/uncertainty visualization, and global/local visualization. 

Hassan {\etal} excelled at classifying papers based on data types and considering how different types of data could be visualized. 
Lipsa {\etal} focused more on visualization techniques.  
A data-centered classification is useful for researchers to explore diverse ways to visualize their data, whereas a technique-centered classification can be useful for researchers who want to explore their data using a particular visualization technique. 
Our survey aims to strike a balance between these two classification schemes and classifies the papers primarily based on data tasks and secondarily on visualization techniques, thus allowing researchers to explore how they can best visualize the data at hand based on the analysis tasks they have in mind. 
We also utilize tertiary categories in topical areas in astronomy for cross-references for the astronomy audience.
Namely, we classify papers based on extragalactic, galactic, planetary, and solar astronomy. 
We further label each paper as dealing with simulated or observational astronomical data. 

To the best of our knowledge, no comprehensive survey of visualization in astronomy has been conducted since 2012. 
Advances in both astronomical data and visualization in the past decade present a need for an updated state-of-the-art report. 
In 2011, Hassan \etal~ identified six grand challenges for scientific visualization in astronomy in the era of peta-scale data.
Our survey discusses how the community has responded to these challenges in the past decade. 
The unique contribution of this survey is the cross-discipline discussion between visualization experts and astronomers via two workshops (a mini-workshop in April 2020 and an IEEE VIS workshop in October 2020), where researchers from both fields worked together in identifying progress, challenges, and opportunities in astronomical visualization. 
This survey aims to become a reference point for building connections and collaborations between two communities: data-rich, but technique-hungry, astronomers and data-hungry, but technique-rich, visualization experts. We further discuss datasets in astronomy in need of new approaches and
methodologies, visualization techniques that have not been applied to astronomical datasets, and visualization techniques that can enhance the educational value of astronomical datasets. 

In \autoref{sec:categories} we define our primary, secondary, and tertiary categories of approaches based on data analysis task, visualization technique, and topical area in astronomy, respectively. In \autoref{sec:data-wrangling}, \ref{sec:data-exploration}, \ref{sec:feature-identification}, \ref{sec:object-reconstruction}, and \ref{sec:data-accessibility} we discuss and group papers based on the primary categories of data wrangling, data exploration, feature identification, object reconstruction, education and outreach, respectively. In \autoref{sec:challenges} we identify challenges and opportunities for astronomy visualization. 
We provide a navigation tool of the surveyed papers in \autoref{sec:navigation-tool}, and we summarize our conclusions in \autoref{sec:conclusions}.

To make the survey results more accessible and actionable to the research community, all papers surveyed, including associated metadata, can be explored online with a visual literature browser ({\small \url{https://tdavislab.github.io/astrovis-survis}}) developed with the SurVis~\cite{BeckKochWeiskopf2015} framework.

%% file: sec-categories.tex
\section{Literature Research Procedure and Classification} 
\label{sec:categories}
\begin{figure}[t]
\centering
\includegraphics[width=1.0\columnwidth]{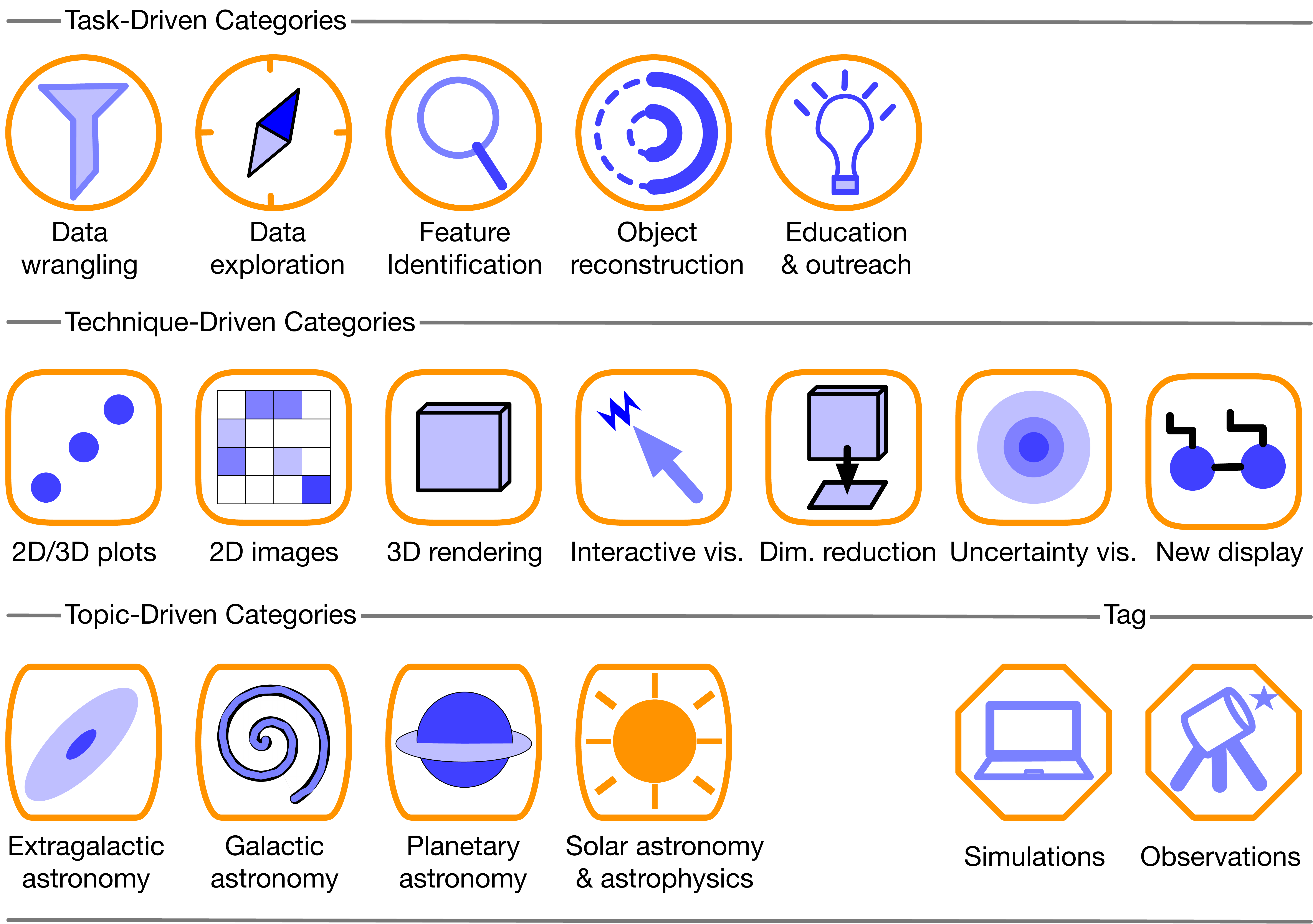}
\caption{A typology of task-driven (primary), technique-driven (secondary), and topic-driven (tertiary) categories used in this survey paper.}
\vspace{-3mm}
\label{fig:typology}
\end{figure}

We reviewed representative papers over the past 10 years (between 2010 and 2020) in the fields of astronomy and visualization that contain strong visualization components for astronomical data. 
The annotation of each paper was guided primarily by a set of \textit{data analysis tasks}; secondarily by a set of \textit{visualization techniques}; and finally by a set of \textit{topical areas in astronomy}.
We view these three categories as being on equal footing and not necessarily hierarchical. Instead, they are considered as \emph{orthogonal} dimensions and provide complementary viewpoints. 
We organize the literature according to these three categories to provide a means of navigation from task-driven, technique-driven, and topic-driven  perspectives.

The literature surveyed spans venues in visualization such as \textit{IEEE Transactions on Visualization and Computer Graphics}, \textit{Computer Graphics Forum}, and \textit{IEEE Computer Graphics and Applications}; and astronomy such as \textit{Astrophysical Journal} and \textit{Astrophysical Journal Letters}, \textit{Monthly Notices of the Royal Astronomical Society}, \textit{Astronomy and Computing}, \textit{.Astronomy} (Dot Astronomy), \textit{ADASS} Conference Series, \textit{PASP} (Publications of the Astronomical Society of the Pacific), \textit{Research Notes of the AAS}.  
We also discuss data types that include simulation data and observation data, with the latter encompassing both image data and tabular data. 
\autoref{fig:typology} shows a typology of primary, secondary, and tertiary categories used in this survey.

\subsection{Task-Driven Categories: Data Analysis Tasks}
\label{sec:primary}
Our literature review allowed us to identify five primary categories of approaches based on data analysis tasks:
\begin{itemize}
\item \img{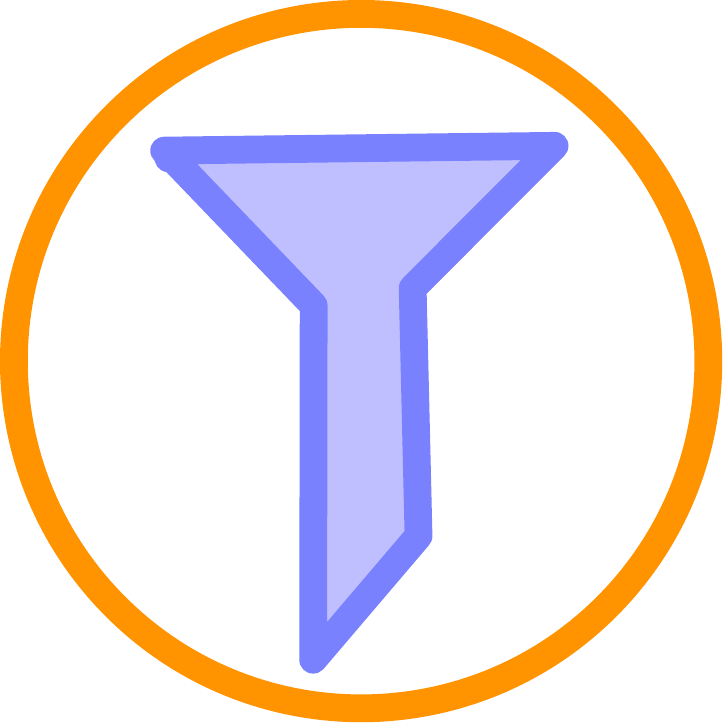} \textbf{Data wrangling}, which transforms astronomy data into formats that are appropriate for general purpose visualization tools; 
\item  \img{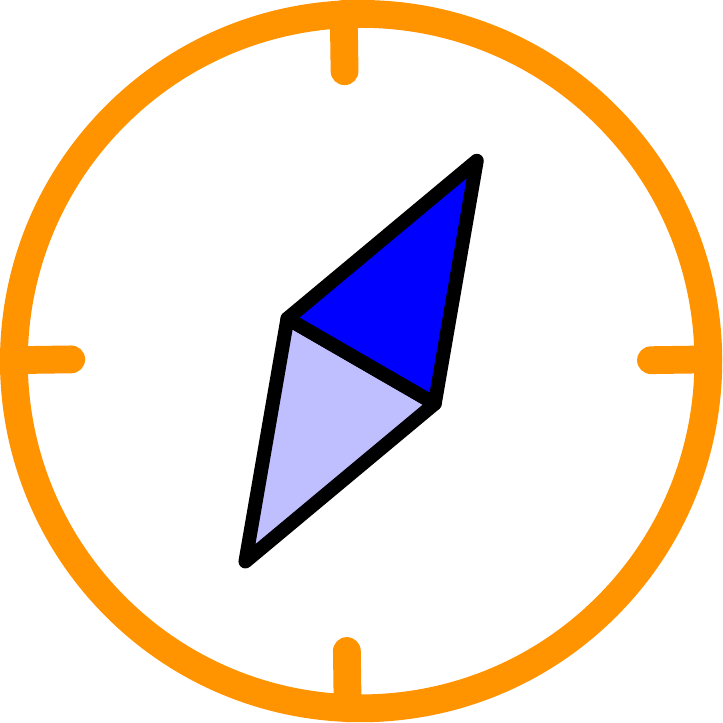} \textbf{Data exploration}, where users explore a dataset in an unstructured way to discover patterns of interest; 
\item  \img{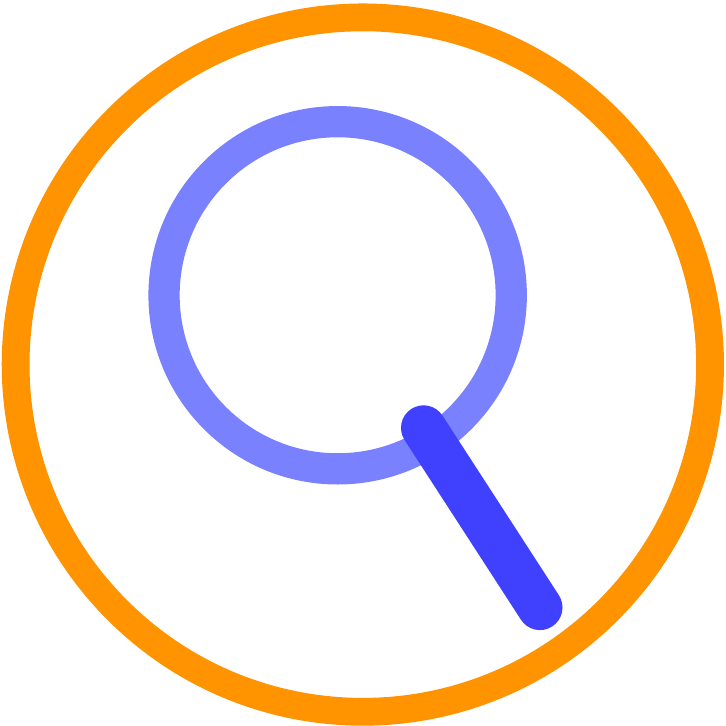} \textbf{Feature identification}, which visually guides the identification and extraction of features of interest; 
\item \img{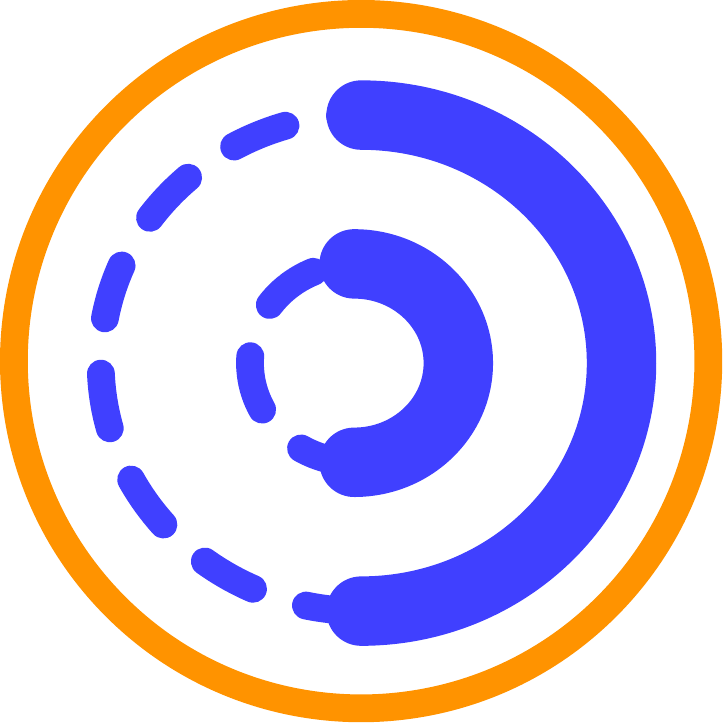} \textbf{Object reconstruction}, which provides an informative visual representation of an astronomical object; 
\item \img{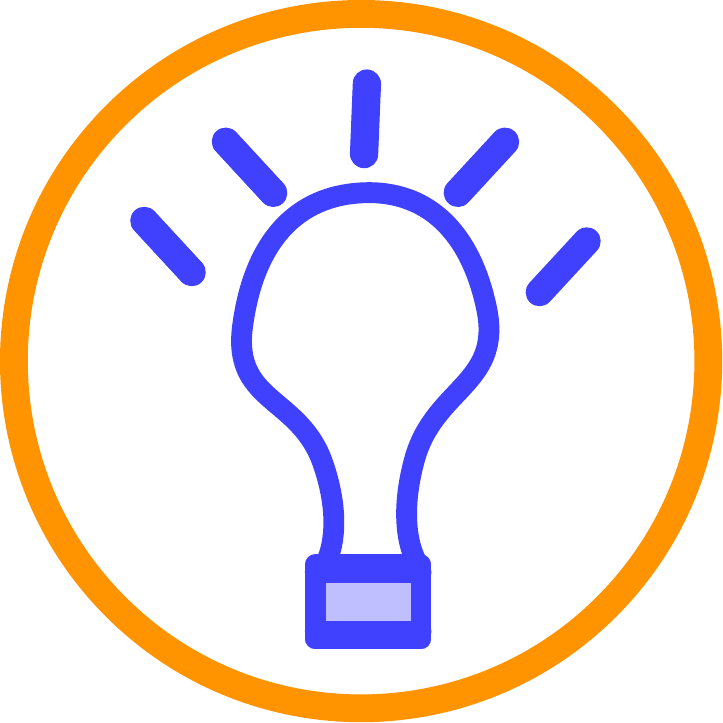} \textbf{Education and outreach}, where astronomical data or data products are made accessible to the general public. 
\end{itemize}

In an on-going paradigm shift in scientific outreach, technological advances are enabling data-driven and interactive exploration of astronomical data in museums and science centers. 
Hence, we include ``education and outreach'' as a data analysis category.   
The word ``feature'' generally means a measurable piece of data that can be used for analysis, whereas the word ``object'' may be considered as a ``feature'' with sharp and/or discontinuous contrast in a dimension of scientific interest.  
Whether a specific aspect of a dataset is considered an ``object'' or a ``feature'' depends on the scientific question at hand. 
We separate object reconstruction from feature identification to be compatible with the literature, but we envision a future in which these entities are recognized as a continuum.

\subsection{Technique-Driven Categories: Visualization Techniques} 
\label{sec:secondary}

Our secondary categories of approaches are based on visualization techniques employed for astronomical data: 
\begin{itemize}
     \item \img{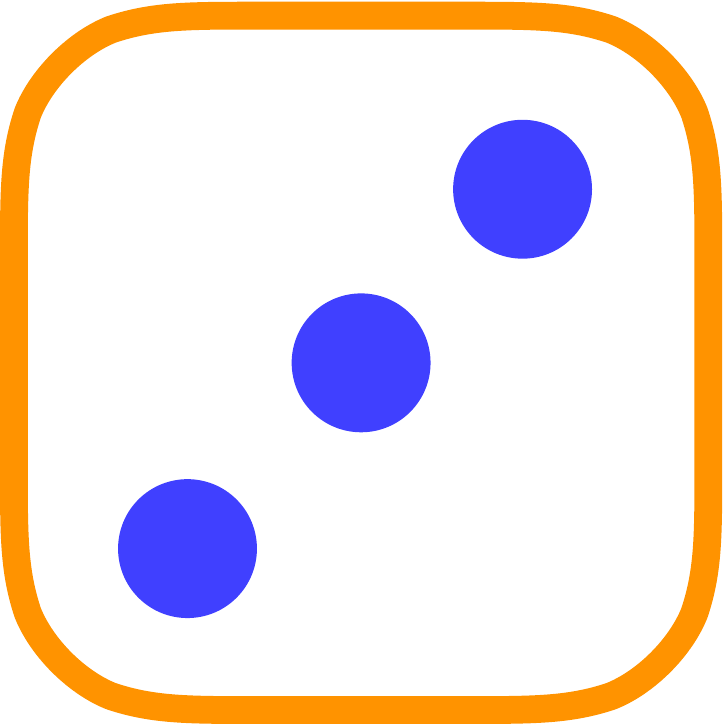} \textbf{2D/3D plots} that encompass classic 2D/3D plots such as histograms, scatter plots, pie chars, pie, bar, and line plots; 
     \item  \img{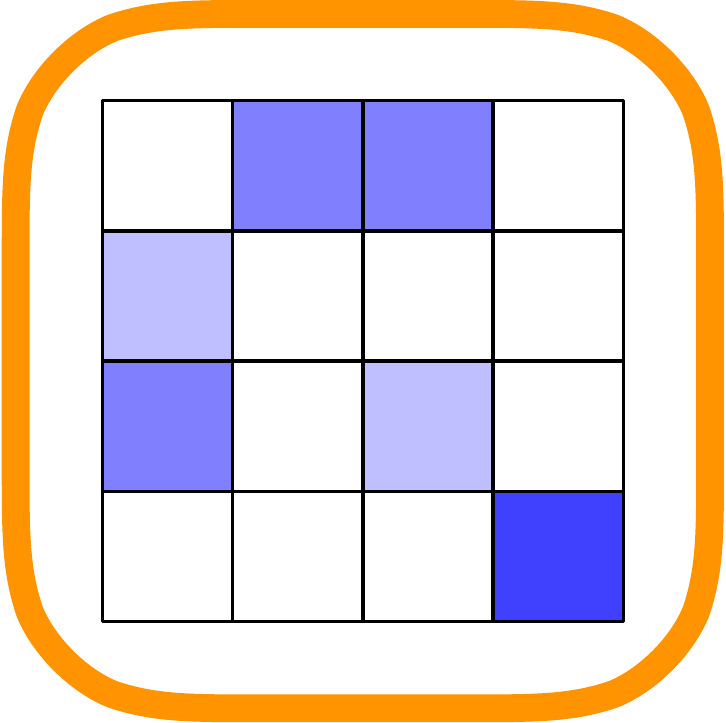} \textbf{2D images} that utilize image processing techniques to generate images of astronomy data; 
    \item \img{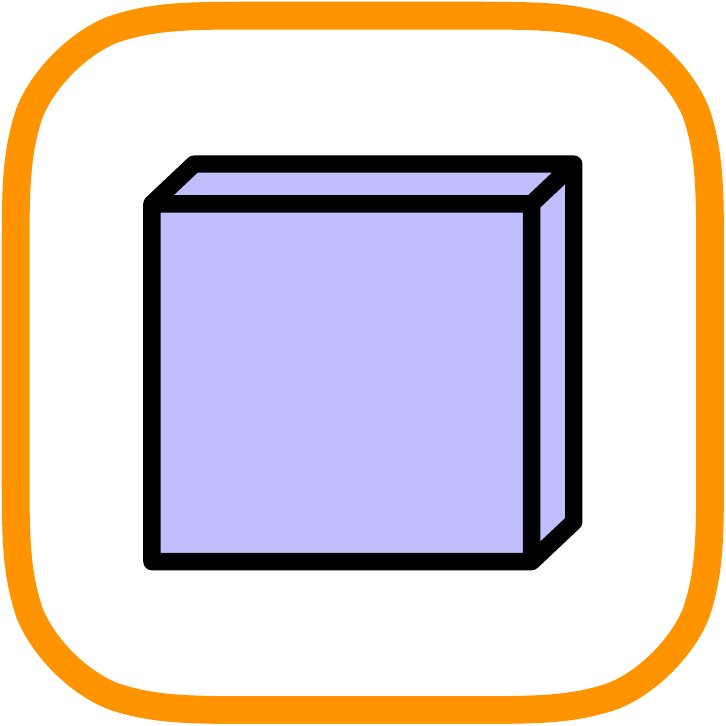} \textbf{3D rendering} that generates representations of 3D volumetric data of interest;    
     \item \img{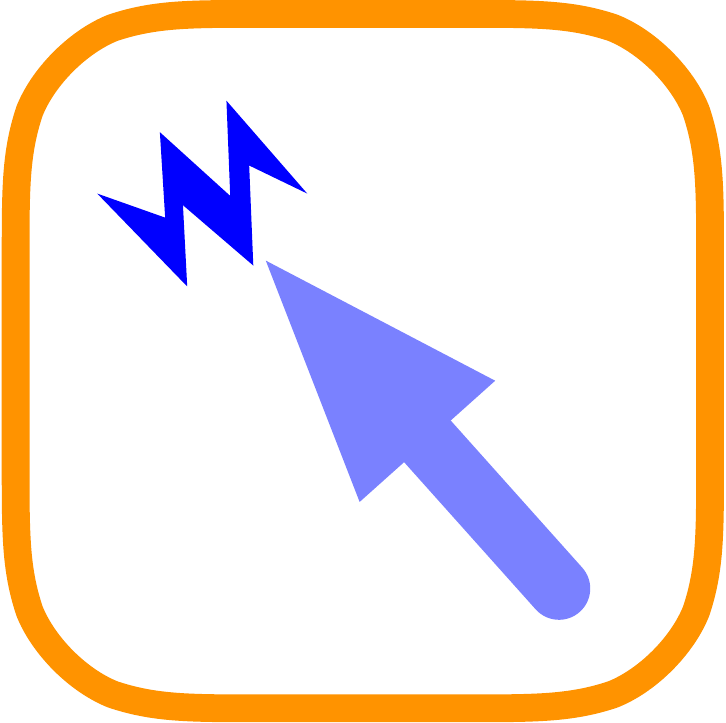} \textbf{Interactive visualization} that includes techniques such as linked views, detail on demand, visual filtering, and querying; 
    \item \img{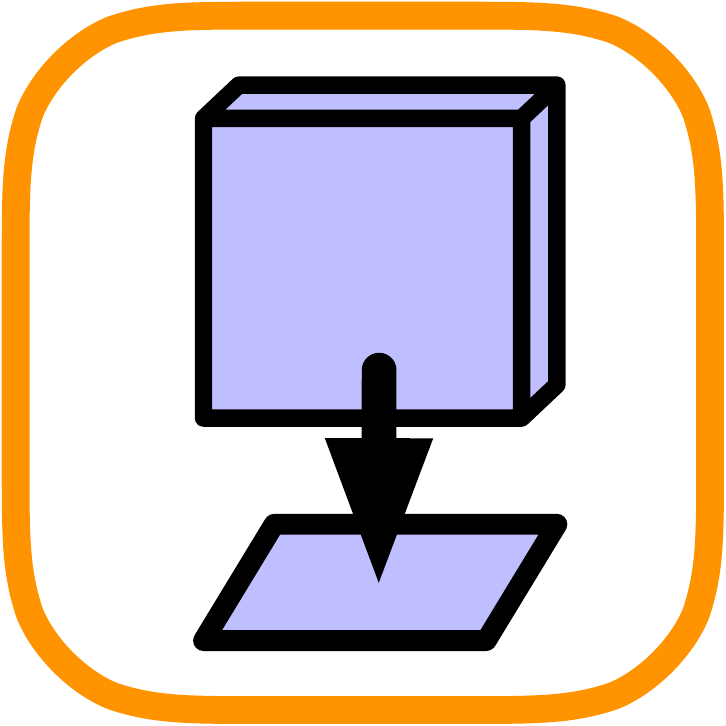} \textbf{Dimensionality reduction} that transforms data from a high-dimensional into a property-preserving low-dimensional space as part of the visualization pipeline;  
    \item \img{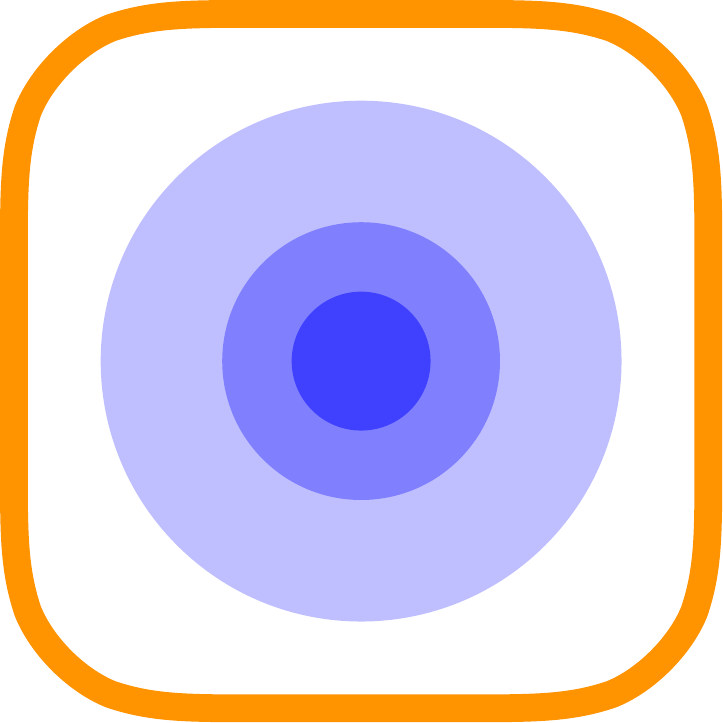} \textbf{Uncertainty visualization} that improves our ability to reason  about the data by communicating  their certainties that arise due to  randomness  in  data  acquisition and processing; 
    \item \img{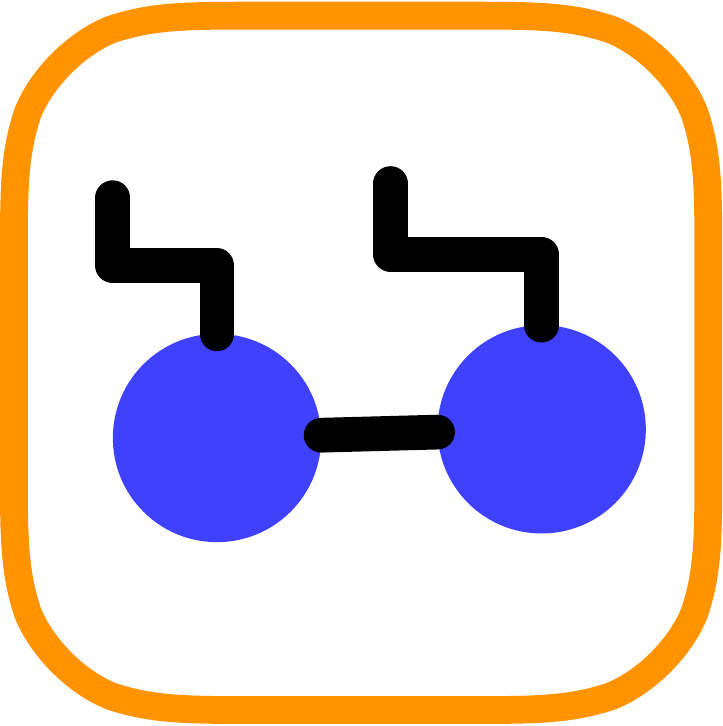} \textbf{New display platforms} that communicate data via techniques such as data physicalization and virtual reality.
\end{itemize}

Although dimensionality reduction can be used as a purely data analysis strategy for noise reduction, clustering, or downstream analysis, it also serves as an integrated part of the visualization pipeline to facilitate data exploration and understanding. 
In this survey, we focus on the use of dimensionally reduction in the context of visualization.  
Dimensionality reduction and clustering may be both considered as data preprocessing techniques, but we choose to exclude clustering as a category as it is a generic class of techniques implicitly implemented within many toolboxes and does not typically represent a main innovation of the surveyed research.

We highlight the new display platforms as a category based on our experiences and workshops held among a growing ``visualization in astrophysics'' community. We believe there is a strong motivation for this research direction as the community as a whole is ready for the next stage of scientific discovery and science communications enabled by new displays.

We also acknowledge that there are additional ways to think about categories based on visualization techniques. For instance, scalable, multi-field, comparative, and time-dependent visualization are all categories mentioned in the 2012 survey of Lipsa {\etal} 
However, as technology has evolved over the past decade, certain visualization techniques (e.g., scalable and comparative   visualization) have become commonplace and thus lack specificity. 
Time-dependent visualization (\autoref{sec:time-series}), in particular, the interplay between spatial and temporal dimensions, will be crucial as more time series astronomy data become available in the near future.  
In this survey, we choose specific visualization techniques that capture the state of the art and lead to informative categorization.

\subsection{Topic-Driven Categories: Topical Areas in Astronomy} 
\label{sec:tertiary}

Our tertiary categories are based upon the list of topics from the National Science Foundation (NSF) Astronomy \& Astrophysics directorate. 
These categories are used as a cross-reference for an astrophysics audience. 
We also investigated a curated list of research topics in astronomy and astrophysics provided by the American Astronomical Society (AAS) ({\small \url{https://aas.org/meetings/aas237/abstracts}}). 
We decided to work with the coarser classification from NSF since the AAS list is overly refined and specialized for the purposes of this survey. 
Our tertiary categories are:
\begin{itemize}
    \item \img{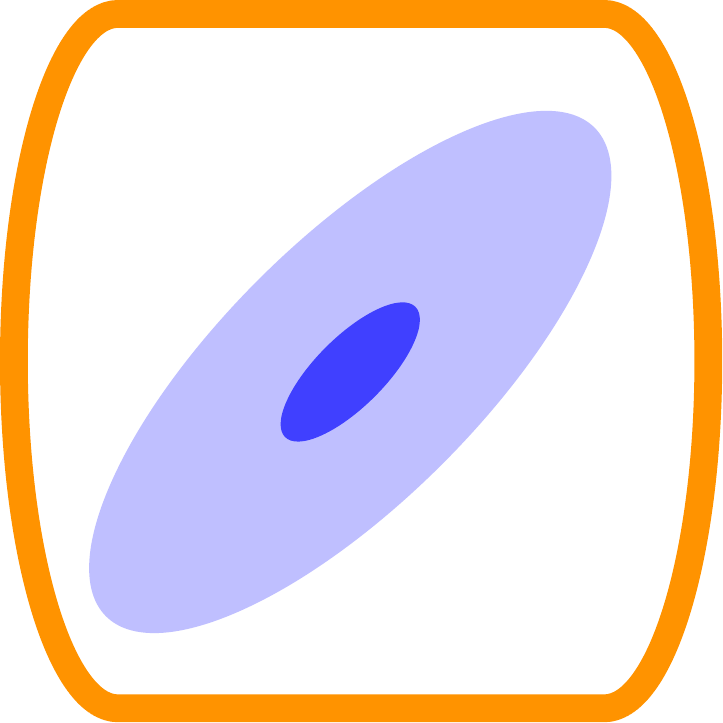} \textbf{Extragalactic astronomy}
    \item \img{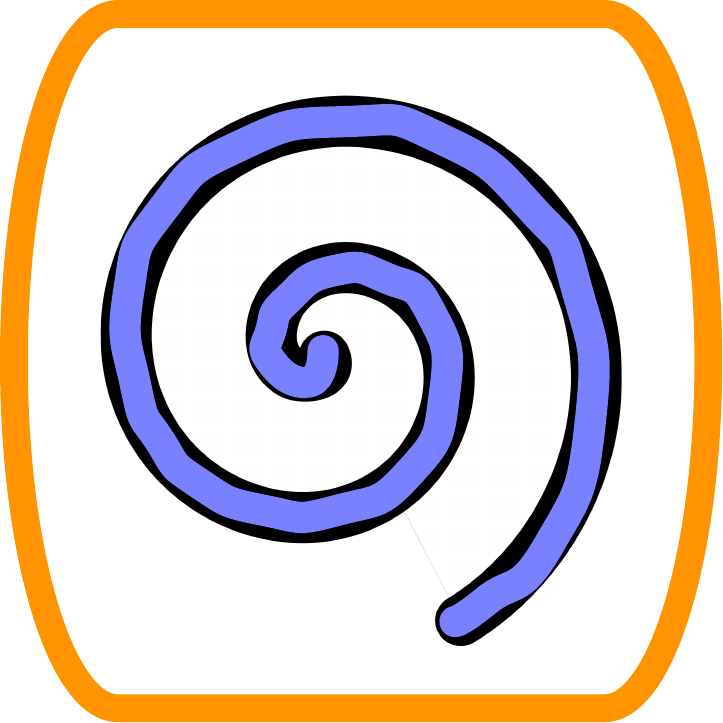} \textbf{Galactic astronomy}
    \item \img{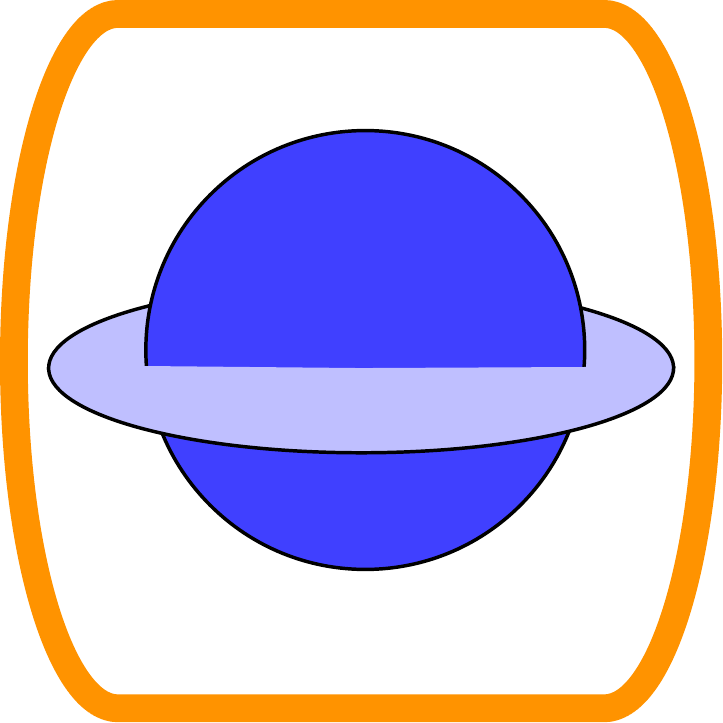} \textbf{Planetary astronomy}
    \item \img{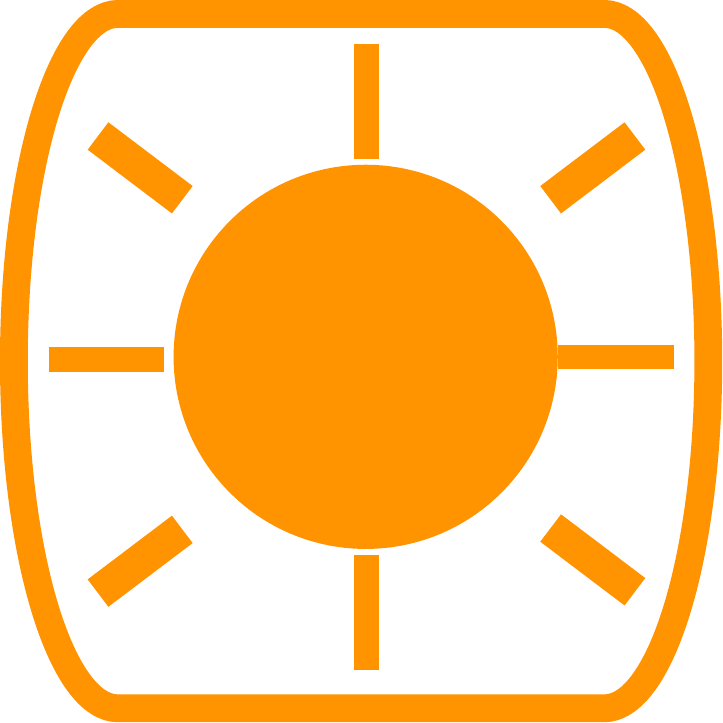} \textbf{Solar astronomy and astrophysics}
\end{itemize}
In addition, we have labeled each paper with two tags:
\begin{itemize}
    \item \img{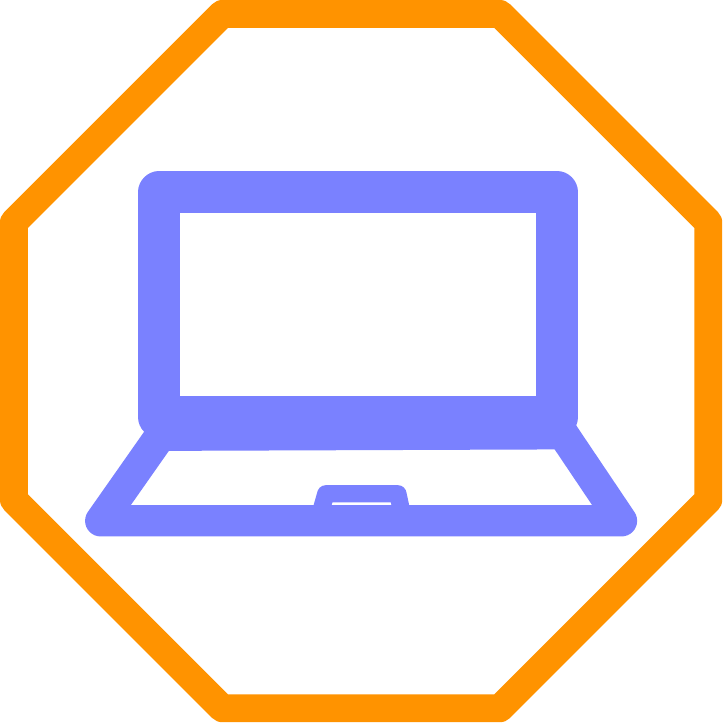} \textbf{Simulated astronomical data} 
    \item \img{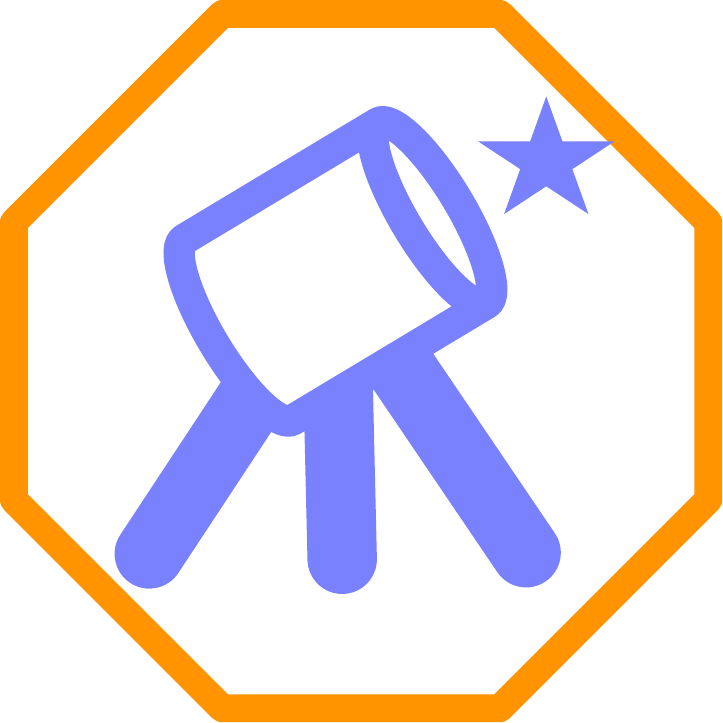} \textbf{Observational astronomical data} 
\end{itemize}

For readers unfamiliar with certain terminology in astronomy or astrophysics, we recommend the astrophysics glossaries from the National Aeronautics and Space Administration (NASA) ({\small\url{https://science.nasa.gov/glossary/}}) or the LEVEL5 Knowledge Base on Extragalactic Astronomy and Cosmology ({\small\url{https://ned.ipac.caltech.edu/level5/}}). 
Meanwhile, we try our best to describe relevant terms the first time they are introduced in the survey. 
We would like to point out that even though certain terminology may appear to be rather straightforward, in some cases, definitions vary within the field, and thus some attention must be given to the precise work in question. For example, the term \emph{halo} typically refers to overdensities in the dark matter but the exact boundary of a \emph{halo} in a specific calculation may vary (e.g.,~\cite{KnebePearceLux2013}).

\para{Overview.} 
One of the main contributions of this paper is the classification of existing works, which are summarized in~\autoref{sec:data-wrangling} to~\autoref{sec:data-accessibility}.
The methods of classification reflect the authors' experience that comes from several meetings with experts in the astronomical visualization community.
For each surveyed paper, we use our best judgment to infer its primary and secondary 
categories, although such classification may not be perfect; many papers span multiple categories. 
The best way to explore our classification is to use the table for each section (from~\autoref{table:data-wrangle} to~\autoref{table:data-access}) as a roadmap.  

We acknowledge that many effective tools were actively used in astronomy research published prior to 2010. 
We emphasize that this paper is \emph{not} a comprehensive catalog of all tools used in astronomy, nor does it include pre-2010 works. 
Rather, this paper surveys active areas of visualization research in astronomy as identified in publications in the last decade (2010--2021). 
We also note that whereas ``astronomy'' has previously meant the cataloging of the positions and motions of objects in the sky, and ``astrophysics'' the physical understanding of those objects, in this survey, we consider ``astronomy'' and ``astrophysics'' to be synonymous since few astronomers make the above distinction.  In fact, by ``visualization in astrophysics'', we consider the intersection of  visualization with astronomy, astrophysics, and space exploration.

%% file: sec-data-wrangling.tex
\begin{table*}[!ht]
	\centering
\resizebox{1.0\textwidth}{!}{	
	\begin{tabular}
	{|>{\centering\arraybackslash}m{0.1\textwidth}
	||>{\centering\arraybackslash}m{0.06\textwidth}
	>{\centering\arraybackslash}m{0.06\textwidth}
	>{\centering\arraybackslash}m{0.06\textwidth}
	>{\centering\arraybackslash}m{0.06\textwidth}
	>{\centering\arraybackslash}m{0.06\textwidth}
	>{\centering\arraybackslash}m{0.06\textwidth}
	>{\centering\arraybackslash}m{0.06\textwidth}
	||>{\centering\arraybackslash}m{0.06\textwidth}
	>{\centering\arraybackslash}m{0.06\textwidth}
	>{\centering\arraybackslash}m{0.06\textwidth}
	>{\centering\arraybackslash}m{0.06\textwidth}
	|>{\centering\arraybackslash}m{0.06\textwidth}
	>{\centering\arraybackslash}m{0.06\textwidth}|}

\mc{\imglarger{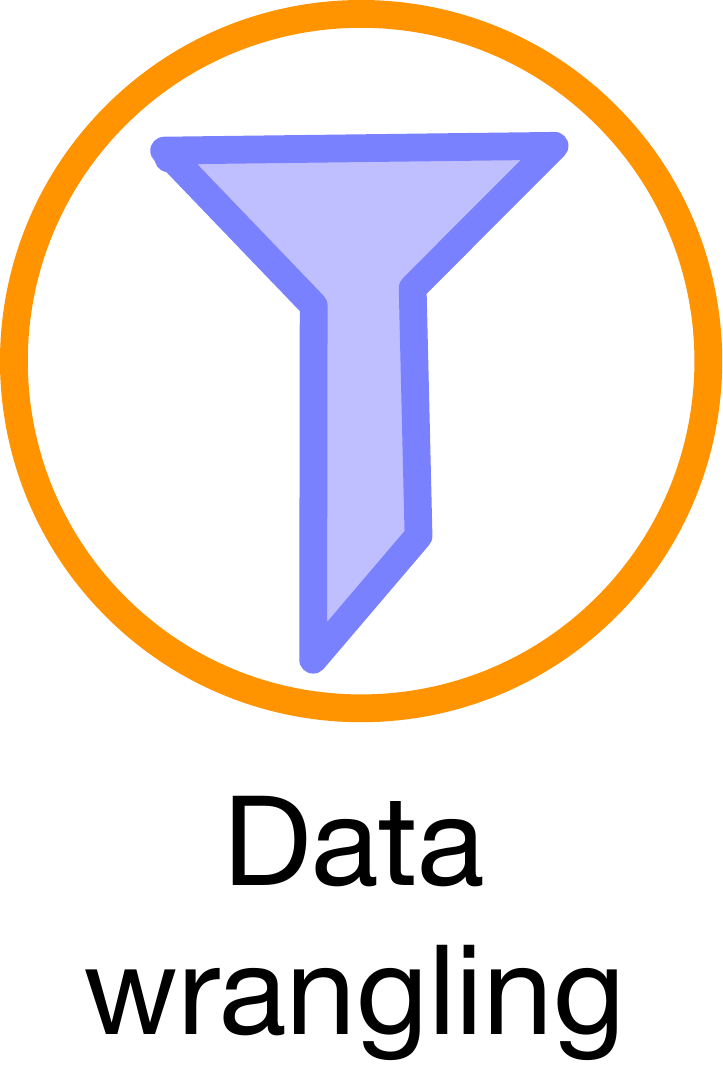}}
& \mc{\imglarger{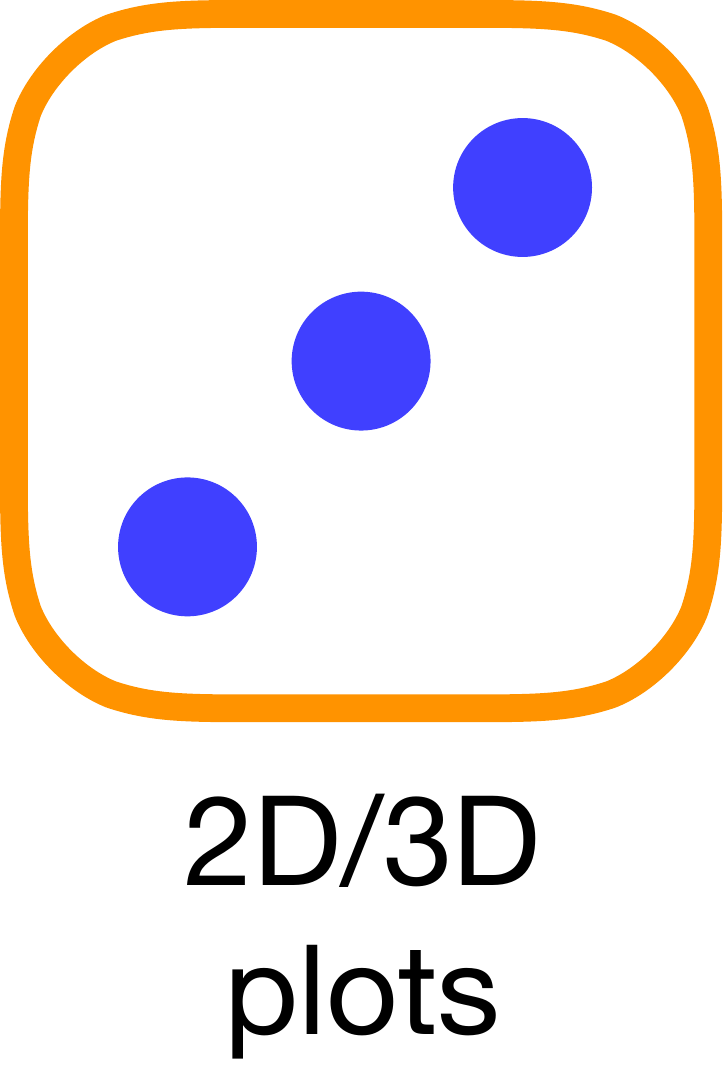}}
& \mc{\imglarger{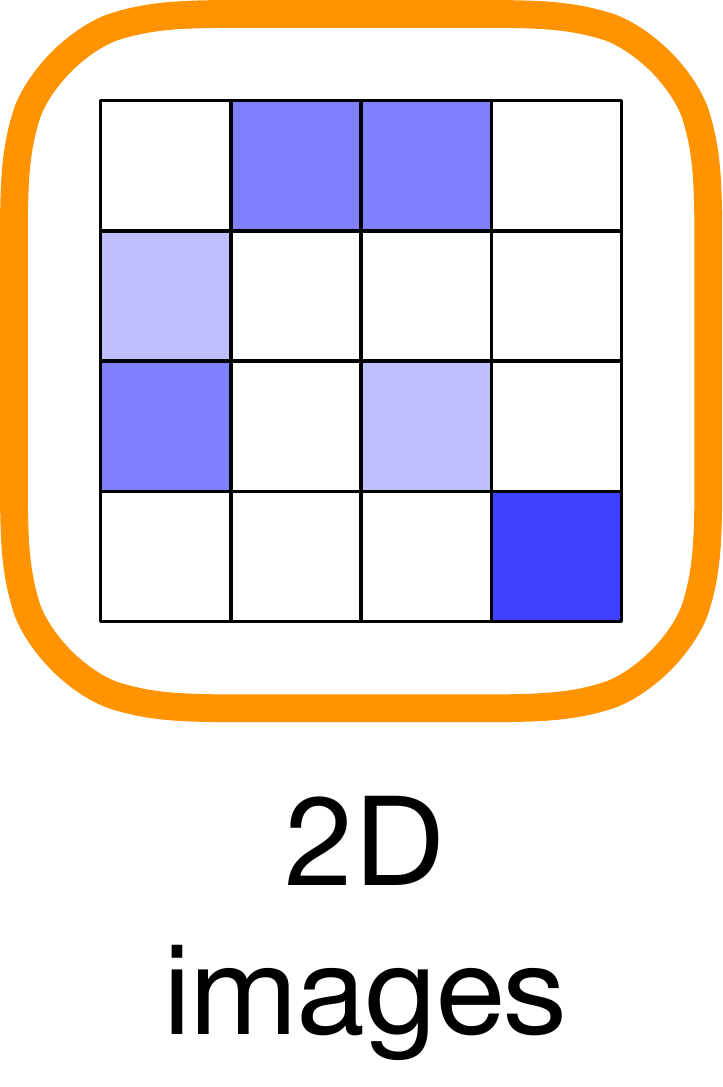}}
& \mc{\imglarger{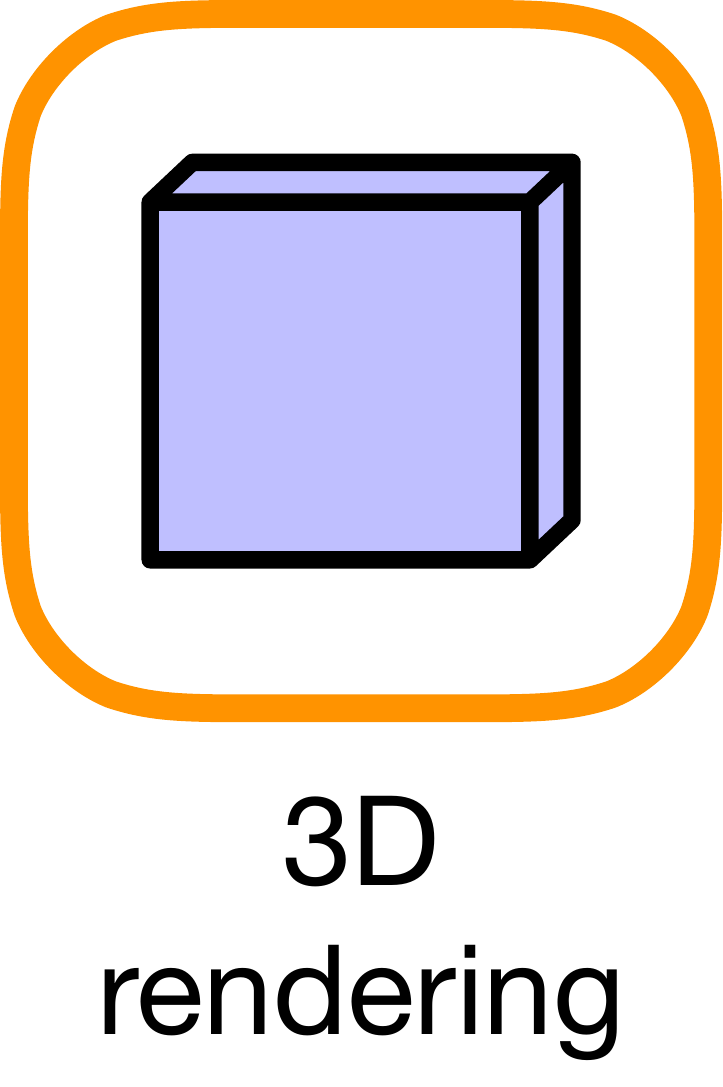}}
& \mc{\imglarger{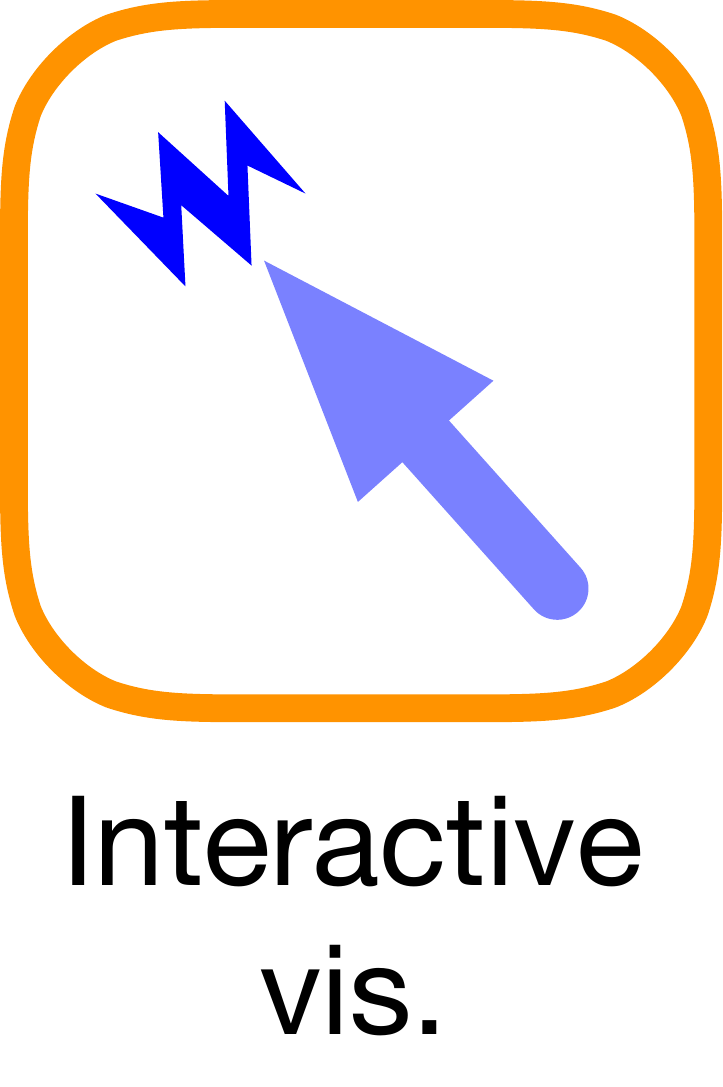}}
& \mc{\imglarger{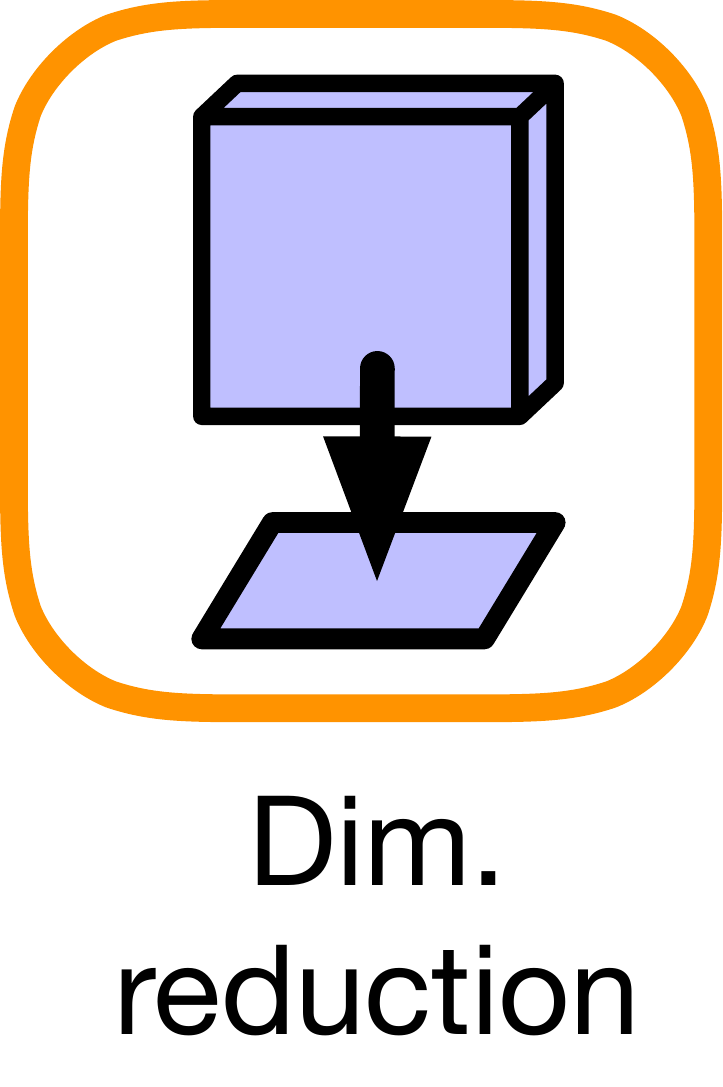}}
& \mc{\imglarger{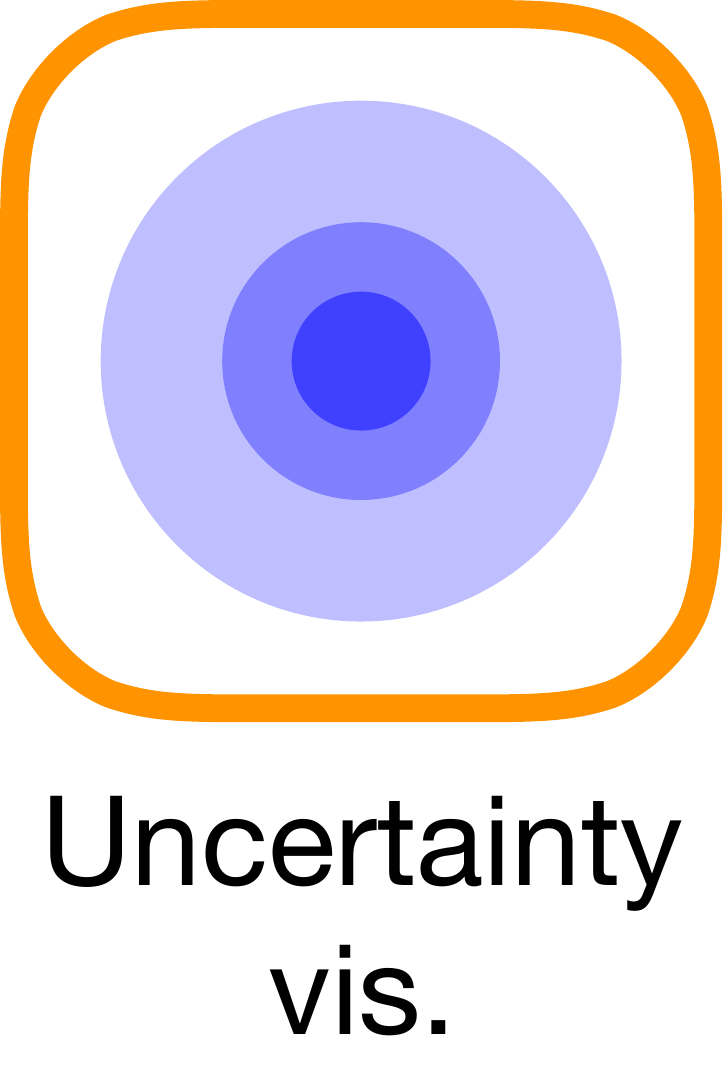}}
& \mc{\imglarger{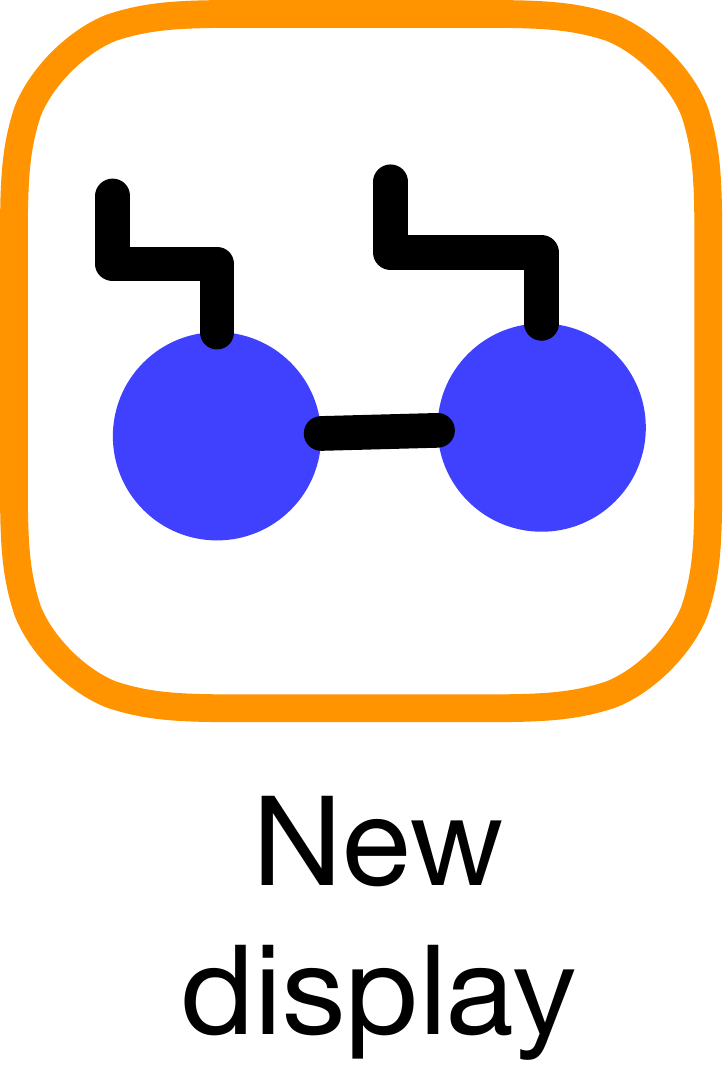}}
& \mc{\imglarger{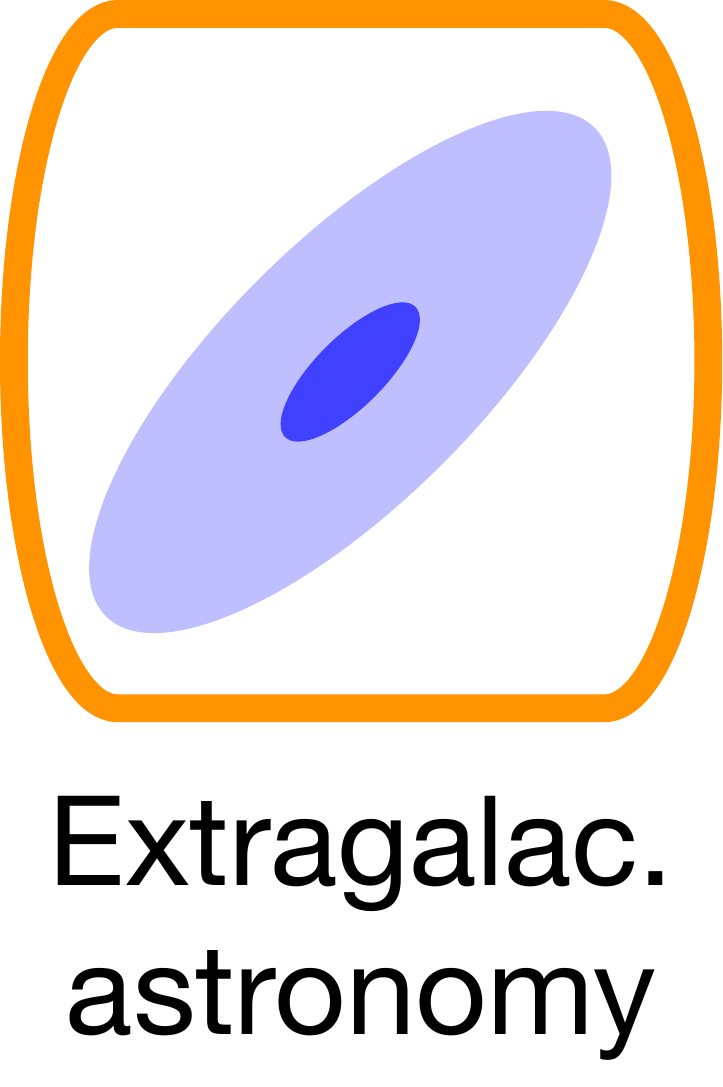}}
& \mc{\imglarger{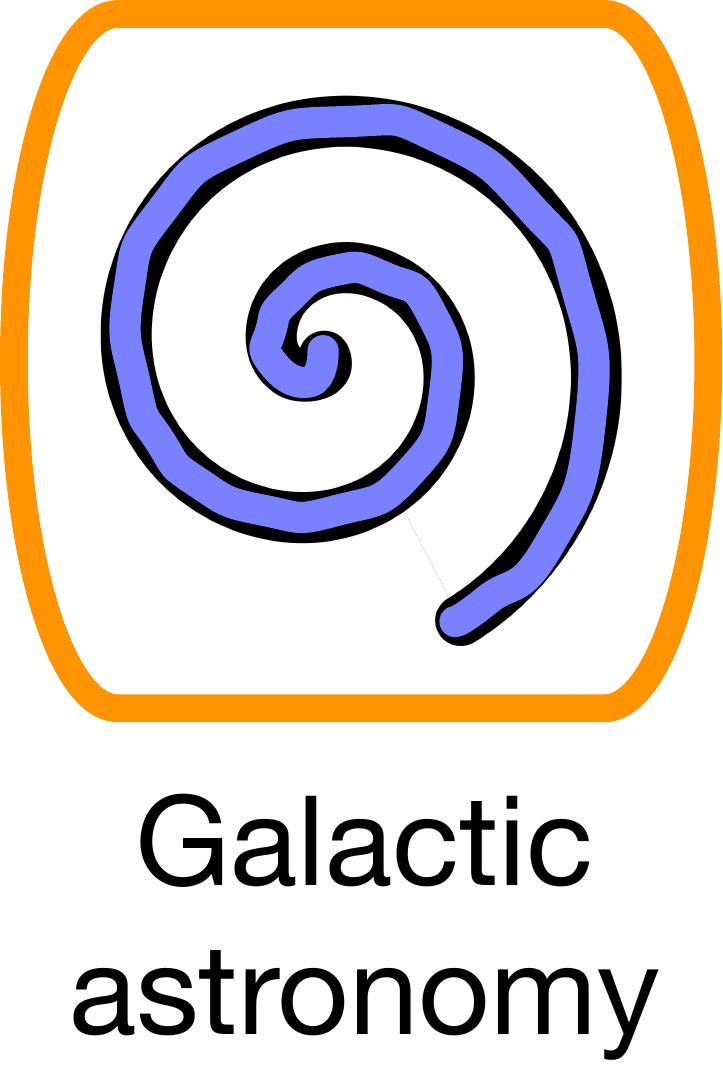}}
& \mc{\imglarger{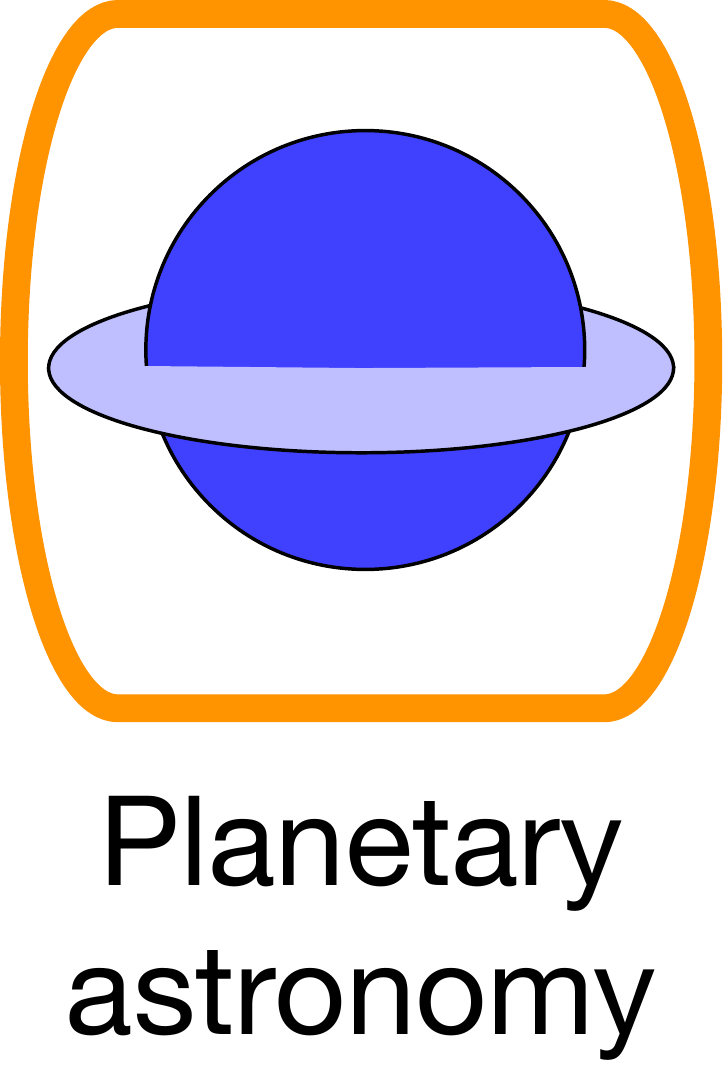}}
& \mc{\imglarger{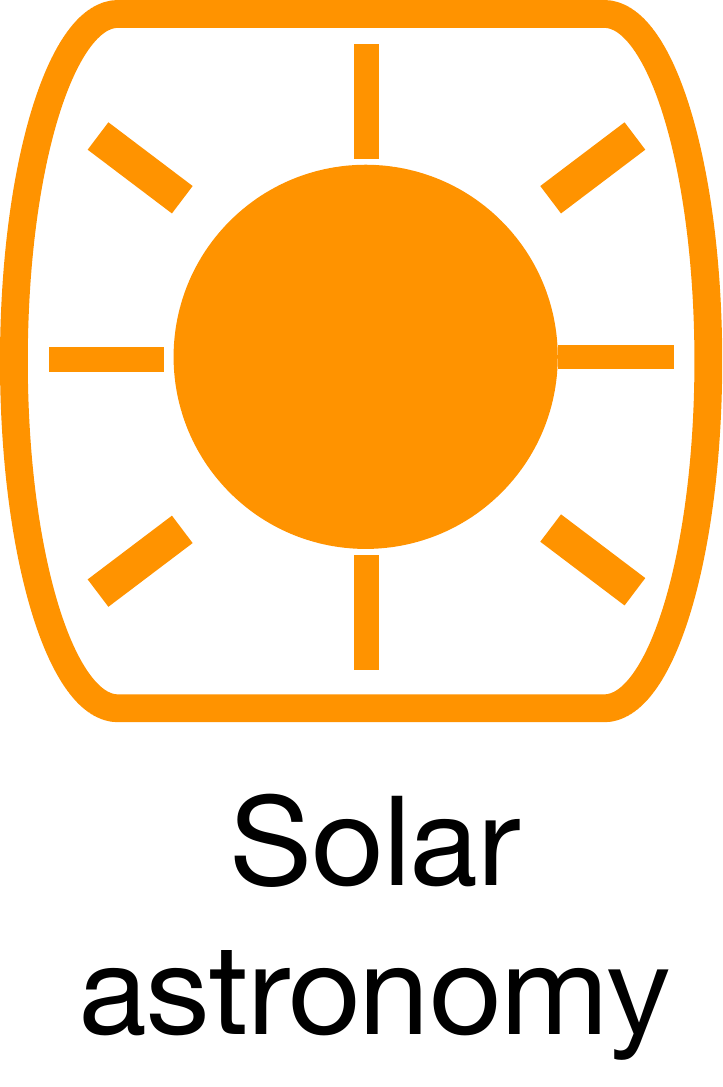}}
& \mc{\imglarger{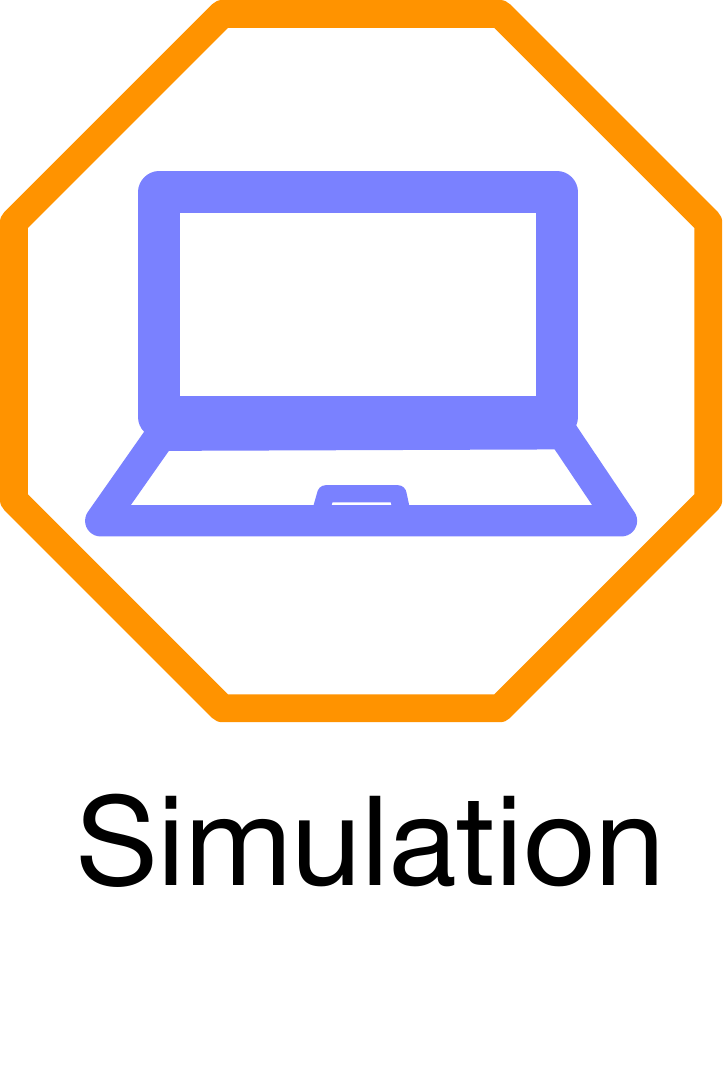}}
& \mc{\imglarger{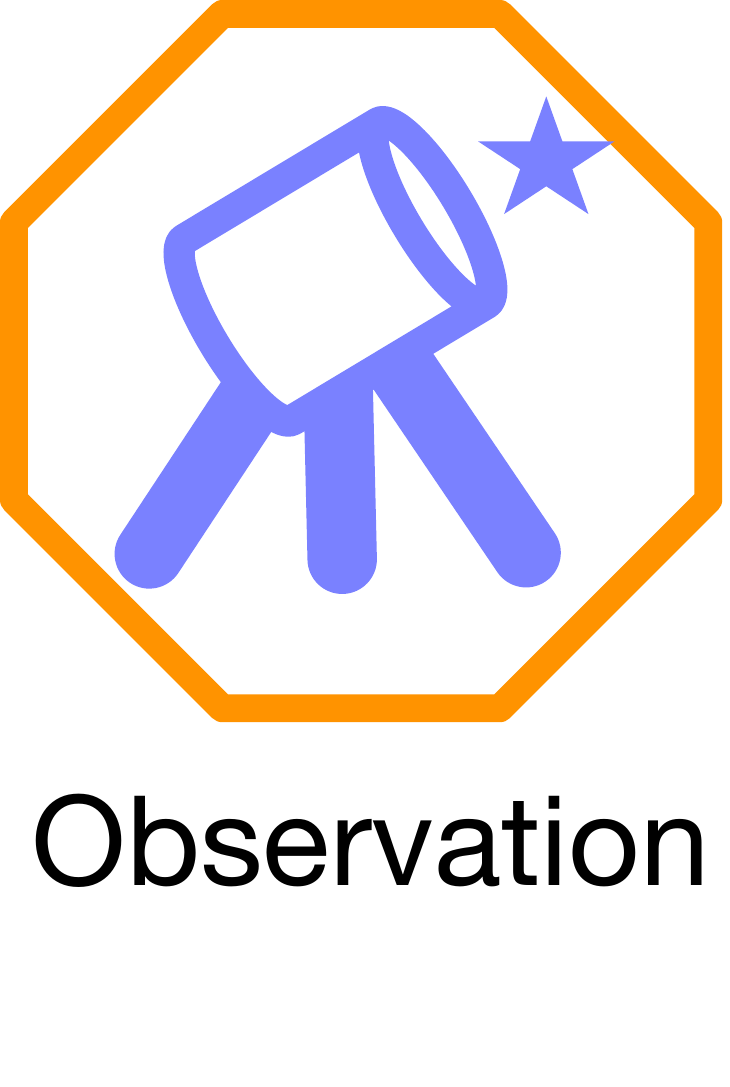}}
\\ \hline

{\cite{Kent2013}}
&           &           & $\bullet$ &           &           &           &           & $\bullet$ &           & $\bullet$ &           & $\bullet$ & $\bullet$ \\ \hline
{\cite{Taylor2015}\cite{Taylor2017}}
&           &           & $\bullet$ & $\bullet$ &           &           &           & $\bullet$ &           & $\bullet$ &           & $\bullet$ & $\bullet$ \\ \hline
{\cite{Garate2017}} 
&           &           & $\bullet$ & $\bullet$ &           &           &           & $\bullet$ &           & $\bullet$ &           & $\bullet$ &           \\ \hline
{\cite{Kent2017}}
&           &           & $\bullet$ &           &           &           & $\bullet$ &           & $\bullet$ & $\bullet$ &           & $\bullet$ & $\bullet$ \\ \hline
{\cite{Naiman2016}}
&           &           & $\bullet$ & $\bullet$ &           &           &           & $\bullet$ & $\bullet$ & $\bullet$ &           & $\bullet$ &           \\ \hline
{\cite{NaimanBorkiewiczChristensen2017}}
&           &           & $\bullet$ & $\bullet$ &           &           &           & $\bullet$ & $\bullet$ & $\bullet$ &           &           & $\bullet$ \\ \hline
{\cite{BorkiewiczNaimanLai2019}}
& $\bullet$ &           & $\bullet$ & $\bullet$ &           &           &           & $\bullet$ &           &           & $\bullet$ & $\bullet$ &           \\ \hline
\cite{WoodringHeitmannAhrens2011}
& $\bullet$ &           & $\bullet$ & $\bullet$ &           &           &           & $\bullet$ &           &           &           & $\bullet$ &           \\ \hline
{\cite{BerrimanGood2017}}
&           & $\bullet$ &           &           & $\bullet$ &           &           & $\bullet$ & $\bullet$ &           &           &           & $\bullet$ \\ \hline
{\cite{VogtOwenVerdesMontenegro2016}\cite{VogtSeitenzahlDopita2017}}
&           &           & $\bullet$ & $\bullet$ &           &           &           & $\bullet$ & $\bullet$ &           &           &           & $\bullet$ \\ \hline
{\cite{ComriePiriskaSimmonds2019}}
& $\bullet$ &           &           & $\bullet$ &           &           &           & $\bullet$ & $\bullet$ &           &           &           & $\bullet$ \\ \hline
\cite{EmontsRabaMoellenbrock2019}\cite{ott2020carta}
& $\bullet$ &           & $\bullet$ & $\bullet$ &           &           &           & $\bullet$ & $\bullet$ &           &           &           & $\bullet$ \\ \hline
\cite{VogtWagner2011}
&           &           &           &           &           &           & $\bullet$ & $\bullet$ & $\bullet$ &           &           & $\bullet$ &$ \bullet$ \\ \hline
\cite{VerbraeckEisemann2021}
&           &           & $\bullet$ &   $\bullet$        &           &           &           &         $\bullet$  & $\bullet$ &  &           &  & $\bullet$ \\ \hline

\end{tabular}
}	
\caption{Classifying papers under data wrangling based on secondary and tertiary categories.
Top row, from left to right: (primary category) Data wrangling; (secondary categories) 2D/3D plots, 2D images, 3D rendering, interactive visualization, dimensionality reduction, uncertainty visualization, and new display platforms; (tertiary categories) extragalactic, galactic, planetary, and solar astronomy; (tags) simulated, and observational data.
}
\label{table:data-wrangle}
\end{table*}

\section{Data Wrangling}
\label{sec:data-wrangling}

Data wrangling is the process of transforming raw data into forms that more effectively support downstream analysis~\cite{kandel2011research}. This process is an important step for astronomy visualization because raw simulation or observational data require significant wrangling into a suitable form for visualization tasks.
In this section, we categorize papers that present novel work in data wrangling for astronomy visualization. Many established tools are available for data wrangling across specific areas of astronomy, but a full survey of such tools is not within the scope of this survey. High-dimensional data abstractions such as data cubes are commonly used in astrophysical sciences and are often stored in the FITS format. Many of the papers placed in this category focus on transforming raw astrophysical data cubes into suitable data formats that can be ingested into open-source visualization tools, such as \textsf{Blender} and \textsf{Houdini}. 
Others introduce new formats that can be used to support various tools for data representation and data analysis. Authors of data wrangling papers have often made significant efforts to introduce astronomers to the visualization pipelines using these tools. We further classify these papers using our secondary categorization on visualization techniques (\autoref{sec:secondary}). 
\autoref{table:data-wrangle} presents an overview of our categorization of data wrangling papers.

\para{Using Blender to visualize astrophysics data.}
\textsf{Blender}~\cite{Blender} is an open-source 3D graphics and visualization tool that supports a wide range of modeling, animation, and rendering functionality. A range of papers have discussed its usefulness for presenting astronomy data, and described pipelines for transforming raw data into scientific visualizations. 
Kent~\cite{Kent2013} demonstrated how Blender can be used to visualize galaxy catalogs, astronomical data cubes, and particle simulations. Taylor~\cite{Taylor2015} introduced \textsf{FRELLED}, a \textsf{Python}-based FITS viewer for exploring 3D spectral line data using \textsf{Blender} that visualizes 3D volumetric data with arbitrary (non-Cartesian) coordinates~\cite{Taylor2017} and is designed for real time and interactive content. 
Using this viewer, astronomers are able to speed up visual cataloging by as much as $50\times$. G\'{a}rate~\cite{Garate2017} described the process of importing simulation outputs from astrophysical hydrodynamic experiments into \textsf{Blender} using the voxel data format.  In order to facilitate immersive data exploration, Kent~\cite{Kent2017} presented a technique for creating 360\textdegree\ spherical panoramas using \textsf{Blender} and \textsf{Google Spatial Media module}. The method supports static spherical panoramas, single pass fly-throughs, and orbit flyovers on browsers or mobile operating systems.

\begin{figure}[b]
    \centering
    \includegraphics[width=0.98\columnwidth]{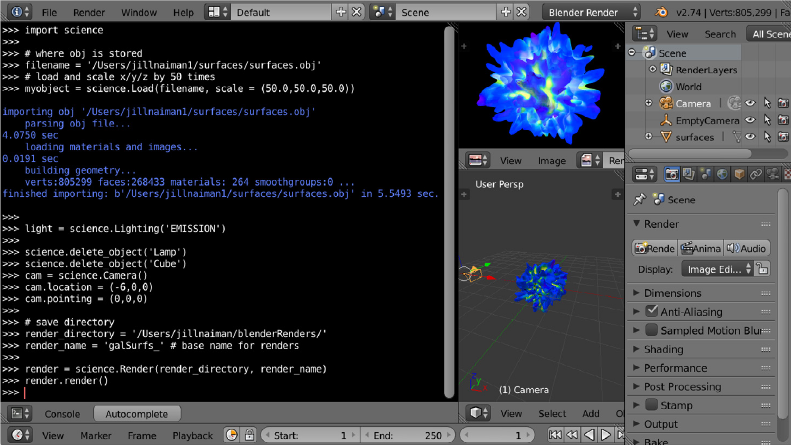}
    \caption{A screenshot from a visualization session in \textsf{AstroBlend}, a \textsf{Blender}-based 3D rendering and analysis tool. Image reproduced from Naiman et al.~\cite{Naiman2016}.}
    \label{fig:astroblend}
\end{figure}

\textsf{AstroBlend}~\cite{AstroBlend, Naiman2016} extends \textsf{Blender}, making it possible to import and display various types of astronomical data interactively, see~\autoref{fig:astroblend}. \textsf{AstroBlend} is an open-source \textsf{Python} library that utilizes \textsf{yt} -- an open-source software for analyzing and visualizing volumetric data -- for 3D data visualization (\textsf{yt} is discussed in~\autoref{sec:data-exploration}). \textsf{AstroBlend} effectively bridges the gap between ``exploratory'' and ``explanatory'' visualization, as discussed by Goodman \etal~\cite{GoodmanBorkinRobitaille2018} and Ynnerman~\etal~\cite{YnnermanLowgrenTibell2018}.

\para{Using Houdini to visualize astrophysics data.} 
In another example of adapting existing 3D graphics software, Naimen \etal~\cite{NaimanBorkiewiczChristensen2017} explored how the 3D procedural animation software \textsf{Houdini} can be used for astronomy visualization, producing high-quality volume renderings for a variety of data types. They utilized \textsf{yt} to transform astronomical data into graphics data formats for \textsf{Houdini}, which bridges the astronomical and graphics community. 
\textsf{Houdini} is a compelling alternative to other rendering software (e.g.,~\textsf{Maya} and \textsf{Blender}) for astronomy because it produces high-quality volume renderings and supports a variety of data types.

Borkiewicz \etal~\cite{BorkiewiczNaimanLai2019} presented a method for creating cinematic visualizations and time-evolving representations of astronomy data that are both educational and aesthetically pleasing. The paper also provided a detailed workflow of importing nested, multi-resolution adaptive mesh refinement data into \textsf{Houdini}.

\para{Using ParaView to visualize astrophysics data.}
\textsf{ParaView} is an open-source, general-purpose, multi-platform analysis and visualization tool for scientific datasets. 
It supports scripting (with \textsf{Python}), web-based visualization, and in situ analysis (using \textsf{Catalyst}). 
Woodring \etal~\cite{WoodringHeitmannAhrens2011} used \textsf{ParaView} to analyze and visualize large N-body cosmological simulations. \emph{N-body cosmological simulations} are simulations of large-scale structures that contain particles that interact only via gravity, in contrast to including gas, which also requires hydrodynamics.
\textsf{ParaView} provides particle readers (supporting ``cosmo'' and ``GADGET'' formats) and efficient halo finders, where a \emph{halo} is a gravitationally bound structure on galactic scales.
Together with existing visualization features, \textsf{ParaView} enables efficient and interactive visualization of large-scale cosmological simulations.   
Recent work from the IEEE VIS 2019 SciVis content~\cite{NguyenNguyenPham2019} used \textsf{ParaView} to visualize HACC (Hardware/Hybrid Accelerated Cosmology Code) cosmological simulations~\cite{HabibPopeFinkel2016}. 

\para{Data wrangling to support visualization}. Beyond the integration of visualization techniques into popular 3D software platforms, a range of projects have explored the transformation of astrophysical data into formats suitable for different forms of presentation, immersion, and analysis. Data wrangling is a perennial concern, and as new display formats are introduced or made more widely accessible, researchers investigate how best to target them. For example, predating our survey, Barnes \etal~\cite{BarnesFlukeBourke2006} introduced \textsf{S2PLOT}, a 3D plotting library for astronomy that supports dynamic geometry and time-varying datasets. \textsf{S2PLOT} has been used to construct models of planetary systems and create outputs for viewing on stereoscopic displays and in digital domes~\cite{fluke2006future}. Barnes and Flute~\cite{BarnesFluke2008} described a technique to embed interactive figures created with \textsf{S2PLOT} into Adobe PDF files to augment astronomy research papers, including 3D renderings of cosmological simulations and 3D models of astronomy instrumentation.

Some earlier approaches to data wrangling continue to be useful for more contemporary projects. The \textsf{Montage Image Mosaic Engine}~\cite{Montage} enables users to stitch a ``mosaic'' together from sets of individual FITS images, and supports a range of image manipulation functionality, such as pixel sampling, image projection/rotation, background rectification, and animation. \textsf{Montage} can be used to create sky coverage maps and animations of data cubes, and its data wrangling capabilities have been integrated into other visualization tools. For example, \textsf{mViewer}, which can be scripted using \textsf{Python}, creates multi-color JPEG and PNG representations of FITS images and provides a wide range of functionality to support various types of image overlays, such as coordinate displays, labels, and observation footprints~\cite{BerrimanGood2017}.

Vogt \etal~\cite{VogtOwenVerdesMontenegro2016} introduced the \textsf{X3D pathway} for improving access to data visualization by promoting the use of interactive 3D astrophysics diagrams based on the \textsf{X3D} format, which can be shared online or incorporated into online publications. Vogt \etal~\cite{VogtSeitenzahlDopita2017} demonstrated the potential of this ``pathway'' by interactively visualizing integral field spectrographs observed in a young supernova remnant in the Small Magellanic Cloud. First, they created an interactive diagram of a reconstructed 3D map of the O-rich ejecta and exported it to the \textsf{X3D} file format. Second, they utilized (and extended) the visualization tools provided by \textsf{X3D} to make the diagram interactive, such as the ability to toggle views, ``peel'' intensity layers to focus on particular ranges of data, and modify clip planes to slice the 3D model at certain locations or angles.

Although the most common format for distributing astronomy images is FITS\cite{wells1979fits}, Comrie \etal~\cite{ComriePiriskaSimmonds2019} suggested that the \textsf{HDF5} format~\cite{folk2011overview} is better suited for hierarchical data and for facilitating efficient visualizations of large data cubes. They identified various common visualization tasks, including the rendering of 2D slices; generating single-pixel profiles, region profiles, and statistics; and interactive panning and zooming, and introduced a HDF5 hierarchical data schema to store precomputed data to facilitate these tasks. After integrating the \textsf{HDF5} schema with the image viewer \textsf{CARTA}~\cite{ott2020carta}, they demonstrated that their schema was able to obtain up to $10^3$ speed-ups for certain tasks. For example, precomputing and storing a dataset of histograms for each channel of a Stokes cube enables \textsf{CARTA} to display the histograms for an entire data cube with minimal delay. 
\textsf{CARTA} is part of \textsf{CASA} -- the Common Astronomy Software Applications package -- a primary data processing software for radio telescopes, including the Atacama Large Millimeter/submillimeter Array (ALMA) and the Karl G. Jansky Very Large Array (VLA). 
\textsf{CASA}~\cite{Jaeger2008} supports data formats from ALMA and VLA, and is equipped with functionalities such as automatic flagging of bad data, data calibration, and image manipulation. 
It has also been used to simulate observations. 
It comes with a graphic user interfaces with viewer, plotter, logger, and table browser~\cite{Jaeger2008}. 
\textsf{CASA} has some recent developments that enhance user experience~\cite{EmontsRabaMoellenbrock2019}, including increased flexibility in \textsf{Python} and data visualization with \textsf{CARTA}. 

Vogt and Wagner advocated for the use of stereoscopy visualization, or ``stereo pairs'', to enhance the perception of depth in multi-dimensional astrophysics data~\cite{VogtWagner2011}. Their technique involves sending distinct images to each eye, and supports both parallel and cross-eyed viewing techniques. They described a straightforward method to construct stereo pairs from data cubes using \textsf{Python}, and used various examples of both observational and theoretical data to demonstrate the potential of stereoscopy for visualizing astrophysical datasets.

\begin{figure}[!h]
    \centering
    \includegraphics[width=0.5\textwidth]{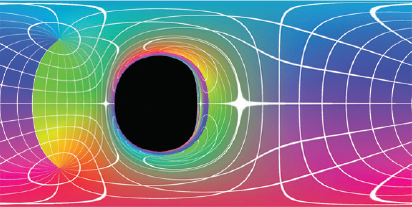}
    \caption{The projection of the distorted celestial sky caused by a Kerr black hole. Image reproduced from Verbraeck and Eisemann~\cite{VerbraeckEisemann2021}.}
    \label{fig:black-hole}
\end{figure}

Verbraeck and Eisemann~\cite{VerbraeckEisemann2021} presented a technique for interactively rendering black holes (see~\autoref{fig:black-hole}), illustrating how a black hole creates spacetime distortions in its environment due to gravitational lensing and redshift. The rendering algorithm first creates an adaptive grid that maps a uniform 360-view surrounding a virtual observer to the distorted view created by the black hole. This mapping is then used to optimize ray tracing through curved spacetime. The rendering solution also includes an interpolation technique that simulates the movement of the observer around the black hole, enabling interactive transitions between multiple sets of adaptive grids.

Data wrangling will continue to be an important component of astrophysics research as new sensors, telescopes, and other space instruments are built that generate datasets at higher resolutions and consisting of new data types. 
New data transformation methods or modifications of existing methods will be required to interoperate with existing visualization tools and to expand the accessibility of the data, making the data available in forms suitable for presentation, collaboration, interactive analysis, and public outreach.

%% file: sec-data-exploration.tex
\begin{table*}[!ht]
	\centering
\resizebox{1.0\textwidth}{!}{	
	\begin{tabular}{|>{\centering\arraybackslash}m{0.1\textwidth}
	||>{\centering\arraybackslash}m{0.06\textwidth}
	>{\centering\arraybackslash}m{0.06\textwidth}
	>{\centering\arraybackslash}m{0.06\textwidth}
	>{\centering\arraybackslash}m{0.06\textwidth}
	>{\centering\arraybackslash}m{0.06\textwidth}
	>{\centering\arraybackslash}m{0.06\textwidth}
	>{\centering\arraybackslash}m{0.06\textwidth}
	||>{\centering\arraybackslash}m{0.06\textwidth}
	>{\centering\arraybackslash}m{0.06\textwidth}
	>{\centering\arraybackslash}m{0.06\textwidth}
	>{\centering\arraybackslash}m{0.06\textwidth}
	|>{\centering\arraybackslash}m{0.06\textwidth}
	>{\centering\arraybackslash}m{0.06\textwidth}|}

\mc{\imglarger{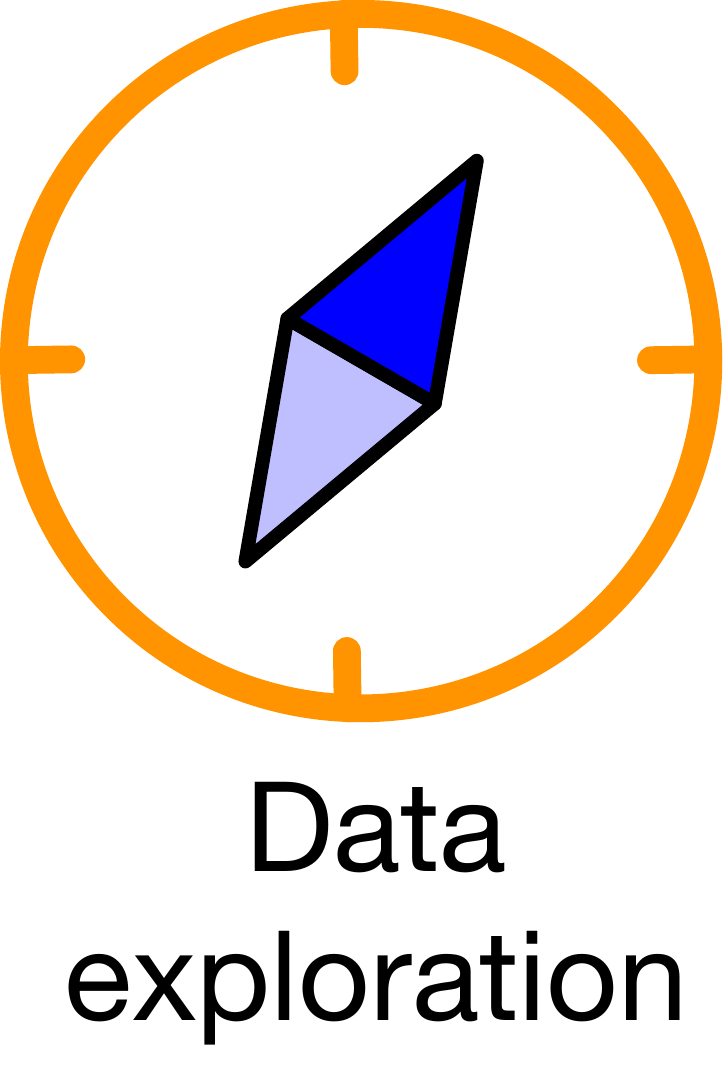}}
& \mc{\imglarger{symbol-plots-txt.pdf}}
& \mc{\imglarger{symbol-images-txt.pdf}}
& \mc{\imglarger{symbol-render-txt.pdf}}
& \mc{\imglarger{symbol-interactive-txt.pdf}}
& \mc{\imglarger{symbol-dr-txt.pdf}}
& \mc{\imglarger{symbol-uv-txt.pdf}}
& \mc{\imglarger{symbol-vr-txt.pdf}}
& \mc{\imglarger{symbol-exc-txt.pdf}}
& \mc{\imglarger{symbol-gal-txt.pdf}}
& \mc{\imglarger{symbol-pla-txt.pdf}}
& \mc{\imglarger{symbol-saa-txt.pdf}}
& \mc{\imglarger{symbol-simulations-txt.pdf}}
& \mc{\imglarger{symbol-observations-txt.pdf}}
\\ \hline	
	
\cite{TurkSmithOishi2010}
& $\bullet$ &           & $\bullet$ & $\bullet$ & $\bullet$ &           &           &     $\bullet$       &    $\bullet$        &           &           &     $\bullet$       &           \\ \hline
{\cite{BurgerStassunPepper2013}}
& $\bullet$ &           &           & $\bullet$ &           &           &           &           & $\bullet$ &           &           &           & $\bullet$ \\ \hline
\cite{LucianiCherinkaOliphant2014}
& $\bullet$ & $\bullet$ &           & $\bullet$ &           &           &           & $\bullet$ & $\bullet$ &           &           &           & $\bullet$ \\ \hline
\cite{BainesGiordanoRacero2016}
& $\bullet$ & $\bullet$ &           & $\bullet$ &           &           &           & $\bullet$ & $\bullet$ &           &           &           & $\bullet$ \\ \hline
\cite{ArgudoFernandezPuertasRuiz2017}
& $\bullet$ &           &           & $\bullet$ & $\bullet$ &           &           & $\bullet$ &           &           &           &           & $\bullet$ \\ \hline
\cite{PomaredeCourtoisHoffman2017}
& $\bullet$ &           & $\bullet$ & $\bullet$ &           &           &           &      $\bullet$     &           &           &           &           &  $\bullet$         \\ \hline
{\cite{AxelssonCostaSilva2017}\cite{KlashedHemingssonEmmart2010}\cite{FuHanson2007}}
&           &           & $\bullet$ &           &           &           &           &           & $\bullet$ & $\bullet$ &           & $\bullet$ & $\bullet$ \\ \hline
{\cite{BockAxelssonCosta2020}\cite{BladinAxelssonBroberg2018}\cite{CostaBockEmmart2021}}
& $\bullet$ &           & $\bullet$ & $\bullet$ &           &           &           &  $\bullet$      &     $\bullet$      &   $\bullet$      &     $\bullet$      &  $\bullet$      &  $\bullet$         \\ \hline
\cite{SagristaJordanMuller2019}
& $\bullet$ & $\bullet$ & $\bullet$ & $\bullet$ &           &           &           &           &   $\bullet$        &     $\bullet$      &           &           &   $\bullet$        \\ \hline
\cite{VohlBarnesFluke2016}
& $\bullet$ &           & $\bullet$ & $\bullet$ &           &           &           &    $\bullet$       &           &           &           &           &     $\bullet$      \\ \hline
\cite{Muna2017}
& $\bullet$ & $\bullet$ &           & $\bullet$ &           &           &           &           &   $\bullet$        &           &           &           &    $\bullet$       \\ \hline
\cite{Taylor2005} \cite{Taylor2014b} \cite{Taylor2017b}
& $\bullet$ & $\bullet$ &           & $\bullet$ &           &           &           &     $\bullet$      &    $\bullet$       &           &           &     $\bullet$      &  $\bullet$         \\ \hline
\cite{ScherzingerBrixDrees2017} 
\cite{MeyerSpradowRopinskiMensmann2009}
& $\bullet$ &           & $\bullet$ & $\bullet$ &           &           &           &     $\bullet$      &           &           &        &     $\bullet$      &           \\ \hline
\cite{hazarika2015visualizing}\cite{almryde2015halos}\cite{hanula2015cavern}
& $\bullet$ &           & $\bullet$ & $\bullet$ &           &           &           &     $\bullet$      &           &           &        &     $\bullet$      &           \\ \hline
\cite{ZhangSunTang2011}
& $\bullet$ &           &           & $\bullet$ &           &           &           &           &           &           &       $\bullet$    &           &     $\bullet$      \\ \hline
\cite{BreddelsVeljanoski2018}
& $\bullet$ &           & $\bullet$ & $\bullet$ &           &           &           &    $\bullet$       &    $\bullet$       &     $\bullet$      &           &     $\bullet$      &    $\bullet$       \\ \hline
\cite{YuEfstathiouIsenberg2012}\cite{YuEfstathiouIsenberg2016}
&          &           & $\bullet$ & $\bullet$ &           &           &           &    $\bullet$       &    $\bullet$       &     $\bullet$      &   $\bullet$  &     $\bullet$      &    $\bullet$       \\ \hline
\cite{FritschiRojoGunther2019}\cite{HesseEdenfeldSantalidisHuesmann2019}\cite{NguyenNguyenPham2019}\cite{SchatzMullerGralka2019}
& $\bullet$ & $\bullet$  & $\bullet$ & $\bullet$ &           &           &           &    $\bullet$       &         &         &     &     $\bullet$      &        \\ \hline
\end{tabular}
}	
\caption{Classifying papers under data exploration based on secondary and tertiary categories.
Top row, from left to right: (primary category) Data exploration; (secondary categories) 2D/3D plots, 2D images, 3D rendering, interactive visualization, dimensionality reduction, uncertainty visualization, and new display platforms; (tertiary categories) extragalactic, galactic, planetary, and solar astronomy; (tags) simulated, and observational data.
}
\label{table:data-explore}
\end{table*}

\section{Data Exploration} \label{sec:data-exploration}
In this section, we summarize research efforts that use visualization to focus on exploratory data analysis~\cite{tukey1977exploratory}. Broadly speaking, the defining attribute of data exploration papers is a focus on facilitating the unstructured investigation of a dataset in order to discover patterns of interest and formulate hypotheses. 
Our interpretation of data exploration follows Goodman's perspective on studying high-dimensional data in astronomy, where ``interactive exploratory data visualization can give far more insight than an approach where data processing and statistical analysis are followed, rather than accompanied, by visualization.''~\cite{Goodman2012}.
We distinguish between ``heterogeneous'' and ``hierarchical'' data exploration to highlight the different methodologies employed, where heterogeneous refers to drawing together disparate datasets and hierarchical refers to a deep exploration of fundamentally similar datasets (perhaps at different resolutions). \autoref{table:data-explore} presents an overview of our categorization of data exploration papers.

\subsection{Heterogeneous Data Exploration}
\label{sec:data-exploration-heterogeneous}

A number of astrophysics visualization software frameworks and tools have emphasized the value of exploring multiple datasets simultaneously in order to generate new insight, often requiring (or facilitating) data transformation pre-processing steps.

\textsf{yt}~\cite{TurkSmithOishi2010} is an open-source, flexible, and multi-code data analysis and visualization tool for astrophysics. Earlier versions of \textsf{yt} focused on making it possible to examine slices and projected regions within deeply
nested adaptive mesh refinement simulations~\cite{bryan2014enzo}. Although still widely used for its data wrangling capabilities, \textsf{yt} now also includes a range of data exploration and feature identification functionalities, providing off-screen rendering, interactive plotting capabilities, and scripting interfaces. It efficiently processes large and diverse astrophysics data, creates 2D visualization with an adaptive projection process and volume rendering by a direct ray casting method. Its cross-code support enables analysis for heterogeneous data types, and facilitates cross-platform collaborations between different astrophysics communities. In order to reduce processing time, \textsf{yt} adopts parallelism and is able to run multiple independent processing units on a single dataset in parallel. Apart from being easily customizable, \textsf{yt} presents a number of pre-defined analysis modules for halo finding, halo analysis, merger tree creation, and time series analysis, among others, and a recent project makes it possible to use \textsf{yt} for interactive data exploration within Jupyter notebooks~\cite{munk2020widgyts}. \textsf{yt} is also notable for its large, active community of users and developers.

\textsf{Filtergraph}~\cite{BurgerStassunPepper2013} is a web application that generates a range of 2D and 3D figures. It is designed to reduce the ``activation energy'' of the visualization process to flexibly and rapidly visualize large and complex astronomy datasets. 
It accepts numerous file formats without meta-data specifications, from text files to FITS images to Numpy files. The interface enables users to plot their data as high-dimensional scatter plots, histograms, and tables. Users can extensively explore the datasets and switch between different representations without cognitive interruption. Users can also customize the visualization through various interactive capabilities, such as panning, zooming, data querying, and filtering. \textsf{Filtergraph} also facilitates the sharing and collaboration of visualizations.

Luciani \etal~\cite{LucianiCherinkaOliphant2014} introduced a web-based computing infrastructure that supports the visual integration and efficient mining of large-scale astronomy observations. 
The infrastructure overlays image data from three complementary sky surveys (SDSS, FIRST, and simulated LSST results) and provides real-time interactive capabilities to navigate the integrated datasets, analyze the spatial distribution of objects, and cross-correlate image fields. 
Additionally, Luciani \etal\ described \emph{interactive trend images}, which are pixel-based, compact visual representations that help users identify trends and outliers among large collections of spatial objects.

\textsf{ESASky}~\cite{BainesGiordanoRacero2016}, developed by the ESA Center Science Data Center, is a web application designed for three use cases: the exploration of multi-wavelength skies, the search and retrieval of data for single or multiple targets, and the visualization of sky coverage for all ESA missions. The end result is a ``Google Earth for space'', effectively combining the vast collection of data hosted by the ESA and providing an annotated map of the Universe that facilitates data querying and exploration across multiple data sources.

\textsf{LSSGalPy}~\cite{ArgudoFernandezPuertasRuiz2017} emphasizes the exploration of the large-scale structures surrounding galaxies and visualizes isolated galaxies, isolated pairs, and isolated triplets in relation to other galaxies within their large-scale structures. The paper describes one use case that investigates the effect of local and large-scale environments on nuclear
activity and star formation, and another use case that visualizes galaxies with kinematically decoupled stellar and gaseous components, including an estimation of the tidal strength that affects each galaxy.

The \textsf{Cosmicflows} project aims to reconstruct and map the structure of the local universe, providing a series of catalogs that measure galaxy distances and velocities~\cite{tully2013cosmicflows}. Supporting this project, Pomarede \etal~\cite{PomaredeCourtoisHoffman2017} provided four ``cosmography'' use cases for the \textsf{SDvision} visualization software, focusing on the creation of animations and interactive 2D and 3D visualizations of scalar and vector fields found in catalogs of galaxies, mapping cosmic flows, representing basins of attraction, and viewing the Cosmic V-web~\cite{pomarede2017cosmic}. 
Pomarede \etalnocite also explored the use of \textsf{Sketchfab}, a web-based interface that enables the uploading and sharing of 3D models that can be viewed in virtual reality.

The vast scales present in astronomical datasets can be difficult to render and present simultaneously. Klashed \etal~\cite{KlashedHemingssonEmmart2010} introduced the ``ScaleGraph'' concept to deal with imprecision in rendering in the \textsf{Uniview} software. Hansen \etal~\cite{FuHanson2007} utilized power-scaled coordinates to cover the distance ranges. More recently, Axelsson \etal~\cite{AxelssonCostaSilva2017} presented a way to enable fast and accurate scaling, positioning, and navigation without a significant loss of precision, which they call the \emph{dynamic scene graph}.  At the core of this technique is the dynamic reassignment of the camera to focus on the object of interest, which then becomes the origin of the new coordinate system, ensuring the highest possible precision. Axelsson \etal\ applied this technique in the open-source software \textsf{OpenSpace}.

\begin{figure}[ht]
    \centering
    \includegraphics[width=0.98\columnwidth]{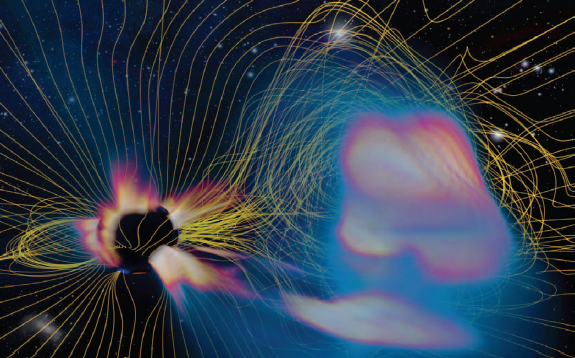}
    \caption{\textsf{OpenSpace}: time-varying corona mass ejection simulation with 3D rendering and field lines. Image reproduced from Bock et al.~\cite{BockAxelssonCosta2020}.}
    \label{fig:openspace}
\end{figure}

\textsf{OpenSpace}~\cite{BockAxelssonCosta2020} is a software system that enables the interactive exploration of a multitude of available astronomy datasets (\autoref{fig:openspace}). It is designed to be robust enough to support educational and outreach activities as well as adaptable enough to allow for the incorporation of new data or analysis tools to support scientific research. For the first task, \textsf{Openspace} has already demonstrated success in science communication at museums and in planetariums. For the second task, \textsf{OpenSpace}'s ability to interface with tools such as \textsf{Glue}~\cite{GoodmanBorkinRobitaille2018} or \textsf{Aladin} exemplifies a growing paradigm in astronomy visualization: the combination of multiple available tools to complete a task rather than building a bespoke system from the ground up. \textsf{OpenSpace} exhibits several novel features, including multi-resolution globe browsing~\cite{BladinAxelssonBroberg2018}, which enables dynamic loading of high-resolution planetary surface textures and physically based rendering of planetary atmospheres~\cite{CostaBockEmmart2021}.

\textsf{Gaia Sky}~\cite{SagristaJordanMuller2019} is an open-source, 3D universe explorer that enables users to navigate the stars of our galaxy from the Gaia Catalog (Gaia data release 2). It also aids in the production of outreach material. 
The system embeds stars in a multi-scale octree structure, where, at different levels, stars with various absolute brightness values are present. 
The system contains a floating camera for space traversal, integrated visualization of relativistic effects, real-time star movement, and simulates the visual effects of gravitational waves. The main strength of \textsf{Gaia Sky} is its capability to provide real-time interactive exploration for hundreds of millions of stars. Its efficient handling of the data allows it to manage a large range of scales with sufficient numerical precision.

\begin{figure}[b]
    \centering
    \includegraphics[width=0.98\columnwidth]{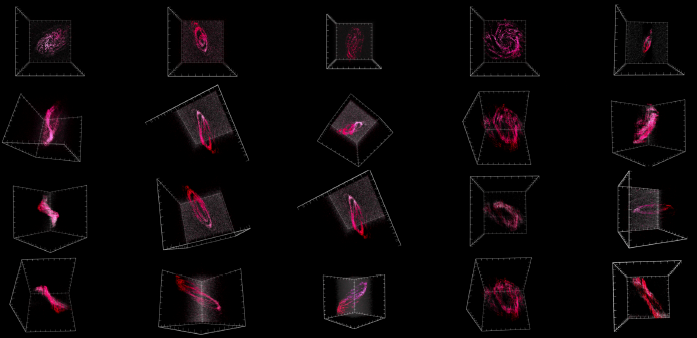}
    \caption{Comparative visualization of 20 galaxy morphologies with \textsf{Encube}~\cite{VohlBarnesFluke2016}. Image reproduced from ``Large-scale comparative visualization of sets of multidimensonal data'', written by Dany Vohl, David G. Barnes, Christopher J. Fluke, Govinda Poudel, Nellie Georgiou-Karistianis, Amr H. Hassan, Yuri Benovitski, Tsz Ho Wong, Owen L. Kaluza, Toan D. Nguyen, and C. Paul Bonnington, published in the PeerJ Computer Science journal. Link to article: {\small\url{https://peerj.com/articles/cs-88/}}.
     }
     \label{fig:encube}
\end{figure}

Vohl \etal~\cite{VohlBarnesFluke2016} presented \textsf{Encube} to accelerate the visual discovery and analysis process of large data cubes in medical imaging and astronomy (\autoref{fig:encube}). \textsf{Encube} can be used on a single desktop as well as the CAVE2 immersive virtual reality display environment. In the CAVE2 environment, \textsf{Encube} enables users to control and interact with a visualization of over 100 data cubes across 80 screens. The design focuses on comparative visualization and related user interactions, such as swapping screens and requesting quantitative information from the selected screens. It uses a distributed model to seamlessly process and render visualization and analysis tasks on multiple data cubes simultaneously. Additionally, \textsf{Encube} serializes the workflow and stores the data in the JSON format, so that the discovery process can be reviewed and re-examined later. A desktop version of \textsf{Encube} supports many of the same functionalities as it does in the CAVE2 environment. Combined with the recording of the discovery process, researchers can continue with their workflow when they return to their desktops.

Recognizing that FITS images were inherently complex, and that existing FITS viewers were not built with an optimal user experience in mind, Muna~\cite{Muna2017} introduced \textsf{Nightlight}, an ``easy to use, general purpose, high-quality'' viewer. \textsf{Nightlight} uses detail-on-demand to provide a high-level view of the file structure upon loading, and allows quick exploration of the data. Instead of reducing the dynamic range of astronomical data while visualizing FITS images, \textsf{Nightlight} leverages its approach on the fact that the input image is likely astronomical data. It provides two modes for the astronomers --- hyperbolic sine function scaling for bright features (e.g. stars), and linear scaling for faint features (e.g., nebulae). For FITS tables, \textsf{Nightlight} provides two views. The first is a grid of ``cards'', where each card represents the metadata of a single column in the table. The ``cards'' view is complemented by a second view in which the user can find the details of the full table.

\begin{figure}[b]
    \centering
    \includegraphics[width=0.80\columnwidth]{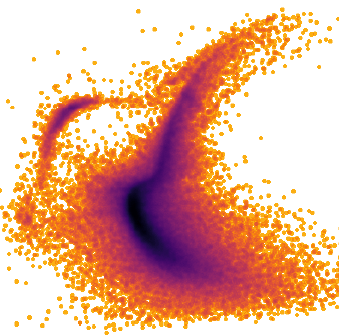}
    \caption{\textsf{TOPCAT}: Hybrid scatter plot/density map~\cite{Taylor2017b}. Image reproduced from ``TOPCAT: Desktop Exploration of Tabular Data for Astronomy and Beyond'', written by Mark Taylor, and published in the Informatics journal. Link to article: {\small \url{https://doi.org/10.3390/informatics4030018}}.}
    \label{fig:topcat}
\end{figure}

Since its introduction, \textsf{TOPCAT}~\cite{Taylor2005} has been widely used to view, analyze, and edit tabular data in the astronomy community. In additional to the generic tasks such as sorting rows, computing statistics of columns, and cross-matching between tables, \textsf{TOPCAT} also provides astronomy specific functionalities including the access to Virtual Observatory data, handling of various coordinate systems, and joining tables based on sky positions \cite{Taylor2017b}. Over the past decade, the developers of \textsf{TOPCAT} have continued to improve its capabilities. Taylor~\cite{Taylor2014b} described a rewrite of the plotting library added to \textsf{TOPCAT} v4, which is designed to improve responsiveness and performance of the visualization of large datasets. One important new feature is the hybrid scatter plot/density map, see \autoref{fig:topcat}, that enables users to navigate interactively between the high- and low-density regions without changing plot types.

Taylor~\cite{Taylor2017b} described the exploratory visualization capabilities of \textsf{TOPCAT}, which include high-dimensional plotting, high-density plotting, subset selection, row highlighting, linked views, and responsive visual feedback. Apart from the GUI application, users can also access \textsf{TOPCAT} from a set of command-line tools. 

\subsection{Hierarchical Data Exploration}

Scherzinger \etal~\cite{ScherzingerBrixDrees2017} proposed a unified visualization tool based on \textsf{Voreen}~\cite{MeyerSpradowRopinskiMensmann2009} that supports the interactive exploration of multiple data layers contained within dark matter simulations. These simulations contain only dark matter particles, in contrast to also including gas and stars. Scherzinger's visualization enables users to view the global structure of the data through 2D and 3D volume rendering and particle rendering, and the time-varying properties of the data through a merger tree visualization. Local structures are explored further through local particles visualization and the integration with \textsf{Galacticus}, an open-source semi-analytic model that computes information about galaxy formation based on merger tree hierarchies of dark matter halos~\cite{Benson2012}. An important aspect of their approach is scalable volume rendering, where the distribution of dark matter is visualized at interactive frame rates based on a pre-processing conversion. During such a conversion, attributes of large-scale particle data are distributed over a voxel grid, and maximum intensity projection in the 3D view is computed to highlight high-density regions of the data for volume rendering. 

Other tools also focus on exploring the evolution of galaxy halos within simulation datasets. Hazarika \etal~\cite{hazarika2015visualizing} presented a series of visualizations to provide insight into halos, including a 3D volume rendering of simulation data and a particle rendering that identifies halo sub-structures. Almryde and Forbes~\cite{almryde2015halos} introduced an interactive web application to created animated ``traces'' of dark matter halos as they move in relation to each other over time, and Hanula \etal~\cite{hanula2015cavern} presented the \textsf{Cavern Halos} project that enables the exploration of halos in virtual reality using the CAVE2 immersive collaboration space (this project was later extended and renamed \textsf{DarkSky Halos}~\cite{hanula2019darksky}). See also the discussion of work by Preston \etal~\cite{PrestonGhodsXie2016} in~\autoref{sec:feature-identification}.

In order to better investigate the nature of solar wind ion data (SWID), which is typically visualized using 1D and 2D methods, Zhang \etal~\cite{ZhangSunTang2011} developed a 3D visualization method for SWID based on the Selenocentric Solar Ecliptic coordinate system, and integrated this method into an interactive tool called \textsf{vtSWIDs}. \textsf{vtSWIDs} enables researchers to browse through numerous records and provides statistical analysis capabilities.

Breddels \etal~\cite{BreddelsVeljanoski2018} introduced \textsf{Vaex}, a \textsf{Python} library that handles large tabular datasets such as the \textsf{Gaia} catalogue. 
Many packages in \textsf{Vaex} are developed with specific visualization challenges in mind, and they overcome the scalability issues with methods such as efficient binning of the data, lazy expressions, and just-in-time compilation. 
For example, \textsf{vaex-core} provides visualization using the \textsf{matplotlib}  library, with 1D histograms and 2D density plots; \textsf{vaex-jupyter} embeds the visualization tools in a web browser, which offers more user interactions such as zooming, panning, and on-plot selection. 
It also enables 3D volume and iso-surface rendering using \textsf{ipyvolume} and connecting to a remote server using \textsf{WebGL}. 
A standalone interface is provided by the \textsf{vaex-ui} package, which supports interactive visualization and analysis. 
The \textsf{vaex-astro} package is specifically designed for astronomical data, supporting the FITS format and the most common coordinate transformations needed for analysis in astronomical data.

To enhance the study of astronomical particle data, the work by Yu \etal~\cite{YuEfstathiouIsenberg2012} was motivated by the need for an enhanced spatial selection mechanism using direct-touch input for particle data such as numerical simulations of the gravitational processes of stars or galaxies. 
They introduced two new techniques, \emph{TeddySelection} and \emph{CloudLasso}, to support efficient, interactive spatial selection in large particle 3D datasets. 
Their selection techniques automatically identify bounding selection surfaces surrounding the selected particles based on the density.  
They applied their techniques to particle datasets from a galaxy collision simulation ({\small{\url{http://www.galaxydynamics.org}}}) and an N-body mass simulation from the Aquarius Project~\cite{SpringelWangVogelsberger2008}, thus reducing the need for complex Boolean operations that are part of traditional multi-step selection processes. 
In a follow-up work~\cite{YuEfstathiouIsenberg2016}, Yu \etal\ further enhanced their 3D selection techniques to aid the exploratory analysis of astronomical data. 
They proposed a collection of context-aware selection techniques (\emph{CAST}) that improve the usability and speed of spatial selection, and applied their methods to a cosmological N-Body simulation and Millennium-II dataset~\cite{SpringelWhiteJenkins2005}.  

The 2019 SciVis contest proposed a visual analysis challenge to explore the structure formation in the cosmic evolution. 
The dataset was from a CRK-HACC (HACC: Hardware/Hybrid Accelerated Cosmology Code) cosmological simulation containing dark matter plus baryon particles in a cubic box, where the particles contain multiple fields such as position, velocity, and temperature.  
The simulations were used to study the impact that the feedback from AGN (Active Galactic Nuclei) has on their surrounding matter distribution.
The entries from the contest (e.g.,\cite{FritschiRojoGunther2019, HesseEdenfeldSantalidisHuesmann2019, NguyenNguyenPham2019, SchatzMullerGralka2019}) represented a diverse collection of visualizations, made possible by these new forms of simulation datasets.

%% file: sec-feature-identification.tex
\begin{table*}[!ht]
	\centering
\resizebox{1.0\textwidth}{!}{	
	\begin{tabular}{|>{\centering\arraybackslash}m{0.1\textwidth}
	||>{\centering\arraybackslash}m{0.06\textwidth}
	>{\centering\arraybackslash}m{0.06\textwidth}
	>{\centering\arraybackslash}m{0.06\textwidth}
	>{\centering\arraybackslash}m{0.06\textwidth}
	>{\centering\arraybackslash}m{0.06\textwidth}
	>{\centering\arraybackslash}m{0.06\textwidth}
	>{\centering\arraybackslash}m{0.06\textwidth}
	||>{\centering\arraybackslash}m{0.06\textwidth}
	>{\centering\arraybackslash}m{0.06\textwidth}
	>{\centering\arraybackslash}m{0.06\textwidth}
	>{\centering\arraybackslash}m{0.06\textwidth}
	|>{\centering\arraybackslash}m{0.06\textwidth}
	>{\centering\arraybackslash}m{0.06\textwidth}|}

\mc{\imglarger{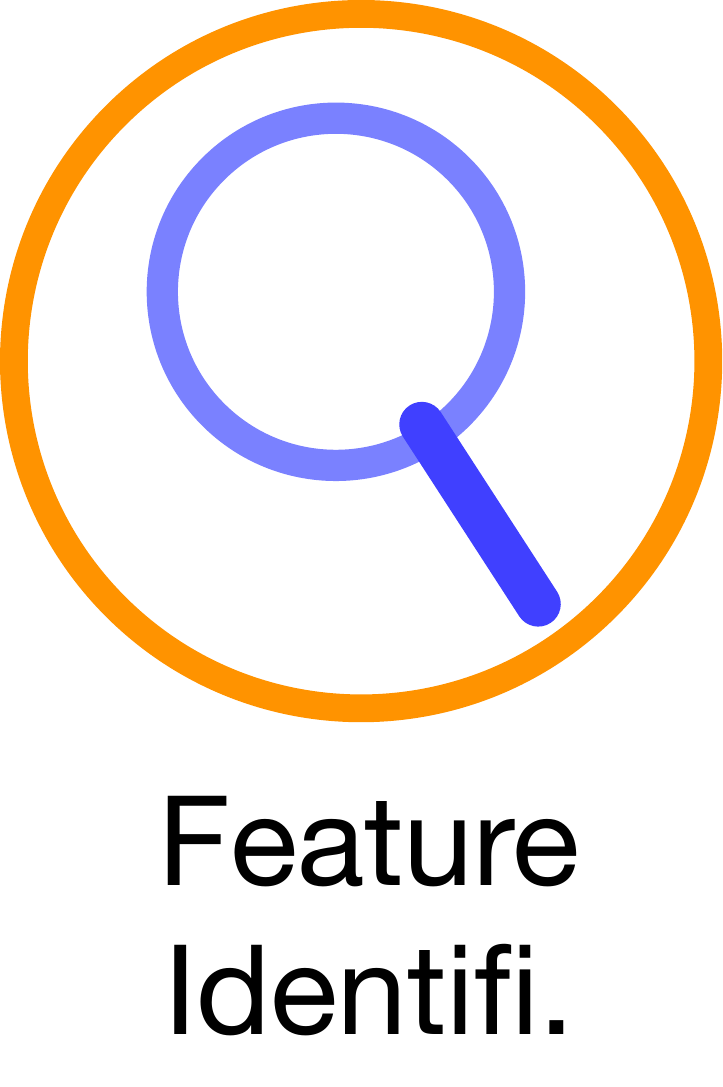}}
& \mc{\imglarger{symbol-plots-txt.pdf}}
& \mc{\imglarger{symbol-images-txt.pdf}}
& \mc{\imglarger{symbol-render-txt.pdf}}
& \mc{\imglarger{symbol-interactive-txt.pdf}}
& \mc{\imglarger{symbol-dr-txt.pdf}}
& \mc{\imglarger{symbol-uv-txt.pdf}}
& \mc{\imglarger{symbol-vr-txt.pdf}}
& \mc{\imglarger{symbol-exc-txt.pdf}}
& \mc{\imglarger{symbol-gal-txt.pdf}}
& \mc{\imglarger{symbol-pla-txt.pdf}}
& \mc{\imglarger{symbol-saa-txt.pdf}}
& \mc{\imglarger{symbol-simulations-txt.pdf}}
& \mc{\imglarger{symbol-observations-txt.pdf}}
\\ \hline
	
\cite{KaehlerHahnAbel2012}
&           & $\bullet$ &           &           & $\bullet$ &           &           &     $\bullet$      &           &           &           &      $\bullet$     &           \\ \hline
\cite{RavouxArmengaudWalther2020}
& $\bullet$ &           &           & $\bullet$ &           &           & $\bullet$ &    $\bullet$       &           &           &           &           &     $\bullet$      \\ \hline
\cite{ShanXieLi2014}
&           &           & $\bullet$ & $\bullet$ & $\bullet$ &           &           &     $\bullet$       &           &           &           &    $\bullet$        &           \\ \hline
\cite{PrestonGhodsXie2016}
& $\bullet$ &           & $\bullet$ & $\bullet$ &           &           &           &      $\bullet$     &           &           &           &     $\bullet$      &           \\ \hline
\cite{PillepichSpringelNelson2017}
& $\bullet$ & $\bullet$ &         &        &           &    $\bullet$  &           &      $\bullet$     &           &           &           &     $\bullet$      &           \\ \hline
\cite{BurchettAbramovOtto2019}
& $\bullet$ &           & $\bullet$ & $\bullet$ &           &           &           &     $\bullet$      &           &           &           &           &     $\bullet$      \\ \hline
\cite{XuNakayamaWu2016}\cite{SawadaUemuraBeyer2020}
& $\bullet$ &   &    &      $\bullet$       &         &           &           & $\bullet$           &           &           &           &           & $\bullet$           \\ \hline
\cite{LibeskindvandeWeygaertCantun2017}
& $\bullet$ &   & $\bullet$  &        &         &           &           & $\bullet$           &           &           &           &   $\bullet$  &            \\ \hline
\cite{Sousbie2011}\cite{SousbiePichonKawahara2011}
& $\bullet$ & $\bullet$ & $\bullet$ &           &           &           &           &      $\bullet$      &           &           &           &     $\bullet$       &     $\bullet$       \\ \hline
\cite{ShivashankarPranavNatarajan2016}
&           &           & $\bullet$ & $\bullet$ &           &           &           &      $\bullet$       &           &           &           &      $\bullet$       &           \\ \hline
\cite{TricocheSchleiHowell2021}
& $\bullet$ &           &           &  $\bullet$  & $\bullet$ &           &           &           &                   &  $\bullet$ &           & $\bullet$ &                   \\ \hline
\cite{RosenSethMills2019}
&           &  $\bullet$   &  &  &           &           &           &      $\bullet$       &           &           &           &            &    $\bullet$        \\ \hline
\cite{CampbellKjarAmico2012}
&           &           & $\bullet$ &  $\bullet$  &  $\bullet$     &           &           &  $\bullet$   &     $\bullet$      &           &           &           &     $\bullet$      \\ \hline
\cite{CiurloCampbellMorris2020}
& $\bullet$ &           & $\bullet$ &           &           &           &           &           &     $\bullet$      &           &           &           &     $\bullet$      \\ \hline
\cite{AragonCalvo2019}
&  &  $\bullet$  & $\bullet$ &           &           &           &           &  $\bullet$   &           &           &           &  $\bullet$  &           \\ \hline
\cite{KhanHuertaWang2019}
&           &           &           &           & $\bullet$ &           &           &    $\bullet$       &           &           &           &           &    $\bullet$       \\ \hline
\cite{NtampakaZuHoneEisenstein2019}
&           & $\bullet$ &           &           &           &           &           &     $\bullet$       &           &           &           &           &     $\bullet$       \\ \hline

\end{tabular}
}
\caption{Classifying papers under feature identification based on secondary and tertiary categories. 
Top row, from left to right: (primary category) feature identification; (secondary categories) 2D/3D plots, 2D images, 3D rendering, interactive visualization, dimensionality reduction, uncertainty visualization, and new display platforms; (tertiary categories) extragalactic, galactic, planetary, and solar astronomy; (tags) simulated, and observational data.}
\label{table:feature-identification}
\end{table*}

\section{Feature Identification}
\label{sec:feature-identification}
Research efforts in this category visually guide the identification and extraction of features of interest.  The term ``feature" is broad and can be used in a number of different astrophysical contexts.  The detection of features in an astrophysical datastream is of critical importance since many interesting phenomena are diffuse or observed with a low signal-to-noise ratio.  For example, physical phenomena may be subtle to detect (or may be detected for the first time), and distinguishing between what is signal and what is noise is critical. Teasing out a tiny signal is so common in astronomy that feature detection is a generically important element of astrophysical progress.  Furthermore, astrophysicists are often looking for diffuse physical contrasts in multiple dimensions (e.g. spatial, chemical, magnetic, density). For these phenomena, methods that establish robust criteria in multiple dimensions for identification and subsequent analysis are crucial.
The majority of these papers focus on dark matter simulations and the cosmic web, in particular voids, filaments, and dark matter halos, as summarized in~\autoref{table:feature-identification}. The \emph{cosmic web} refers to the large-scale structure of matter, distributed in filaments, the gravitationally collapsed structures that tend to connect galaxy halos, and voids, the low-density areas of the Universe.

\para{Visualizing dark matter simulations and cosmic web.}
Papers in this subsection employ various visualization techniques to visualize dark matter simulations and cosmic web, including GPU-assisted rendering with a tailored tessellation mesh~\cite{KaehlerHahnAbel2012}, tomographic map~\cite{RavouxArmengaudWalther2020}, and interactive visual exploration of cosmic objects~\cite{PrestonGhodsXie2016, ShanXieLi2014}.

Dark matter generates small-scale density fluctuations and plays a key role in the formation of structures in the Universe. 
Kaehler \etal~\cite{KaehlerHahnAbel2012} visualized N-body particle dark matter simulation data using GPU-assisted rendering approaches. 
Their method leverages the phase-space information of an ensemble of dark matter tracer particles to build a tetrahedral decomposition of the computational domain that allows a physically accurate estimation of the mass density between the particles~\cite{KaehlerHahnAbel2012}. 
During the simulation, vertices of a tessellation mesh are defined by the dark matter particles in an N-body simulation, whereas tetrahedral cells contain equal amounts of mass.
The connectivity within the mesh is generated once and is kept constant over the simulation as the cells warp and overlap.  
The density of a given location in the simulation is obtained by considering the density contribution from overlapping cells in the region of interest. 
Their new approaches are shown to be effective in revealing the structure of the cosmic web, in particular, voids, filaments, and dark matter halos. 

The Ly$\alpha$ forest, which is a series of individual over-densities of neutral hydrogen within the \emph{intergalactic medium} (IGM, the space between galaxies), provides a 1D measurement of information in the IGM, which is largely correlated with the distribution of matter in the Universe. 
Ravoux \etal~\cite{RavouxArmengaudWalther2020} used a tomographic reconstruction algorithm called the Wiener filtering to create a 3D tomographic map with the \emph{eBoss Strip p82 Ly$\alpha$ forest}  datasets. 
The map is used as a representation of the associated matter fluctuation to identify over- and under-densities in the cosmic web. 
Extended over-densities can be detected with the tomographic map by searching for the large deficit in the Ly$\alpha$ forest flux contrast. 
The authors adopt a simple-spherical algorithm to identify large voids. 
In order to further investigate the regions of interest, the paper provides 3D representations of the tomographic map over the entire strip. Users can interactively explore the map through rotating, panning, and zooming.

\begin{figure}[b]
    \centering
    \includegraphics[width=1\columnwidth]{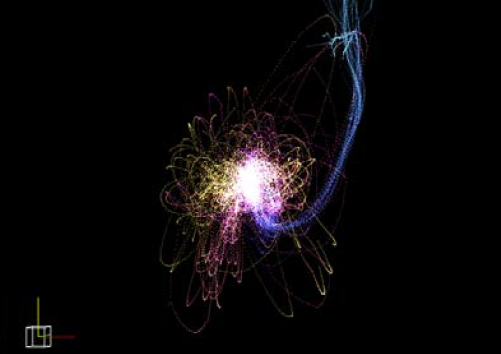}
    \caption{An example of a particle trace path. Image reproduced from Shan et al.~\cite{ShanXieLi2014}.}
    \label{fig:particle-trace-path}
\end{figure}

Gravity causes dark matter particles to collapse into larger structures over time. The individual groups of particles formed during this process are called halos, one of the most common elements in the dark matter simulation \cite{PrestonGhodsXie2016}. Their evolution process and behaviors are often the focus of astronomical discoveries. 
Two recent tools facilitate the visual exploration of halos. 
Shan \etal~\cite{ShanXieLi2014} built an interactive visual analysis system that focuses on exploring the evolutionary histories of halos.
The interface allows the user to manually select regions of interest in 3D space. It then uses the marching cubes algorithm to perform iso-surface extraction and cluster separation based on the region's density distribution. 
To prevent overlaps in the 3D space, the system employs multi-dimensional scaling (MDS) to project the halos into 2D space. 
Multiple linked views are generated to support the exploration through time.
In addition to a merger tree view that is commonly used to visualize evolution of objects over time, Shan {\etal} proposed a unique particle trace path graph (see \autoref{fig:particle-trace-path}), which encodes the evolution history of selected particles. 

Preston \etal~\cite{PrestonGhodsXie2016}, on the other hand, aimed to increase the efficiency and interactions in studying the evolution of halos, described by merger trees. 
Their integrated visualization system consists of a \emph{merger tree view}, a \emph{3D rendering view}, and a \emph{quantitative analysis view}. 
Their \emph{merger tree view} is an enhancement from~\cite{ShanXieLi2014} with more interactive capabilities. 
The system allows users to select specific halos through the merger tree and organize the tree based on other physical variables such as velocity and mass.
The \emph{3D rendering view} displays the particles' physical behaviors over a number of time steps, providing additional contextual information for the merger tree. 
A remote parallel renderer is employed to improve the scalability of the rendering process.
Finally, the \emph{quantitative analysis view} extends the other two views by providing quantitative information of selected particles that reveals additional insights into the behavior of the halo. 
For instance, a chronological plot visualizes the anomalous events automatically detected in the history of a halo. 
An important feature of the system is that it enables simultaneous exploration of heterogeneous cosmology data; see~\autoref{sec:data-exploration} for further discussions. 

The \textsf{IllustrisTNG} project ({\small \url{https://www.tng-project.org/}}) contains collections of large, cosmological magnetohydrodynamical simulations of galaxy formation. 
It is designed to ``illuminate the physical processes that drive galaxy formation".
The tool provides a number of volume rendering capabilities to visually demonstrate the multi-scale, multi-physics nature of the simulations, as well as to perform qualitative inspections~\cite{PillepichSpringelNelson2017}.  

Moving from clusters of galaxies to the spaces between them, the IGM is composed of gas complexes in the spaces between galaxies. Although it has research values on its own, investigating IGM along with quasar sightlines helps put IGM in context. A \emph{quasar} is a supermassive blackhole at the center of a galaxy that is accreting gas at a high rate and is therefore very bright.
It enables scientists to associate certain absorption features with galactic environment, such as the \emph{circumgalactic medium} (CGM), which is the gaseous envelope surrounding a galaxy. 
\textsf{IGM-Vis}~\cite{BurchettAbramovOtto2019} is a visualization software specifically designed to investigate IGM/CGM data. It supports a number of identification, analysis, and presentation tasks with four linked views. 
The \emph{Universe panel} provides a 3D interactive plot of galaxies in circles and quasar sightlines in cylindrical ``skewers''. 
The user can select a galaxy of interest to further examine it in the \emph{galaxy panel}, which contains a list of attributes and corresponding data from SDSS. 
Additionally, quasar sightlines can be explored in the \emph{spectrum panel} where multiple spectral plots can be displayed and stored. 
The final \emph{equivalent width plot panel} facilitates dynamic correlation analysis and helps users discover absorption patterns in the regions of interest. 
The four views complement each other to streamline the discovery processes, including the identification of foreground and sightline features, the measure of absorption properties, and the detection of absorption patterns.

\emph{Blazars} -- similar to quasars, an active galactic nuclei with relativistic jets ejecting toward the Earth -- are one of the most attractive objects for astronomers to observe.
The \textsf{TimeTubes} visualization~\cite{XuNakayamaWu2016} transforms time-varying blazar data and polarization parameters into a series of ellipses arranged along a time line, forming a volumetric tube in 3D space. The most recent iteration of the project, \textsf{TimeTubesX}~\cite{SawadaUemuraBeyer2020}, includes feature identification techniques to detect recurring time variation patterns in blazar datasets. 
It includes an automatic feature extraction functionality to identify time intervals that correspond to well-known blazar behaviors, as well as dynamic visual query-by-example and query-by-sketch functionality.
Such a functionality enables users  to search long-term observations that are similar to a selected time interval of interest, or match a sketch of temporal pattern. 
The technique aims to enhance the reliability of blazar observations, and to identify flares, rotations, and other recurring blazar patterns in order to validate hypotheses about observable, photometric, and polarimetric behaviors.

To study the agreements and disparities of feature identification methods created for classifying the cosmic web, Libeskind \etal\cite{LibeskindvandeWeygaertCantun2017} collected 12 representative methods and applied them to the same GADGET-2 dark matter simulation. They classified the dark matter density field of the cosmic web into knots, filaments, walls, and voids. They used comparative visualization accompanied with a variety of 2D plots to provide intuitive representations of the different structures identified by these methods. 
We introduce one of the topology-based methods with a strong visualization component in the next subsection.

\para{Topology-based feature extraction.}
There are several examples of using topological techniques to extract cosmological features from simulations, in particular, galaxy filaments, voids, and halos. 
Topological methods have also been applied to observational data cubes. 
We believe that the integration of topological techniques in astronomical feature extraction and visualization will be a growing area of interest (see~\autoref{sec:challenges}).

Sousbie \cite{Sousbie2011} presented \textsf{DisPerSE}, a topology-based formalism that is designed to analyze the cosmic web and its filamentary structure. 
It leverages discrete Morse theory and computes a Morse-Smale complex (MSC) on a density field. 
The MSC is then simplified using the theory of persistent homology by canceling the topological features with low persistence values (i.e., those that are likely generated by noise). The relationship between the topological and geometrical features is easily detectable in the MSC, where the ascending 3-manifolds correspond to the voids, ascending 2-manifolds to the walls, and ascending 1-manifolds to the filaments. The technique is scale-free, parameter-free, and robust to noise. Sousbie \etal\ then demonstrated the effectiveness of \textsf{DisPerSE} at tracing cosmological features in 2D and 3D datasets~\cite{SousbiePichonKawahara2011}.

\begin{figure}[b]
    \centering
    \includegraphics[width=0.98\columnwidth]{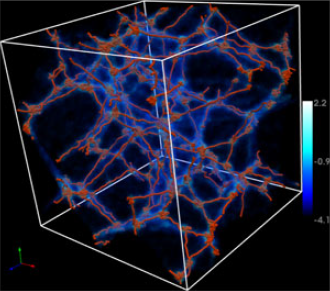}
    \caption{\textsf{Felix:} Extracting filamentary structures (orange) from a Voronoi evolution time-series dataset. Image reproduced from Shivashankar et al.~\cite{ShivashankarPranavNatarajan2016}.}
    \label{fig:felix}
\end{figure}

Following a similar path, Shivashankar \etal~\cite{ShivashankarPranavNatarajan2016} proposed \textsf{Felix}, another topology-based framework that identifies cosmological features (see~\autoref{fig:felix}). 
\textsf{Felix} focuses on extracting the filamentary structures and incorporates a visual exploration component. It also computes a MSC over a density field and simplifies it by iteratively canceling pairs of simplices, which generates a hierarchy of MSCs. 
Realizing that it is nearly impossible to find a version of the MSC within the hierarchy that best separates noise and features for cosmology datasets, \textsf{Felix} allows users to query for specific density ranges across all generated MSCs. 
This process increases user engagement in the parameter selection process and helps preserve filament structures within void-like or cluster-like regions. \textsf{Felix} also utilizes 3D volume rendering to interactively guide the selection of parameters for the query and visualizes the extracted filaments along with the density field. 
Interactive visual exploration of these intricate features remains a challenging and largely unexplored problem. 

Recently, a new method has been proposed by Tricoche \etal~\cite{TricocheSchleiHowell2021} to extract the topology of the Poincar\'e map in the \emph{circular restricted three-body problem} (CR3BP). They created an interactive visualization of the topological skeleton to support spacecraft trajectory designers in their search for energy-efficient paths through the interconnected web of periodic orbits between celestial bodies. The new method extends the existing approach by Schlei \etal~\cite{SchleiHowellTricoche2014}, and significantly improves the results of fixed point extraction and separatrices construction. In order to reduce the high computational cost, Tricoche \etal\ pre-screened for impractical spaceflight structures, and leveraged previous knowledge on the accuracy limitations of sensors and engines to impose restrictions on certain parameters. These adjustments reduce the computational workload of the method and enable  the interactive visualization of the topology. 
The visualization displays the fixed points identified by the system and each individual selected orbit as a closed curve. The visualization also enables a manifold arc selection mechanism to help the trajectory designer to determine the precise path a spacecraft would need to follow from any arbitrary location.

From an observational perspective, current radio and millimeter telescopes, particularly ALMA, are producing data cubes with significantly increased sensitivity, resolution, and spectral bandwidth. 
However, these advances often lead to the detection of structure with increased spatial and spectral complexity. 
Rosen \etal~\cite{RosenSethMills2019} performed a feasibility study for applying topological technique -- in particular, contour trees -- to extract and simplify the complex signals from noisy ALMA data cubes. 
They demonstrated the topological de-noising capabilities on a NGC 404  data cube (also known as Mirach’s Ghost) and a CMZ (Central Molecular Zone) data cube. 
Using topological techniques, Rosen \etal\ sought to improve upon existing analysis and visualization workflows of ALMA data cubes, in terms of accuracy and speed in feature extraction.

\para{Feature extraction from astronomy data cubes.}
In addition to the work by Rosen \etal~\cite{RosenSethMills2019}, other visualizations of integral field spectrometer (IFS) data cubes have been proposed. 
Campbell \etal~\cite{CampbellKjarAmico2012} presented a 3D interactive visualization tool specifically designed to render IFS data cubes. A typical display tool reduces a 3D IFS datacube to 2D images of either the spatial or the wavelength dimension. 
Campbell \etal\ proposed to use volume rendering instead to highlight features and characteristics of astronomical objects that are difficult to detect in lower dimension projections. 
The tool, known as \textsf{OsrsVol}, allows users to easily manipulate the visualized data cube by interactions such as zooming, rotating, and aspect ratio adjustment.

Ciulo \etal~\cite{CiurloCampbellMorris2020} used \textsf{OsrsVol} to identify four objects orbiting the supermassive black hole at the center of our galaxy Sagittarius A\textsuperscript{*}. Two unusual objects have been recently discovered around Sagittarius A\textsuperscript{*}, referred to as the G sources, and their possible tidal interactions with the black hole have generated considerable attention. Ciulo {\etal} selected 24 relevant data cubes and processed them through the \textsf{OSIRIS} pipelines. They analyzed the data cubes with \textsf{OsrsVol}, as well as several conventional 1D/2D visualization tools. \textsf{OsrsVol} helps to disentangle the various dimensions of data cubes and allows more flexible explorations among crowded regions. 
Using \textsf{OsrsVol}, Ciulo {\etal} also characterized the best-fit orbits of the four new objects, and determined that they exhibited many traits in common with the previously discovered G sources.

\para{Feature identification with deep learning.}
We end this section by giving a couple of examples of using neural network models as feature extractors for unsupervised clustering of galaxies.  
These works demonstrate the potential of using deep learning in feature identification tasks, for which both astronomers and visualization experts are cautiously excited.

Aragon-Calvo was the first to apply a deep convolutional neural network to the task of semantic segmentation of the cosmic web \cite{AragonCalvo2019}. He proposed a network with a U-net architecture and trained the model using a state-of-the-art manually guided segmentation method. Two types of training datasets were generated using the standard Voronoid model and an N-body simulation. 
Their method provides exciting results as it efficiently identifies filaments and walls with high accuracy for well-structured data such as the Voronoid model. For more complex datasets such as the N-body simulation, the U-net achieves higher quality segmentation than the state-of-the-art methods.

Khan \etal~\cite{KhanHuertaWang2019} constructed galaxy catalogs using  transfer learning. 
They employed a neural-network-based image classifier \emph{Xception}, pre-trained on ImageNet data, to classify galaxies that overlap both Sloan Digital Sky Survey (SDSS) and Dark Energy Survey (DES) surveys, achieving state-of-the-art accuracy of 99.6\%. 
Khan \etal\ then used their neural network classifier to label and characterize over 10,000 unlabelled DES galaxies, which do not overlap previous surveys.
They further extracted abstract features from one of the last layers of their neural network and clustered them using t-SNE, a dimensionality reduction technique.  
Their clustering results revealed two distinct galaxy classes among the unlabelled DES images based on their morphology. 
The analysis of Khan {\etal} provides a path forward in creating large-scale DES galaxy catalog by using these newly labelled DES galaxies as data for recursive training. 

\emph{Galaxy clusters} are gravitationally bound systems that contain hundreds or thousands of galaxies in dark matter halos~\cite{NtampakaZuHoneEisenstein2019}, with typical masses ranging from $10^{14}$ to $10^{15}$ solar masses. 
Ntampaka {\etal} applied deep learning to estimate galaxy cluster masses from \emph{Chandra} mock -- simulated, low-resolution, single-color X-ray images~\cite{NtampakaZuHoneEisenstein2019}. 
They used a relatively simple convolutional neural network (CNN) with only three convolutional and pooling layers followed by three fully connected layers. 
Despite the simple framework, the resulting estimates exhibit only small biases compared to the true masses. 
The main innovation of the paper is the visual interpretation of the CNN, using an approach inspired by Google's \textsf{DeepDream}, which uses gradient ascent to produce images that maximally activate a given neuron in a network. 
Ntampaka {\etal} used gradient ascent to discover which changes in the input cause the model to predict increased masses. 
They found that the trained model is more sensitive to photons in the outskirts of the clusters, and not in the inner regions; and their observations aligned with other statistical analyses performed on galaxy clusters. 
Their work illustrates the utility of interpreting machine learning ``black boxes'' with visualization since it provides physical reasoning to predicted features.

%% file: sec-object-reconstruction.tex
\begin{table*}[!ht]
	\centering
\resizebox{1.0\textwidth}{!}{	
	\begin{tabular}
	{|>{\centering\arraybackslash}m{0.1\textwidth}
	||>{\centering\arraybackslash}m{0.06\textwidth}
	>{\centering\arraybackslash}m{0.06\textwidth}
	>{\centering\arraybackslash}m{0.06\textwidth}
	>{\centering\arraybackslash}m{0.06\textwidth}
	>{\centering\arraybackslash}m{0.06\textwidth}
	>{\centering\arraybackslash}m{0.06\textwidth}
	>{\centering\arraybackslash}m{0.06\textwidth}
	||>{\centering\arraybackslash}m{0.06\textwidth}
	>{\centering\arraybackslash}m{0.06\textwidth}
	>{\centering\arraybackslash}m{0.06\textwidth}
	>{\centering\arraybackslash}m{0.06\textwidth}
	|>{\centering\arraybackslash}m{0.06\textwidth}
	>{\centering\arraybackslash}m{0.06\textwidth}|}

\mc{\imglarger{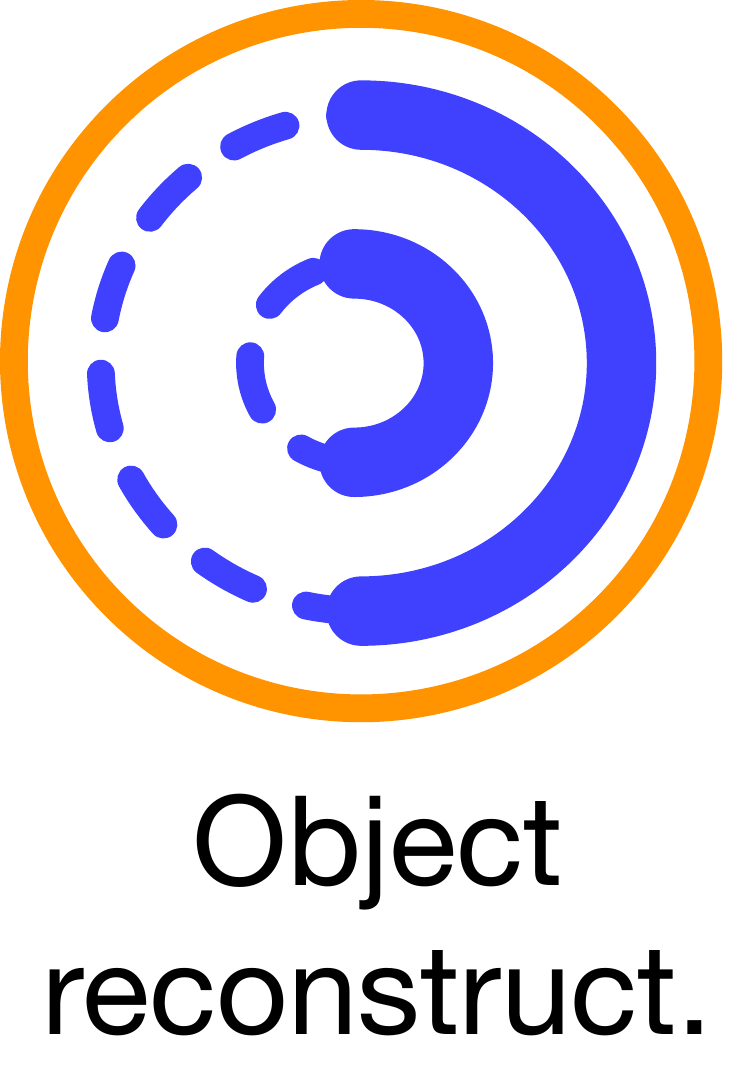}}
& \mc{\imglarger{symbol-plots-txt.pdf}}
& \mc{\imglarger{symbol-images-txt.pdf}}
& \mc{\imglarger{symbol-render-txt.pdf}}
& \mc{\imglarger{symbol-interactive-txt.pdf}}
& \mc{\imglarger{symbol-dr-txt.pdf}}
& \mc{\imglarger{symbol-uv-txt.pdf}}
& \mc{\imglarger{symbol-vr-txt.pdf}}
& \mc{\imglarger{symbol-exc-txt.pdf}}
& \mc{\imglarger{symbol-gal-txt.pdf}}
& \mc{\imglarger{symbol-pla-txt.pdf}}
& \mc{\imglarger{symbol-saa-txt.pdf}}
& \mc{\imglarger{symbol-simulations-txt.pdf}}
& \mc{\imglarger{symbol-observations-txt.pdf}}
	
\\ \hline
\cite{SteffenKoningWenger2011}
& $\bullet$ &           & $\bullet$ & $\bullet$ & $\bullet$ &           &           &           &     $\bullet$       &           &           &    $\bullet$        &     $\bullet$       \\ \hline
\cite{WengerAmentGuthe2012}\cite{WengerLorenzMagnor2013}
&           & $\bullet$ & $\bullet$ &   $\bullet$        &           &           &           &           &     $\bullet$      &           &           &           &  $\bullet$         \\ \hline
\cite{HasenbergerAlves2020}
&           & $\bullet$ & $\bullet$ &           &           &           &           &           &     $\bullet$      &           &           &           &     $\bullet$      \\ \hline
\cite{GrosschedlAlvesMeingast2018}
&           &           & $\bullet$ &           &           & $\bullet$ &           &           &     $\bullet$      &           &           &          &    $\bullet$       \\ \hline
\cite{SkowronSkowronMroz2019}
& $\bullet$ &           &           &           &           &           &           &           &     $\bullet$      &           &           &           &       $\bullet$    \\ \hline
\cite{VogtDopita2010}
& $\bullet$ & $\bullet$ &           &           &           &           &           &     $\bullet$      &           &           &           &           &    $\bullet$       \\ \hline
\cite{ElekBurchettProchaska2021}
&           &           & $\bullet$ & $\bullet$ &           &           &           &     $\bullet$       &           &           &           &     $\bullet$       &     $\bullet$       \\ \hline
\cite{BockPembrokeMays2015}
& $\bullet$ &           & $\bullet$ & $\bullet$ &           & $\bullet$ &           &           &          &        &    $\bullet$      &    $\bullet$       &   $\bullet$         \\ \hline
\cite{OrtnerWalchNowak2020} 
& $\bullet$ &           & $\bullet$ & $\bullet$ &           &      &           &           &          &  $\bullet$  &        &         &   $\bullet$         \\ \hline
		
\end{tabular}
}	
\caption{Classifying papers under object reconstruction based on secondary and tertiary categories.
Top row, from left to right: (primary category) Object reconstruction; (secondary categories) 2D/3D plots, 2D images, 3D rendering, interactive visualization, dimensionality reduction, uncertainty visualization, and new display platforms; (tertiary categories) extragalactic, galactic, planetary, and solar astronomy; (tags) simulated, and observational data.
}
\label{table:object-reconstruction}
\end{table*}

\section{Object Reconstruction}
\label{sec:object-reconstruction}

Research works in this category provide informative visual representation of  astronomical objects; see~\autoref{table:object-reconstruction} for their fine-grained classifications under secondary and tertiary categories, where there is a strong focus on observational data. 
Object reconstruction utilizes and is also constrained by imagery and other observational data obtainable via our vantage point -- the Earth  and the solar system. 
The works surveyed here cover 3D object reconstructions using 2D images~\cite{SteffenKoningWenger2011, WengerAmentGuthe2012, WengerLorenzMagnor2013, HasenbergerAlves2020}, distances of young stellar objects~\cite{GrosschedlAlvesMeingast2018}, spectroscopic data~\cite{VogtDopita2010}, and extrapolation from sparse datasets such as SDSS~\cite{ElekBurchettProchaska2021}, where visualization helps produce plausible reconstructions that provide structural insights for analysis and modeling. 
Important challenges include scalable computation, trade-off between automatic reconstruction and expert knowledge, and in particular, physically accurate structural inference with limited observations. 

As mentioned previously, we recognize that ``objects'' are, in fact, ``features'' with sharp and/or discontinuous contrast in a dimension of scientific interest.  
Whether a specific aspect of a dataset is considered an ``object'' or a ``feature'' depends on the scientific question posed.
We separate object reconstruction from feature identification to be compatible with the literature, but we envision a future where these entities are recognized as a continuum. 
An example of such a continuum is \textsf{Polyphorm}~\cite{ElekBurchettProchaska2021}, where the filament reconstruction and interactive visualization are intertwined via a fitting session, where structural or visual parameters are adjusted interactively to produce satisfactory reconstruction results.

Object reconstruction employs both images and other observational data, and thus is closely related to image reconstruction in astronomy. 
As discussed in~\autoref{sec:introduction}, we do not consider state-of-the-art image reconstruction methods in astronomy based on optimizations or signal processing techniques, but rather, we will focus on reconstruction with modern visualization techniques, such as 3D object reconstruction, 3D rendering, and interactive visualization.  
There is existing literature on the ``historic account'' of astronomical image reconstruction~\cite{Dainty1985,PuetterGosnellYahil2005}, recent surveys about this field~\cite{TheysAime2016}, and machine learning approaches~\cite{Flamary2017}.

\para{3D object reconstruction from 2D images.}
Steffen \etal~\cite{SteffenKoningWenger2011} presented \textsf{Shape}, one of the first publicly available tools using interactive graphics to model astronomical objects. 
\textsf{Shape} allows astrophysicists to interactively define 3D structural elements using their prior knowledge about the object, such as spatial emissivity and velocity field. 
\textsf{Shape} provides a unified modeling and visualization flow, where physical knowledge from the user is used to construct and iteratively refine the model, and model parameters are automatically optimized to minimize the difference between the model and the observational data.   
The interactive feedback loop helps introduce expert knowledge into the object reconstruction pipeline and has proven to be incredibly useful for many applications, such as rendering hydrodynamical simulations, reconstructing Saturn Nebula, modeling the structure and expansion of nova RS Ophiuchi~\cite{SteffenKoningWenger2011}. 
\textsf{Shape} also comes with educational potential in digital planetariums. 

Wenger \etal~\cite{WengerAmentGuthe2012} developed an automatic 3D visualization of astronomical nebulae from a single image using a tomographic approach.  
Their 3D reconstruction exploits the fact that many astronomical nebulae, interstellar clouds of gas and dust, exhibit approximate spherical or axial symmetry~\cite{MagnorKindlmannDuric2004}. 
This symmetry allows for object reconstruction by replicating multiple virtual viewpoints based on the view from Earth. 
This assemblage of different views results in a tomographic reconstruction problem, which can be solved with an iterative compressed sensing algorithm. 
The reconstruction algorithm relies on a constrained optimization and computes a volumetric model of the nebula for interactive volume rendering. 
Wenger \etal\ demonstrated that their method preserves a much higher amount of detail and visual variety than previous approaches. 
However, they also noted that the quality of their reconstruction is limited by the fact that ``the algorithm has no knowledge about the physical processes underlying the objects being reconstructed'', and suggested restricting the search space to solutions compatible with a physical model~\cite{WengerAmentGuthe2012}. 

\begin{figure}[b]
    \centering
    \includegraphics[width=0.98\columnwidth]{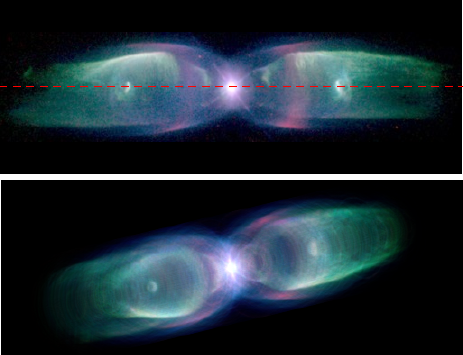}
    \caption{A fast reconstruction algorithm that creates 3D models of nebulae based on their approximate axial symmetry. Image reproduced from Wenger et al.~\cite{WengerLorenzMagnor2013}.}
    \label{fig:reconstruction_nebulae}
\end{figure}

In a follow-up work, Wenger \etal~\cite{WengerLorenzMagnor2013} presented an algorithm based on group sparsity that dramatically improves the computational performance of the previous approach~\cite{WengerAmentGuthe2012} (see~\autoref{fig:reconstruction_nebulae}).
Their method computes a single projection instead of multiple projections and thus reduces memory consumption and computation time. 
It is again inspired by compressed sensing: an $\ell_\infty$ group sparsity regularizer is used to suppress noise, and an $\ell_2$ data term is used to ensure that the output is consistent with the observational data~\cite{WengerLorenzMagnor2013}. 
This method enables astronomers and end users in planetariums or educational facilities to reconstruct stellar objects without the need for specialized hardware.

Hasenberger \etal~\cite{HasenbergerAlves2020} added to the hallowed pantheon of automatic object reconstruction algorithms with \textsf{AVIATOR}: a Vienna inverse-Abel-transform-based object reconstruction algorithm. 
Existing reconstruction techniques (e.g.,~\cite{WengerAmentGuthe2012, WengerLorenzMagnor2013}) contain potentially problematic requirements such as symmetry in the plane of projection. 
\textsf{AVIATOR}'s reconstruction algorithm assumes that, for the object of interest, its morphology ``along the line of sight is similar to its morphology in the plane of the projection and that it is mirror symmetric with respect to this plane''~\cite{HasenbergerAlves2020}. 
Hasenberger \etal\ applied \textsf{AVIATOR} to dense molecular cloud cores and found that their models agreed well with profiles reported in the literature.

\para{3D object reconstruction using stellar object distances.}
The Gaia data release 2 (Gaia DR2) contains a wealth of information about the night sky. Gro{\ss}schedl \etal~\cite{GrosschedlAlvesMeingast2018} used the distances of 700 stellar objects from this dataset to infer a model of Orion A that describes its 3D shape and orientation. 
This 3D model leads to many insights, among them that the nebulae is longer than previously thought and that it has a cometary shape pointing toward the Galactic plane, where the majority of the Milky Way's disk mass lies. 
The authors pointed out that Gaia is bringing the critical third spatial dimension to infer cloud structures and to study start-form interstellar medium.   

In a similar manner, Skowron \etal~\cite{SkowronSkowronMroz2019} constructed a 3D map of the Milky Way galaxy, using the positions and distances of thousands of classical Cepheid variable stars, which in turn are obtained through observations and accounting of the stars' pulsating periods coupled with luminosity. \emph{Cepheid variable}  are regularly pulsating stars, where their regular pulsations allow us to calculate their distances precisely. 
Skowron \etal\ used $2341$ such stars to sketch the Milky Way galaxy and observe the warped shape of the galactic disk, and they were able to define the characteristics of this warping with some precision. 
They visualized and performed additional analysis on this 3D map using a combination of static 2D/3D plots. 

\para{3D object reconstruction using spectroscopic data.}
Vogt \etal~\cite{VogtDopita2010} aimed to characterize the 3D shape of a young oxygen-rich supernova remnant (N132D) in the Large Magellenic Cloud, a satellite dwarf galaxy of the Milky Way. Using spectroscopic data from the Wide Field Spectrograph along with sophisticated data reduction techniques, they produced a data cube, which they used to construct a 3D map of the oxygen-rich ejecta of the supernova remnant of interest. 
They provided several different 2D and 3D plots showing unique views of this 3D map. 
Their visual analysis has led to insights about the structure of this supernova remnant beyond what was previously known.

\begin{figure}[b]
    \centering
    \includegraphics[width=0.98\columnwidth]{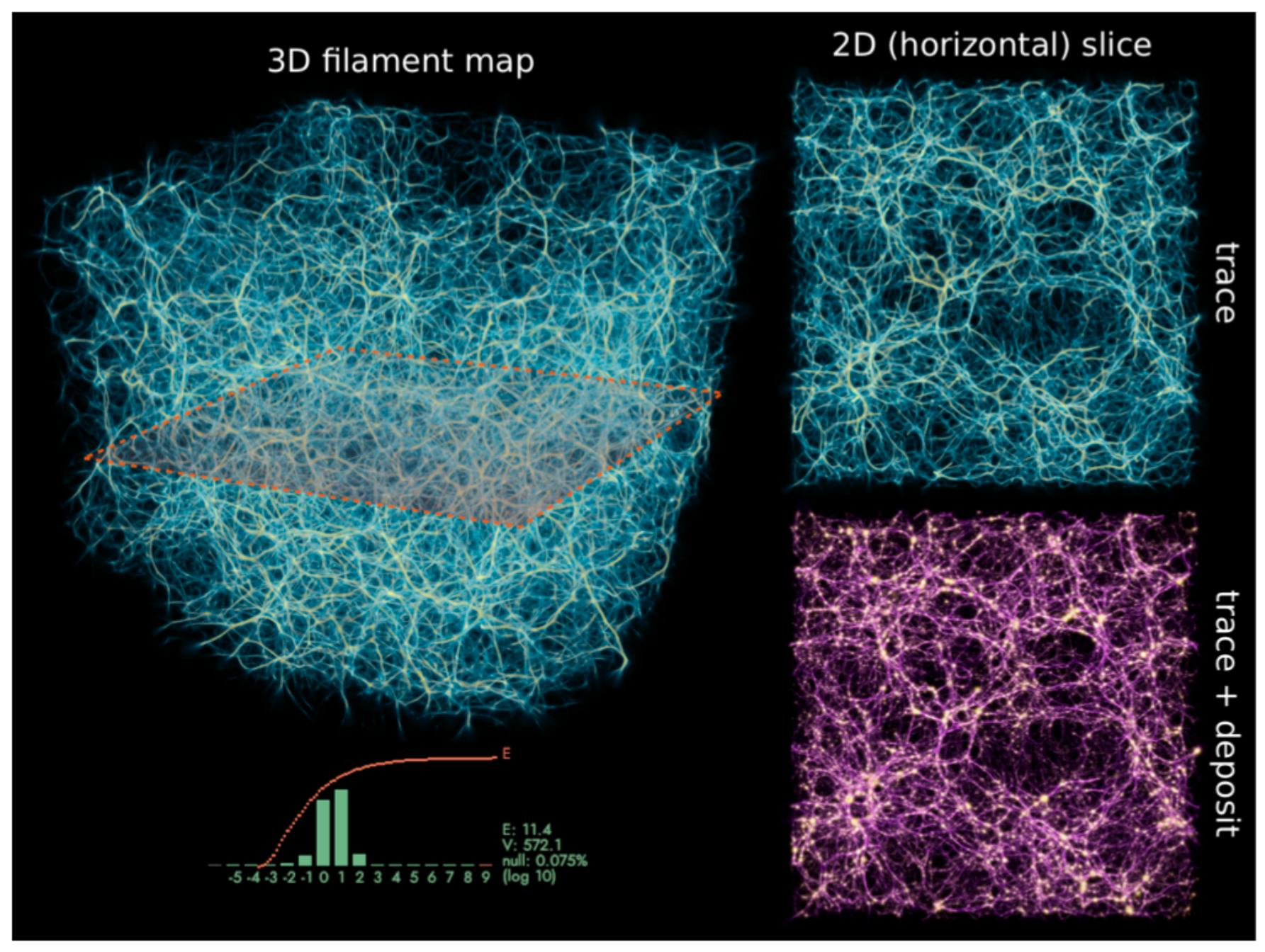}
    \caption{\textsf{Polyphorm:} reconstruction of dark matter filaments in the simulated BolshoiPlanck dataset where \textsf{Polyphorm} yields a consistent 3D structure, enabling its calibration to cosmic over density values. Thin slices of the filament map are shown on the right. Image reproduced from Elek et al.~\cite{ElekBurchettProchaska2021}.}
    \label{fig:polyphorm}
\end{figure}

\para{Dark matter filament reconstruction.}
\textsf{Polyphorm}~\cite{ElekBurchettProchaska2021} is an interactive visualization and filament reconstruction tool that enables the investigation of cosmological datasets (see~\autoref{fig:polyphorm}). Through a fast computational simulation method inspired by the foraging behavior of Physarum polycephalum, astrophysicists are able to extrapolate from sparse datasets, such as galaxy maps archived in the SDSS, and then use these extrapolations to inform analyses of a wide range of other data, such as spectroscopic observations captured by the Hubble Space Telescope. Researchers can update the simulation at interactive rates by a wide range of adjusting model parameters. \textsf{Polyphorm} has been used to reconstruct the cosmic web from galaxy observations~\cite{Burchett_RevealingDarkThreads_2020} and to infer the ionized intergalactic medium contribution to the dispersion measure of a fast radio burst~\cite{Simha_Disentangling2020_ApJ}.

\para{Visual verification of simulations.}
Currently, predictions of the Sun's Coronal mass ejections (CMEs) rely on simulations generated from observed satellite data. CMEs are powerful eruptions from the surface of the sun. These simulations possess inherit uncertainty because that the input parameters are entered manually, and the observed satellite data may contain measurement inaccuracies.  These simulations treat CMEs as  singular objects with discrete boundaries that are well defined and thus enable their treatment as entire objects.
In order to mitigate this uncertainty, Bock \etal~\cite{BockPembrokeMays2015} proposed a multi-view visualization system that generates an ensemble of simulations by perturbing the CME input parameters, and enables comparisons between these simulations and ground truth measurements. 
The system has many capabilities useful to domain experts, including integration of 3D rendering of simulations with satellite imagery, comparison of simulation predictions with observed data, and time-dependent analysis.

\para{3D visualization of planetary surfaces.}
Ortner \etal~\cite{OrtnerWalchNowak2020} performed 3D reconstruction visualization for planetary geology. 
Their geological analysis of 3D Digital Outcrop Models is used to reconstruct ancient habitable environments, which serves as an important aspect of the upcoming ESA ExoMars 2022 Rosalind Franklin Rover and the NASA 2020 Rover Perseverance missions on Mars. 
They conducted a design study to create \textsf{InCorr} (Interactive data-driven Correlations), which includes a 3D geological logging tool and an interactive data-driven correlation panel that evolves with the stratigraphic analysis. 
See~\cite[Section 2.2.2]{Gerndt2014} for more references on Mars geology and geodesy data and tools.  
Bladin \etal~\cite{BladinAxelssonBroberg2018} integrated multiple data sources and processing and visualization methods to interactively contextualize geospatial surface data of celestial bodies for use in science communication.

%% file: sec-data-accessibility.tex
\begin{table*}[!ht]
	\centering
\resizebox{1.0\textwidth}{!}{	
	\begin{tabular}{|>{\centering\arraybackslash}m{0.1\textwidth}
	||>{\centering\arraybackslash}m{0.06\textwidth}
	>{\centering\arraybackslash}m{0.06\textwidth}
	>{\centering\arraybackslash}m{0.06\textwidth}
	>{\centering\arraybackslash}m{0.06\textwidth}
	>{\centering\arraybackslash}m{0.06\textwidth}
	>{\centering\arraybackslash}m{0.06\textwidth}
	>{\centering\arraybackslash}m{0.06\textwidth}
	||>{\centering\arraybackslash}m{0.06\textwidth}
	>{\centering\arraybackslash}m{0.06\textwidth}
	>{\centering\arraybackslash}m{0.06\textwidth}
	>{\centering\arraybackslash}m{0.06\textwidth}
	|>{\centering\arraybackslash}m{0.06\textwidth}
	>{\centering\arraybackslash}m{0.06\textwidth}|}
	
\mc{\imglarger{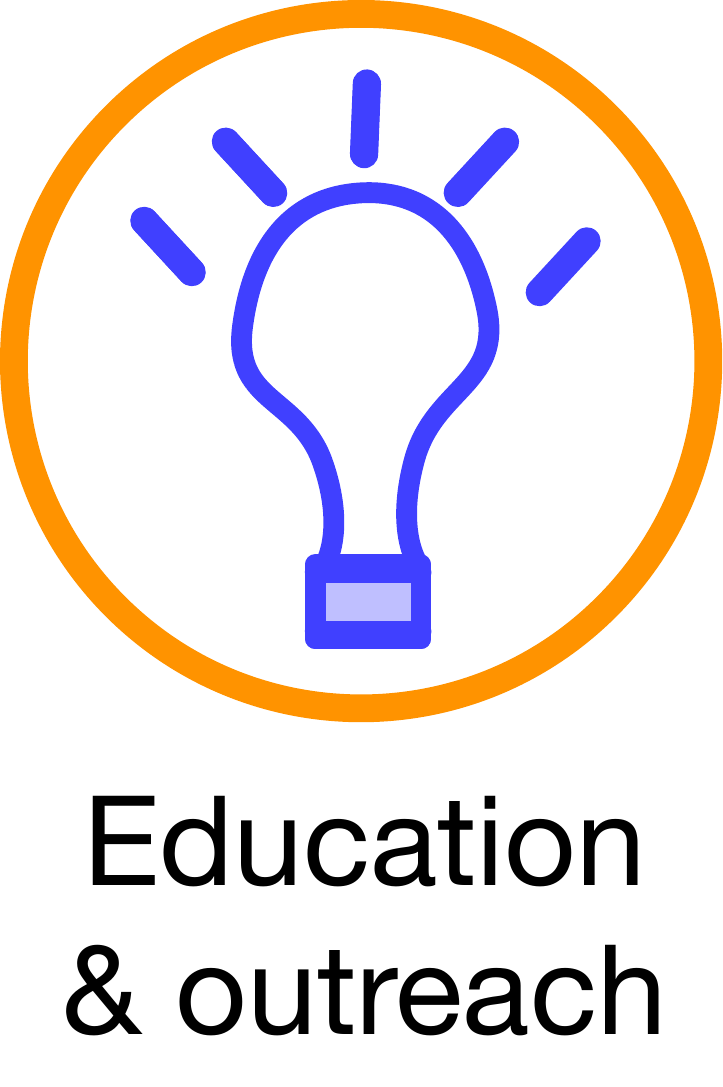}}
& \mc{\imglarger{symbol-plots-txt.pdf}}
& \mc{\imglarger{symbol-images-txt.pdf}}
& \mc{\imglarger{symbol-render-txt.pdf}}
& \mc{\imglarger{symbol-interactive-txt.pdf}}
& \mc{\imglarger{symbol-dr-txt.pdf}}
& \mc{\imglarger{symbol-uv-txt.pdf}}
& \mc{\imglarger{symbol-vr-txt.pdf}}
& \mc{\imglarger{symbol-exc-txt.pdf}}
& \mc{\imglarger{symbol-gal-txt.pdf}}
& \mc{\imglarger{symbol-pla-txt.pdf}}
& \mc{\imglarger{symbol-saa-txt.pdf}}
& \mc{\imglarger{symbol-simulations-txt.pdf}}
& \mc{\imglarger{symbol-observations-txt.pdf}}
\\ \hline	
	
\cite{BockHansenYnnerman2018}
&           & $\bullet$ & $\bullet$ & $\bullet$ &           &           & $\bullet$ &           &                    &       $\bullet$     &                    &       $\bullet$&           \\ \hline
\cite{FahertySubbaRaoWyatt2019}
&           &           & $\bullet$ &        &           &           & $\bullet$ &  $\bullet$    &      $\bullet$     &      $\bullet$     &       $\bullet$    &   $\bullet$        &  $\bullet$          \\ \hline
\cite{RosenfieldFayGilchrist2018}
&           &   $\bullet$  & $\bullet$ & $\bullet$ &           &           & $\bullet$ &  $\bullet$  &     $\bullet$      &     $\bullet$      &  $\bullet$  &   $\bullet$  &      $\bullet$     \\ \hline
\cite{UsudaSatoTsuzukiYamaoka2018}
&           &           & $\bullet$ & $\bullet$ &           &           &           &     $\bullet$      &       $\bullet$    &           &           &           &    $\bullet$       \\ \hline
\cite{OrlandoPillitteriBocchino2019}
&           &           & $\bullet$ & $\bullet$ &           &           & $\bullet$ &           &      $\bullet$     &      $\bullet$     &       $\bullet$    &   $\bullet$        &           \\ \hline
\cite{ArcandJiangPrice2018}
&           &           & $\bullet$ & $\bullet$ &           &           & $\bullet$ &           &     $\bullet$      &           &           &           &      $\bullet$     \\ \hline
\cite{VogtShingles2013}
&           &           &           & $\bullet$ &           &           & $\bullet$ &           &      $\bullet$       &           &           &           &       $\bullet$      \\ \hline
\cite{Madura2017}
&           &           &           & $\bullet$ &           &           & $\bullet$ &           &     $\bullet$      &           &           &    $\bullet$      &     $\bullet$      \\ \hline
\cite{DiemerFacio2017}
&           &           & $\bullet$ &           &           &           & $\bullet$ &     $\bullet$       &           &           &           &   $\bullet$         &           \\ \hline
\cite{DykesHassanGheller2018}
&           &           & $\bullet$ & $\bullet$ & $\bullet$ &           &           &     $\bullet$      &           &           &           &     $\bullet$      &           \\ \hline

\end{tabular}
}	
\caption{Classifying papers under education and outreach based on secondary and tertiary categories.
Top row, from left to right: (primary category) Education and outreach; (secondary categories) 2D/3D plots, 2D images, 3D rendering, interactive visualization, dimensionality reduction, uncertainty visualization, and new display platforms; (tertiary categories) extragalactic, galactic, planetary, and solar astronomy; (tags) simulated, and observational data.
}
\label{table:data-access}
\end{table*}

\section{Education and Outreach}
\label{sec:data-accessibility}

Currently, an on-going paradigm shift is occurring in scientific outreach. Technological advances are enabling data-driven and interactive exploration to be possible in public environments such as museums and science centers, increasing their availability to the general public. These advances are shortening the distance between research and outreach material, and enriching the scientific exploration process with new perspectives. Ynnerman \etal~\cite{YnnermanLowgrenTibell2018} and Goodman \etal~\cite{GoodmanHansenWeiskopf2019} introduced the \emph{Exploranation} concept, a euphemism encapsulating this confluence of explanation and exploration.  

Scientific storytelling of astrophysical findings using visualization has a deep history.  Ma \etal~\cite{MaLiaoFrazier2011} described how visualization can aid scientific storytelling using the NASA Scientific Visualization Studio.  
Borkiewicz \etal~described storytelling based on data-driven cinematic visualization in a SIGGRAPH course~\cite{BorkiewiczChristensenKostis2019}. 
More recently, and with a greater focus on interactive methods where the user becomes part of the exploration, Bock \etal~\cite{BockHansenYnnerman2018} described the challenge of presenting the details of NASA space missions to the public.

Research efforts in this category (as summarized in~\autoref{table:data-access}) are related \emph{w.r.t.} important aspects of education outreach and/or public accessibility. In addition, several are concerned with large-scale immersive visualization in planetariums and also with personal virtual reality (VR) and augmented reality (AR) experiences.   
Absent from the current literature, to the best of our knowledge, is a comprehensive analysis of the effect of VR for scientific exploration in astronomy. 

\begin{figure}[b]
    \centering
    \includegraphics[width=0.98\columnwidth]{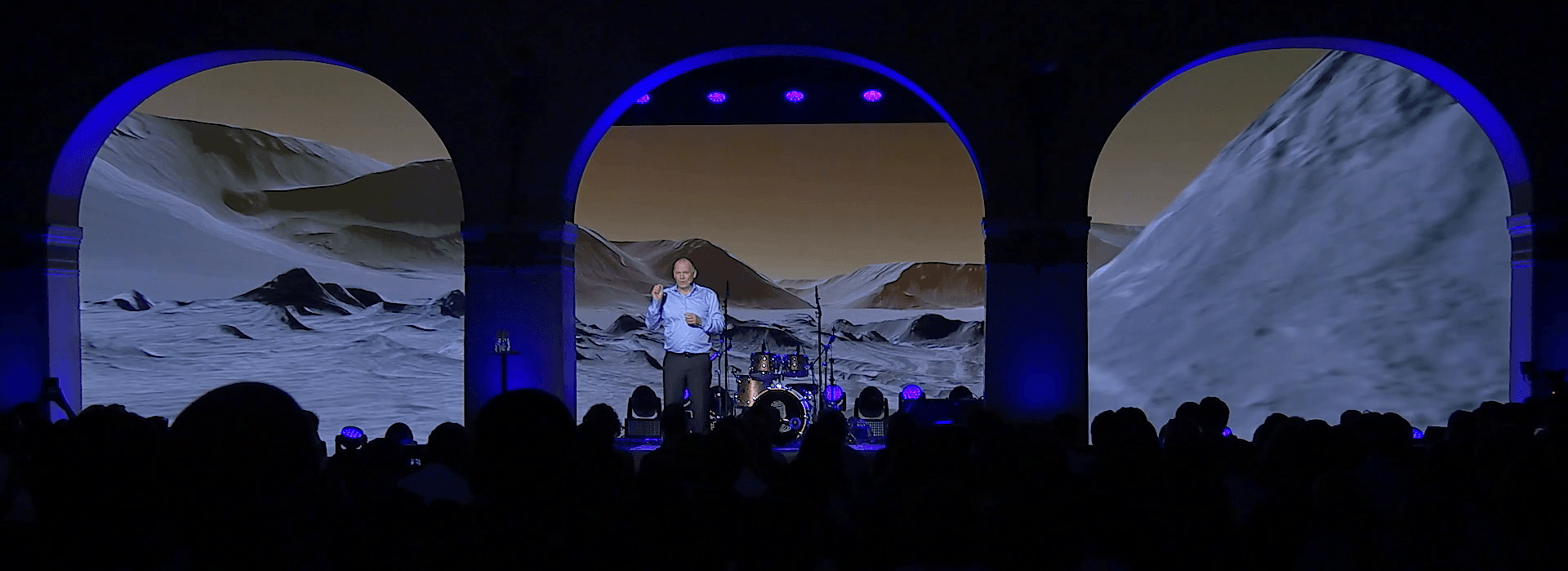}
    \caption{An example of an interaction presentation in a large-scale immersive environment to the general public, in this case of topographical features on the surface of Mars present at the \emph{Brilliant Minds} conference.}
    \label{fig:outreach-planetarium}
\end{figure}

\para{Planetarium and other large-scale immersive environments.}
Immersive visualization in planetarium dome theaters (see~\autoref{fig:outreach-planetarium}) has been the primary outreach mechanism for astronomy from their initial conception.  The immersive nature of the display system plays an important role in the contextualization of the available data, which is one of the unique challenges of astronomical datasets.  The birth of the usage of interactive visualization software in planetarium can be traced to the \textsf{Uniview} software~\cite{KlashedHemingssonEmmart2010}, which pioneered many of the interaction paradigms that are still in use today.  To a large extent, these live presentations based on interactive visualization are enabled by software provided by the planetarium vendors, which are, in general, commercial solutions and thus fall outside the scope of this survey.  Our focus here is instead on the large number of open-source initiatives, which are easily accessible to the academic community, targeting planetariums and other large-scale immersive environments.  Although aimed at the use in immersive environments, these initiatives also constitute a bridge between outreach and research-driven data exploration, as described by Faherty \etal~\cite{FahertySubbaRaoWyatt2019}, which is increasingly gaining momentum.

Among the most widely used software packages tailored to astrophysical data in immersive environments is \textsf{WorldWide Telescope}~\cite{RosenfieldFayGilchrist2018}, which is a virtual observatory for the public to share and view data from major observatories and telescopes.  The software provides the capability to visualize the solar system and stars, and show observational material in context; however it focuses on the data as displayed from the Earth's viewpoint.  
\textsf{Celestia} is an open-source initiative that shows the objects of the solar system and the greater universe in a 3D environment.  It provides high-resolution imagery and accurate positioning of the planetary bodies of the solar system and the ability to show other datasets in their context outside the solar system.  \textsf{OpenSpace}, \textsf{Gaia Sky}, and \textsf {ESASky}, as described in~\autoref{sec:data-exploration}, also provide contextualization of astronomical data, but with a stronger emphasis on the ability for domain experts to import their data into an immersive environment for public presentations. The \textsf{Stellarium} software can be used by the general public to look at a virtual nights sky from the surface of any planet.  The data contained in the software include a star catalog, known deep space objects, satellite positions, and other datasets that can be added dynamically by the user.  \textsf{NASA Eyes} is a suite of web-based visualization tools that enable the user to learn about the Earth, Solar System, Exoplanets, and ongoing NASA missions.  While providing a rich experience for the end user, the avenues for extension are limited.  The \textsf{Mitaka} software~\cite{UsudaSatoTsuzukiYamaoka2018} enables users to explore the observable universe and makes it easy for them to create custom presentations that can be shown on a large variety of immersive display environments.

\para{Personal virtual and augmented reality.}
In stark contrast to the immersive display environments described thus far, a large amount of work has been presented in the realm of virtual reality~(VR)~and augmented reality~(AR).  VR in this context refers to a \emph{personal} experience, rather than one that is \emph{shared} with other participants.  These two fields overlap in some areas, but they present distinct research challenges.

With \textsf{3DMAP-VR}, Orlando \etal~\cite{OrlandoPillitteriBocchino2019} sought to unite the excessively data-rich world of astronomy with VR with the hopes of facilitating productive engagement with traditionally inaccessible data. Specifically, they visualized 3D magnetohydrodynamic (MHD) models of astronomical simulations, which include the effects of gravity and hydrodynamics as well as magnetic fields. 
Their workflow consisted of two steps: obtaining accurate simulations and then converting these simulations to navigable 3D~VR environments using available data analysis and visualization software. 
In additional to providing a method to explore these dense data cubes in VR, they allowed these VR environments to be explored by anyone with a VR~rig by uploading them to \textsf{Sketchfab}, a popular open platform for sharing VR content. 
Orlando \etal~excelled at meeting two emerging goals in astronomical visualization: first, using existing software to achieve their goals rather than creating something from scratch; and second, making the visualizations widely accessible.

Arcand \etal~\cite{ArcandJiangPrice2018} developed a 3D VR~and~AR~program to visualize the \emph{Cassiopeia A} (\ie, \emph{Cas A}) supernova remnant, the resulting structure from an exploded star. They aimed to make the best use of the high-resolution, multi-wavelength, multi-dimensional astronomical data and give users the experience of walking inside the remains of a stellar explosion. They first performed 3D reconstruction of \emph{Cas A} and then employed volume and surface rendering to display the model in the VR~system \textsf{YURT} (\ie, Yurt Ultimate Reality Theatre). The user can select a specific part of the supernova model and access the annotations. These interactive features not only help non-experts engage in the story of the star, but also assist researchers observe changes in its properties.

Vogt \etal~\cite{VogtShingles2013} explored the potential of AR in astrophysics research/education by introducing Augmented Posters and Augmented Articles. 
The authors included Augmented Posters at the Astronomical Society of Australia Annual General meeting in 2012. Incorporating AR into posters allowed attendees to use their smartphones to engage with the posters in a virtual space and easily save and share the poster information if they found it interesting. 
Through tracking of engagement and feedback, they discovered that the majority of conference attendees found the technology to ``have some potential.'' As mentioned, the authors also experimented with Augmented Articles. They showed how results from an earlier work (the 3D structure of super nova remnants) can be viewed in 3D interactively within the article using a smartphone. Vogt \etal\ concluded by speculating on the future of AR in astrophysics. They were optimistic about the potential for AR to be an effective supplementary technology, but cited long-term stability and backwards compatibility in terms of AR apps and technology as a major limitation to AR moving forward. They suggested that a dedicated service for AR used in scientific publishing and outreach may be an effective way to handle this limitation.

\begin{figure}[!h]
    \centering
    \includegraphics[width=0.98\columnwidth]{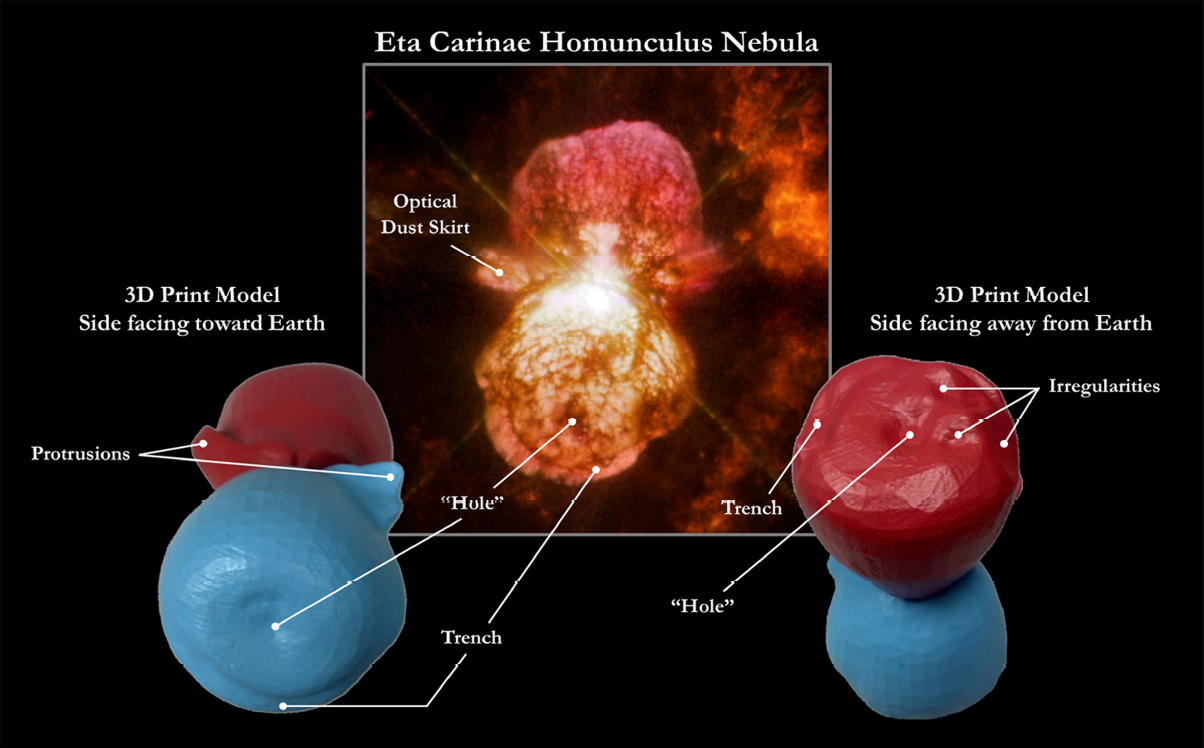}
    \caption{Dual-color 3D print of the $\eta$ Car Homunculus nebula. Image reproduced from Madura~\cite{Madura2017}.}
    \label{fig:3d-printing}
\end{figure}

\para{Novel interfaces.}
Madura \cite{Madura2017} presented a case study using 3D printing to visualize the \emph{$\eta$ Car Homunculus nebula}, see~\autoref{fig:3d-printing}. Extending the traditional monochromatic 3D prints, Madura proposed to use full-color sandstone prints to generate more informative models. Although the sandstone material is not as sturdy, these printers produce noticeably higher quality prints that preserve smaller details. The colors of the prints can be based on physical properties, which provides additional information to visual learners and helps distinguish different structures. The 3D models not only facilitate research discoveries, but also help communicate scientific discoveries to non-astronomers, especially to the visually impaired. The New Mexico Museum of Space History and the New Mexico School jointly hosted the first week-long astronomy camp for the visually impaired students across states in the summer of 2015. The camp received overwhelmingly positive feedback. 
Madura also discussed the use of other methods, including audio, tactile sign language, and tactile fingerspelling, to further expand the 3D model interactive experience for tactile learners. Overall, 3D printing could be a useful and effective tool for astronomy outreach and education.

In a work as artistic as it is scientific, Diemer \etal~\cite{DiemerFacio2017}  modeled the cosmic web through dark matter simulations and represented it artistically through 3D sculptures and woven textiles. 
The dark matter simulation is run using the publicly available code \textsf{GADGET2} and the halos and \emph{subhalos} (\ie, the halos within halos), are identified using \textsf{ROCKSTAR}. 
To identify the structural elements of the cosmic web (walls and filaments), they used \textsf{DISPERSE}, which is publicly available code that leverages discrete Morse theory to identify salient topological features (see~\autoref{sec:feature-identification}). 
Converting from the simulation data to their artistic representation, Diemer \etal\ stated that ``we believe that art, as much as science, seeks to say something true about the nature of existence, and that end is best served by artistic representation that grapples with real data and not only with allegorical concepts.'' 
They accomplished this stated ideal through a structured simplification of the model to a form where it can be represented using 3D woven textiles. 
The techniques used in their paper are not novel, but the combination of them is, and the end result is a powerful installation that instills wonder in those who move through it.  This work is able to take scientifically rigorous simulation data and represent it in an accessible form without losing the deep beauty of the underlying science. 

This elegant translation of numbers to forms, of thoughts to feelings, lies at the heart of science communication and outreach. The importance of this translation is especially crucial for astronomy, where the physical embodiment of the things we are studying can really only ever live in our minds and as the modern equivalent of paint on the walls of our caves.

\para{Virtual observatories.} 
Virtual observatories (VOs) are web-based repositories of astronomical data from multiple sources with the goal of improving access to astronomical data for scientists not directly involved in data collection. Many advanced VO applications aid in the mining and exploration of observational data through the use of state-of-the-art visualization techniques, but comparatively few that perform similar functions for theoretical data. In the interactive 3D visualization for theoretical VOs, Dykes \etal~\cite{DykesHassanGheller2018} examined the current capabilities of VOs containing theoretical data, and additionally presented a tool based on \textsf{SPLOTCH}, which is designed to aid in addressing some of the shortcomings they identified with current methods. \textsf{SPLOTCH} is a visualization tool designed for high-performance computing environments that is capable of quickly rendering large particle datasets, which makes it ideal for interactive visualization of 3D particle data. Dykes \etal\ combined their tool with a VO and demonstrated the effectiveness of interactively filtering and quantitatively visualizing the data for identifying features of interest in large particle simulations. Future steps involve comparative visualization, which would consist of methods to generate mock 2D observational images from the 3D simulation data to compare with actual observations.

\para{Broad dissemination through mobile applications.}
Furthermore, recent progress has been made on mobile and commercial applications that provide visualizations of astronomical data. 
Although these applications are not focused on research questions, they serve to broadly disseminate astrophysics visualizations, many of which also have strong educational value for the general public.

For instance, \textsf{SkySafari} ({\small{\url{https://skysafariastronomy.com/}}}), \textsf{StarMap} ({\small{\url{http://www.star-map.fr/}}}), \textsf{NightSky} ({\small{\url{https://icandiapps.com/}}}), or \textsf{SkyView} (on Apple App Store) bring stargazing to everyone's mobile phones, offer AR features that guide the layperson towards interested targets, and can also be used to control amateur telescopes and aid in astrophotography. 
Many JavaScript libraries and tools are available for astronomical visualization, such as \textsf{Asterank} ({\small{\url{https://www.asterank.com/}}}), based on spacekit.js ({\small{\url{https://typpo.github.io/spacekit/}}}), which enables the user to visually explore a database containing over 600,000 asteroids, including estimated costs and rewards of mining asteroids.
 
Web applications that introduce scientific knowledge of astronomy are also easily accessible.
For example, NASA JPL's \textsf{Eyes on the Solar System} ({\small{\url{https://eyes.nasa.gov/}}}) has the ability to show the dynamics of our solar system, but can also be used to show the evolution of space missions, and the discovery of exoplanets.  Another example is an adaptation of the \textsf{Uniview} software to be used in schools on mobile platforms targeting grades 4-6 in the NTA Digital project.

There are also a number of popular applications for VR headsets, such as \textsf{Star Chart} ({\small{\url{http://www.escapistgames.com/}}}) and \textsf{Our Solar System} on Oculus, which provide immersive experiences for users seeking knowledge about the Universe.
\textsf{Merge Cube} ({\small{\url{https://mergeedu.com/cube}}}) accompanied with AR/VR apps such as \textsf{MERGE Explorer} has been used to enable a new way of interactive learning for astronomy and beyond. It allows users to hold digital 3D objects and explore stars and galaxies in their palms.

%% file: sec-challenges.tex
\section{Challenges and Opportunities}
\label{sec:challenges}

In addition to a taxonomy of existing approaches that utilize visualization to study astronomical data from~\autoref{sec:data-wrangling} to \autoref{sec:data-accessibility}, our contribution includes a summary of the current challenges and opportunities in visualization for astronomy. 
We ask the following questions: What are the missing tools in current visualization research that astronomers need to formulate and test hypotheses using modern data? What visualization capabilities are expected to become available for astronomical data over the next decade?

In a Carnegie + SCI mini-workshop conducted in April 2020 and a Visualization in Astrophysics workshop during IEEE VIS in October 2020, astrophysicists and visualization experts discussed recent advances in visualization for astrophysics, as well as the current visualization needs in the astronomical community. 
As a result of these workshops, we have identified the following list of challenges and opportunities: 
\begin{itemize}
    \item Open-source tools: we need more open-source data visualization software that is suitable for astronomical data.  These tools must be flexible, modular, and integrable within a broader ecosystem of workhorse tools; 
    \item Intelligent data querying: we need to enable intelligent data queries for large data;
    \item Discovery: we need ways to turn high-quality renderings of data (observed and simulated) into quantitative information for discovery;
    \item Scalable feature extraction: we need to extract and visualize features from large and physically complex data cubes;
    \item In situ analysis and visualization: we need to interact with simulation data in real time, by utilizing visualization for parameter tuning and simulation steering;
    \item Uncertainty visualization: we need to develop more techniques to mitigate and communicate the effects of data uncertainty on visualization and astronomy;
    \item Benchmarks: we need to develop clear, widely adopted benchmarks or mock data catalogs for comparison with observed data; 
    \item Time and space efficiency: we need to improve upon memory and/or space intensive data analysis tasks. 
\end{itemize}

\subsection{Challenges Identified from Previous Surveys}
We first review the challenges identified from previous surveys\cite{HassanFluke2011,LipsaLarameeCox2012} and describe how the community has responded to these challenges in the past decade.  

Hassan \etal~\cite{HassanFluke2011} identified six grand challenges in their 2011 survey for the peta-scale astronomy era: 
\begin{itemize}
\item Support quantitative visualization;
\item Effectively handle large data sizes;
\item Promote discoveries in low signal-to-noise data;
\item Establish better human-computer interaction and ubiquitous computing;
\item Design better workflow integration;
\item Encourage adoption of 3D scientific visualization techniques. 
\end{itemize}
Lipsa \etal~\cite{LipsaLarameeCox2012} discussed visualization challenges in astronomy in their 2012 survey; however, only a few papers had addressed these challenges at the time of the survey.  These challenges include:
\begin{itemize}
\item Multi-field visualization, feature detection, graphics hardware;
\item Modeling and simulation;
\item Scalable visualization, error and uncertainty visualization, time-dependent visualization, global and local visualization, and comparable visualization. 
\end{itemize}
In the past decade, considerable advances have been made in addressing the challenges identified by Hassan \etal\ and Lipsa \etal\
 With modern computing power, interactive visualizations have become increasingly popular for both scientific explorations and public outreach.
 A variety of scalable visualization tools for large simulation and survey data are now easily accessible (e.g.,~\textsf{ParaView}~\cite{WoodringHeitmannAhrens2011},  \textsf{OpenSpace}~\cite{BockAxelssonCosta2020} and \textsf{Gaia Sky}~\cite{SagristaJordanMuller2019}). Many tools are also adopting graphics hardware and parallelism in their visualization, rendering, and analysis processes to increase efficiency (e.g., \textsf{yt}~\cite{TurkSmithOishi2010} and \cite{MohammedPolysFarrah2020}).
 Scientists and educators are also incorporating novel visual display methods, such as VR~\cite{OrlandoPillitteriBocchino2019} and 3D printing \cite{Madura2017}, for education and outreach services.
 
Visualizations of evolving astronomical systems have also seen advances. Lipsa \etal~\cite{LipsaLarameeCox2012} listed only two papers under time-dependent astronomy visualization in their survey. In contrast, we present a number of research papers with the capabilities of analyzing halo evolution histories~\cite{ShanXieLi2014, PrestonGhodsXie2016, ScherzingerBrixDrees2017} and rendering real-time stellar orbits~\cite{SagristaJordanMuller2019}.
 Volumetric data can now be rendered in 3D with sufficient numerical accuracy to enable a wide range of research in feature detection and extraction (e.g., \textsf{AstroBlend}~\cite{Naiman2016}, \textsf{FRELLED}~\cite{Taylor2017} and \textsf{Houdini} for astrophysics~\cite{NaimanBorkiewiczChristensen2017}), see~\autoref{sec:feature-identification}.

Quantitative analysis for heterogeneous data types (\autoref{sec:data-exploration-heterogeneous}) is often supported as a supplement to the visual analysis (e.g., \textsf{Encube}~\cite{VohlBarnesFluke2016} and \textsf{TOPCAT}~\cite{Taylor2017b}). In order to perform analytic tasks effectively, many of these tools utilize visualization techniques such as linked-views (multi-field visualization~\cite{LipsaLarameeCox2012}, \textsf{Glue}\cite{Glue}), detail-on-demand (global/local visualization~\cite{LipsaLarameeCox2012}), and comparative visualization.
 
 On a higher level, a few techniques and platforms have been developed to provide better visualization workflow in astronomy. The \textsf{Glue} visualization environment described by Goodman \etal~\cite{GoodmanBorkinRobitaille2018} hosts a variety of shared datasests and open-source software. It facilitates flexible data visualization practices, and bridges the gap between scientific discovery and communication. On a similar note, the EU-funded \textsf{CROSS DRIVE}~\cite{Gerndt2014} creates ``collaborative, distributed virtual workspaces'' in order to unite the fragmented experts, data, and tools in European space science. Mohammed \etal~\cite{MohammedPolysFarrah2020} formalized the scientific visualization workflow and brought structure to a visualization designer's decision-making process. The paradigm provided by Mohammed \etal\ divides the visualization process into four steps: processing, computation of derived geometric and appearance properties, rendering, and display. In each of these steps, the workflow systematically incorporates high-performance computing to efficiently work with multi-variate multi-dimensional data.

 However, despite the progress, some of the challenges identified a decade ago, such as uncertainty visualization and time-dependent visualization, remain largely under-explored today or have great potential for improvement. 
A careful inspection of~\autoref{table:data-wrangle} to~\autoref{table:data-access} gives rise to a number of useful observations regarding research gaps for further investigation. 
 In this section, we describe a number of challenges and opportunities that we believe are essential for the development of visualization in astronomy in the years to come.

\subsection{Astronomical Data Volume and Diversity}
A challenge identified by both Hussan \etal~\cite{HassanFluke2011} and Lipsa \etal~\cite{LipsaLarameeCox2012} is the effective handling of large datasets.
Substantial effort and progress has been made in processing large datasets in the past decade in both astronomy and visualization. 
Luciani \etal~\cite{LucianiCherinkaOliphant2014} pre-processed large-scale survey data to ensure efficient query and smooth interactive visualization. 
\textsf{Frelled}~\cite{Taylor2015} accelerates visual source extraction to enable  the visualization of large 3D volumetric datasets. 
\textsf{Filtergraph}~\cite{BurgerStassunPepper2013} supports the rapid visualization and analysis of large datasets using scatter plots and histograms.  
\textsf{Gaia Sky} \cite{SagristaJordanMuller2019} uses the magnitude-space level-of-detail structure to effectively visualize hundreds of millions of stars from the Gaia mission with sufficient numerical precision. 
\textsf{yt} \cite{TurkSmithOishi2010} adopts parallelism to run multiple independent analysis units on a single dataset simultaneously.

Visualizing large datasets remains a challenge for astronomical data, especially because of its multi-dimensional property. 
Visualization researchers recognize that scalability is an immediate obstacle that prevents them from introducing many interactive capabilities~\cite{Taylor2015, PrestonGhodsXie2016, SteffenKoningWenger2011, WengerAmentGuthe2012, ScherzingerBrixDrees2017}. 
For analysis tasks, the developers of \textsf{yt} identified the challenge of load balancing for parallel operations on large simulation data. 
They added support for robust CPU/GPU mixed-mode operation to accelerate numerical computation~\cite{TurkSmithOishi2010}. 
We believe that even more improvements can be achieved by using network data storage and high-performance computing.

As the volume and diversity of data increase rapidly, connecting related heterogeneous datasets has become a priority in astronomy. 
Goodman \etal~\cite{GoodmanBorkinRobitaille2018} identified the growing open-source and collaborative environment as the future of astronomy. 
They described the \textsf{Glue}\cite{Glue} visualization environment, a platform that hosts a large variety of data and numerous open-source modular software. 
The \textsf{Glue} environment allows users to load multiple datasets at once and ``glue'' the related attributes together from different data types. 
Many exploratory astronomy visualization software packages (e.g., \textsf{OpenSpace}, \textsf{ESASky}) are capable of dealing with various data types. Some can integrate with \textsf{Glue}, which further improves their integrability and flexibility. 

Nevertheless, most software packages are still striving to expand the variety of data formats that they can process. 
Naiman \etal~\cite{NaimanBorkiewiczChristensen2017} aimed to use \textsf{Houdini} to render data with non-uniform voxel sizes. 
Baines \etal~\cite{BainesGiordanoRacero2016} retrieved spectroscopic data and aimed to link it to more of the mission catalogs for the next release of \textsf{ESASky}. 
Burchett \etal~\cite{BurchettAbramovOtto2019} incorporated data pre-processing as part of the \textsf{IGM-Vis} application to allow more data formats as inputs.
Vogt \etal~\cite{VogtSeitenzahlDopita2017} provided a unique perspective for simplifying the access to 3D data visualization by promoting the X3D pathway, as the X3D file format lies in the center of various visualization solutions, such as interactive HTML, 3D printing and high-end animations. However, the conversion into X3D file format remains the largest obstacle.

\subsection{Interactive Visualization and Intelligent Querying}

Hassan \etal\ identified ``better human-computer-interaction'' as one of the six grand challenges~\cite{HassanFluke2011}, and visualization experts and astronomers have joined forces to explore the potential of using interactive visualization in astronomy research and public outreach.
We see the overwhelming popularity of interactive visualization in the realm of astronomy research. 
Barnes and Fluke~\cite{BarnesFluke2008} demonstrated the convenience of embedding interactive visualizations in astronomy publications, via 3D PDF and its extension \textsf{S2PLOT} programming library. 
\textsf{Frelled} and \textsf{AstroBlend} leverage the 3D capability of \textsf{Blender} to improve the process of analyzing and visualizing volumetric data~\cite{Taylor2017, Naiman2016}. 
Naiman \etal~\cite{NaimanBorkiewiczChristensen2017} explored the potential use of the graphics tool \textsf{Houdini} in astronomy research. 
\textsf{ESASky}, \textsf{LSSGalpy}, \textsf{SDvision}, \textsf{OpenSpace}, and \textsf{Gaia Sky}~\cite{BainesGiordanoRacero2016, ArgudoFernandezPuertasRuiz2017, PomaredeCourtoisHoffman2017, BockAxelssonCosta2020, SagristaJordanMuller2019} all provide visual exploratory capabilities to large-scale astronomy survey data, each with their own scientific focuses and distinguishing features.

Many of these interactive software tools are expanding their impact in public outreach.  A video produced with the visualizations from \textsf{SDvision} -- titled ``Laniakea: Our home supercluster'' -- gained millions of views on YouTube~\cite{PomaredeCourtoisHoffman2017}. 
The authors are also pursuing the software's integration with VR technology to further contribute to education and public outreach services. 
\textsf{OpenSpace} has already demonstrated its success in museums, planetariums, and a growing library of publicly accessible video material. 
The software is built to be easily accessible to the general public via a simple installation onto any computer. 
AR, VR, and 3D printing are emerging technologies that are used at a greater scale in educational and public outreach services~\cite{ArcandJubettWatzke2019, Madura2017}. 
In order to reach a more artistic audience, Diemer \etal~\cite{DiemerFacio2017} have also explored integrating art and the physical visualization of astronomical objects.

However, many researchers also recognize the limitations of current interactive visualizations and intelligent querying of volumetric data. 
Barnes and Fluke~\cite{BarnesFluke2008} proposed the capturing of mouse clicks on individual elements of a scene to enable 3D selection and queries.  Goodman advocated for the need of 3D selection in astronomy visualization and analysis~\cite{Goodman2012}. 
\textsf{Blender} allows only one side of a transparent spherical mesh to be displayed at a time~\cite{Taylor2017}. 
The selection of pixels in regions of interest could also be a potential problem \cite{Taylor2017}. 
Yu \etal~\cite{YuEfstathiouIsenberg2016} proposed several context-aware selections in 3D particle clouds, which help to alleviate the issues associated with 3D selection and query. 
\textsf{Felix} tackles the challenge of querying by simultaneously satisfying two density ranges, but Shivashankar \etal\cite{ShivashankarPranavNatarajan2016} identified interactive visualization of intricate 3D networks as a ``largely unexplored problem of major significance''.
\textsf{WYSIWYG} creates a marching cube of a 2D selection and finds the cluster with the largest projection area as the cluster of interest \cite{ShanXieLi2014}. 
The technique lacks flexibility as it depends heavily on the assumption that the largest cluster is always of interest.

\subsection{Uncertainty Visualization} 

\begin{figure}[b]
    \centering
    \includegraphics[width=0.3\textwidth]{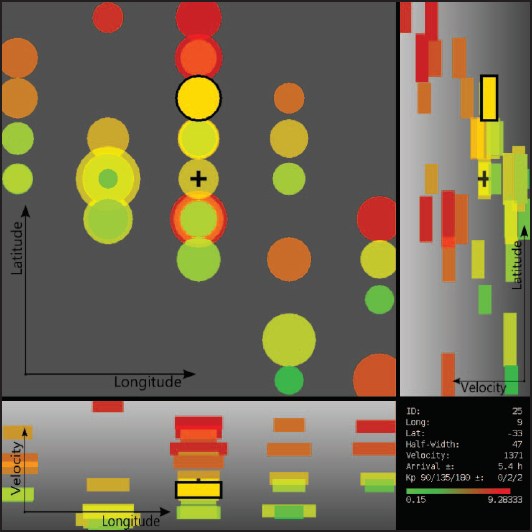}
    \caption{Ensemble selection view that captures the uncertainty of all ensemble runs by displaying the full 4D parameter space. Image reproduced from Bock et al.~\cite{BockPembrokeMays2015}.}
    \label{fig:uncertainty}
\end{figure}

Uncertainty visualization in astronomy remains largely unexplored, even though  errors and uncertainties are introduced due to data acquisition, transformation, and visualization.  
Li \etal~\cite{LiFuLi2007} noticed that uncertainty visualization is seldom available in astronomical simulations and developed techniques that enhance perception and comprehension of uncertainty across a wide range of scales.
Since then, a few studies have considered errors created during the simulation pipeline. 
Gro{\ss}schedl \etal~\cite{GrosschedlAlvesMeingast2018} used uncertainty plots to effectively present the distribution of the data and demonstrated the confidence in their reconstruction results. 
With the direct intention of incorporating uncertainty in the discovery process, Bock \etal~\cite{BockPembrokeMays2015} displayed the uncertainty of space weather simulation by visualizing an ensemble of simulation results with different input parameters (\autoref{fig:uncertainty}). 
Combined with a timeline view and a volumetric rendering of each ensemble member, scientists are able to compare each simulation with measured data, gain an understanding of the parameter sensitivities, and detect correlations between the parameters. 

Applying uncertainty visualization to 3D astronomy data is challenging because we lack the techniques to deal with sparse/far away samples and their large error cones. 
However, the potential exists to display uncertainty in a localized object or regions of interest, and that potential must be developed further.

\subsection{Time Series Data Visualization and Analysis}
\label{sec:time-series}
\begin{figure}[!h]
    \centering
    \includegraphics[width=0.98\linewidth]{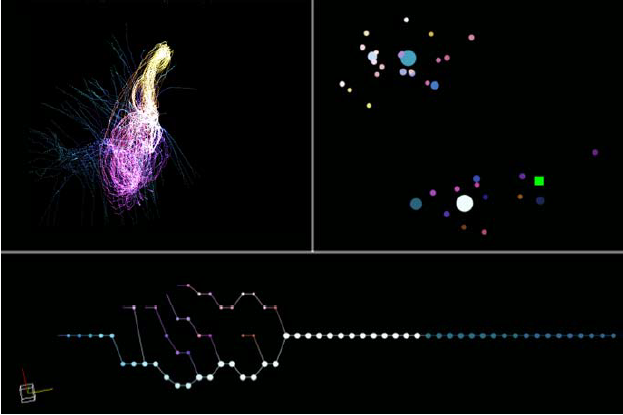}
    \caption{Interactive linked views of halo evolutionary history. Top left shows the evolution path of a selected halo and top right is the halo projected onto a 2D screen. Bottom is the merger tree visualization of the halo evolution. Image reproduced from Shan et al.~\cite{ShanXieLi2014}.}
    \label{fig:time-series}
\end{figure}

Most of the current time series data visualizations are built to display halo evolution.
One common technique is to use a merger tree to visualize the development of halos over time~\cite{ShanXieLi2014, almryde2015halos, PrestonGhodsXie2016, ScherzingerBrixDrees2017}.
Other techniques are often used along with the merger tree to enhance the effectiveness of the visualization. Shan \etal~\cite{ShanXieLi2014} visualized the changes of selected particles as a particle trace path image (see \autoref{fig:time-series}). Preston \etal~\cite{PrestonGhodsXie2016} added interactivity into their software and facilitated more efficient analysis for large, heterogeneous data. Scherzinger \etal~\cite{ScherzingerBrixDrees2017} extended their framework by adding particle visualization and analysis for individual halos.

In general, time series data visualization can also be helpful when tracking star movements. 
However, little effort has been expended in this regard, with \textsf{Gaia Sky} being the only example, to the best of our knowledge. 
Given the instantaneous proper motion vector of stars and simulation time, \textsf{Gaia Sky}~\cite{SagristaJordanMuller2019} computes a representation of proper motions. The software is able to visualize real-time star movements with sufficient numerical accuracy. 

\subsection{Machine Learning}

During our recent astrophysics workshops, many astronomers voiced their desires as well as concerns about using machine learning techniques (ML),  and in particular, deep learning in the astrophysics discovery processes, mostly surrounding the interpretability of ``black box" ML models.  
Active discussions have concerned the maturity of ML in astronomy, and a number of surveys have been created to assess this maturity~\cite{BallBrunner2010, FlukeJacobs2020, NtampakaAvestruzBoada2019}. 

\para{Combine visualization with machine learning.} 
Although the concerns regarding the interpretability of ML models in astronomy are valid in some cases, we believe that combining visualization with ML models has the potential to make the results more accessible to classical theoretical interpretation. Indeed, some of the most successful applications of ML in astrophysics involve cases where the interpretation is straightforward. 

The use of deep learning techniques in astrophysics has mostly been limited to convolutional neural networks (CNNs). 
Khan \etal~\cite{KhanHuertaWang2019} used a combination of deep learning and dimensionality reduction techniques to help construct galaxy catalogs. 
They extracted the features from one of the last layers of their pre-trained CNN and clustered the features using t-SNE. 
Their method not only leads to promising classification results, but also points out errors in the galaxy zoo dataset with the misclassified examples. 
Ntampaka \etal~\cite{NtampakaZuHoneEisenstein2019} presented a CNN that estimated galaxy cluster masses from the \emph{Chandra} mock images. 
They used visualization to interpret the results of learning and to provide physical reasoning. 
Kim and Brunner~\cite{KimBrunner2016} performed star-galaxy classification using CNNs. They studied images of activation maps, which help to explain how the model is performing classification tasks. 
Apart from deep learning, Reis \etal~\cite{ReisPoznanskiBaron2018} used random forests to generate distances between pairs of stars, and then visualize such a distance matrix using t-SNE. Their techniques have been shown to be useful to identify outliers and to learn complex structures with large spectroscopic surveys. 

Many efforts in recent years have focused on interpreting ML models~\cite{Molnar2020}. 
We believe a good starting point to obtain interpretability is to combine visualization with  models that are inherently interpretable~\cite{Rudin2019} (e.g., linear regression, decision tree, decision rules, and naive Bayes) in studying astronomical data. 
Alternatively, we may train an interpretable model as a surrogate to approximate the predictions of a black box model (such as a CNN) and integrate such a surrogate in our visualization. 

\para{Topological data analysis.}
Furthermore, topological data analysis (TDA) is an emerging field that promotes topology-based unsupervised learning techniques. TDA infers insights from the shape of the data, and topology has a reasonably long history in its applications in scientific visualization~\cite{HeineLeitteHlawitschka2016}.
A few researchers have applied TDA to astrophysics. 
Novikov \etal~\cite{NovikovColombiDore2006} were the first to propose the method of extracting the skeleton of the cosmic web using discrete Morse theory~\cite{Forman2002}. 
Both Sousbie \cite{Sousbie2011} and Shivashankar \etal~\cite{ShivashankarPranavNatarajan2016} used discrete Morse theory to develop geometrically intuitive methods that extract features from the cosmic web (e.g., filaments, walls, or voids). They demonstrated the efficiency and effectiveness of topological techniques in astronomical tasks. 
Xua \etal~\cite{XuaCisewski-KeheGreen2019} used TDA techniques to identify cosmic voids and loops of filaments and assign their statistical significance. 
Not many applications of topology have been proposed in de-noising astronomy data, other than the work of Rosen \etal~\cite{RosenSethMills2019}, which uses contour trees in the de-noising and visualization of radio astronomy (ALMA) data cubes.

\subsection{Further Advancements in Education and Outreach} 

A general ambition in science communication is to shorten the distance between research and outreach and make current research results and data available at science centers, museums, and in on-line repositories. This ambition applies to both shortening the time between discovery and dissemination and creating increased public access to research data. Even real-time public participation in scientific endeavors has been shown to be of public interest~\cite{BockHansenYnnerman2018}. This science communication trend is supported by rapid development of commodity computing platforms capable of handling large datasets, availability of open research data, and improved data analysis and visualization tools. These trends now enable visitors to public venues and home users to become ``explorers'' of scientific data. Astrophysics is one of the prime examples of a domain of large public interest and with vast amounts of publicly available data. 

\begin{figure}[!ht]
    \centering
    \includegraphics[width=0.98\columnwidth]{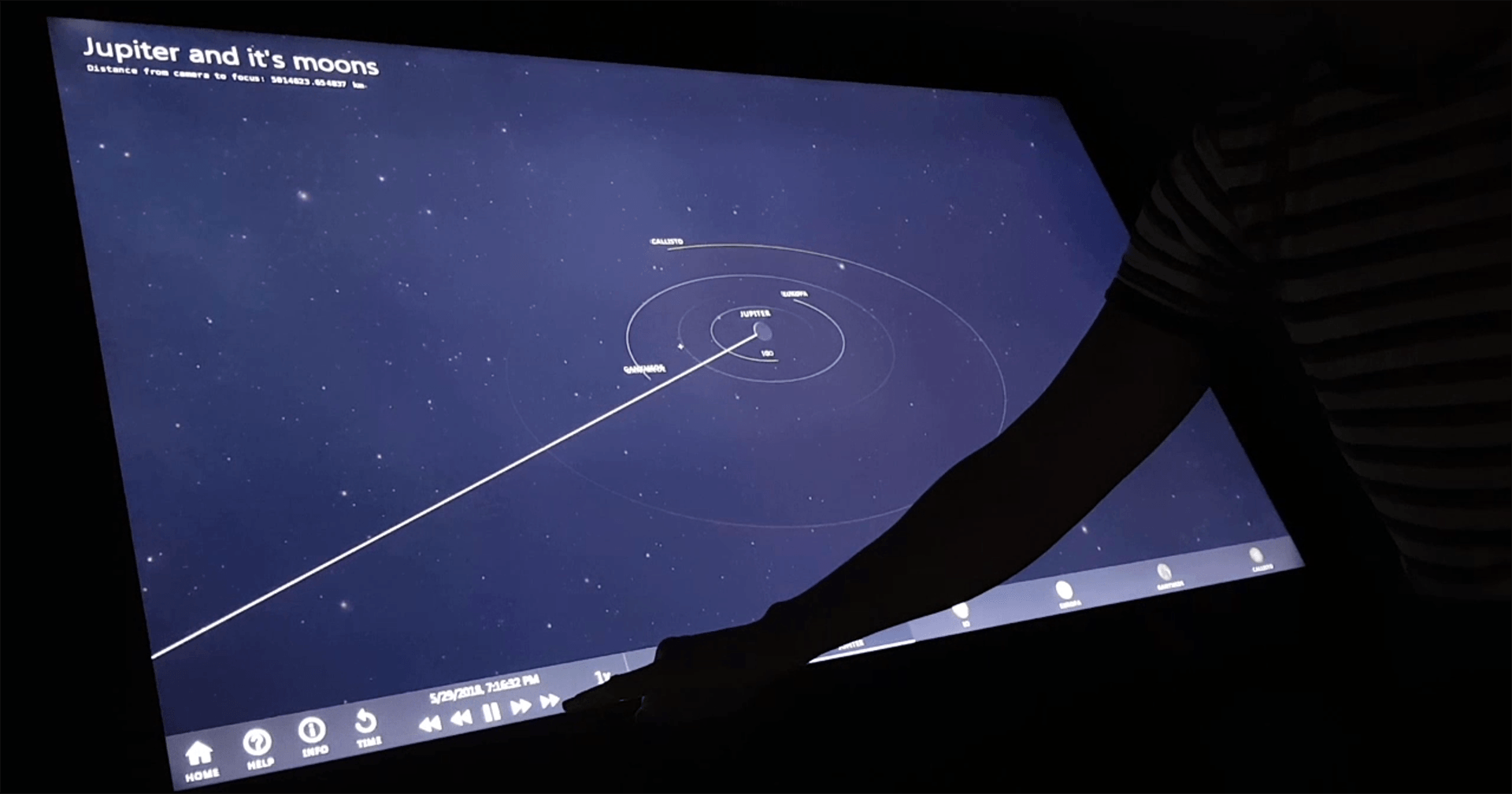}
    \caption{Usage of a touch interface in a museum installation guiding the user using cues built directly into the data exploration.}
    \label{fig:touch-astronomy}
\end{figure}

The trend described above poses several challenges. In a public setting, an interactive exploration has to be curated and guided in such a way that learning and communication goals are reached while not interfering with ``freedom'' of interaction~\cite{YnnermanRydellAntoine2016}. \autoref{fig:touch-astronomy} shows an example of this approach on a touch-capable table used in museum environments for a self-guided learning experience, exemplified on a CT scan of a meteoroid originating from Mars.
Ynnerman~\etal~\cite{YnnermanLowgrenTibell2018} coined the term \emph{Exploranation} to describe the introduction of data exploration in traditional explanatory contexts, and Goodman \etal~\cite{GoodmanBorkinRobitaille2018} described the need for interaction in explanation.  
This endeavor calls for a new generation of authoring and production tools targeting production of interactive non-linear storytelling\cite{WohlfartHauser2007}, with interfaces that interoperate with research tools and repositories. We also see that interactive installations will need to feature several different modes of operation. A first scenario would be a "walk-up-and-use" situation in which the content and the interaction are intuitive. The second scenario is a guided experience with a trained facilitator who can unlock advanced features of the installation and also bring in more data sources for an in-depth science communication session. 

Interaction plays a central role in science communication, and public settings put demands on robust, engaging, intuitive, and interactive visualization interfaces.
Sund\'en ~\etal~\cite{SundenBockJonsson2014} discussed aspects of demand and the potential use of multi-modal interfaces. Yu~\etal~\cite{YuEfstathiouIsenberg2012, YuEfstathiouIsenberg2016} addressed challenges posed by the interactive selections of data using touch surfaces.

In live presentation situations based on interactive software, more advanced tools are needed that support the presenter (and the pilot). Apart from the authoring tools discussed above, research on features such as automatic camera moves during presentations is also needed. An interesting challenge in view of advances in machine learning and natural language processing is the use of voice and gesture based interaction during presentations. Support for the embedding of multi-media data sources and other on-line services is also needed.

In outreach, the key role of visual representations cannot be underestimated, which calls for systems and tools that generate both visually appealing and still scientifically correct representations. The challenge here is a trade-off between artistic and scientific considerations. From an artistic point of view, Rector \etal~\cite{RectorLevayFrattare2017} aimed to strike a balance between the scientific and aesthetic quality of an astronomical image. They pointed out that people have different expectations and misconceptions of colored-image-based factors such as cultural variation and expertise. Therefore, scientists need to carefully consider the choices they make in order to create astronomical images. An example of how this challenge is met is the work on cinematic data visualization by Cox~\etal~\cite{CoxPattersonLevy2019,BorkiewiczChristensenWyatt2020}. Another example is the interactive blackhole visualization \cite{VerbraeckEisemann2021} described in~\autoref{sec:data-wrangling}.

The on-going rapid development of computer hardware creates opportunities and challenges as the users expect visual quality on the same level as the state-of-the-art games. At the same time, new levels of widespread public use are made possible. The challenge is to work with visual quality and performance as well as to create awareness of limited computer capabilities, data size and complexity. Another challenge for outreach is the use of social media and connected services, which entails not only development of tools and availability of data, but also engagement of a large number of domain experts with a science communication mission.

%% file: sec-navigation-tool.tex
\section{Navigation Tool}
\label{sec:navigation-tool}

\begin{figure*}[ht]
     \centering
     \includegraphics[width=1.0\textwidth]{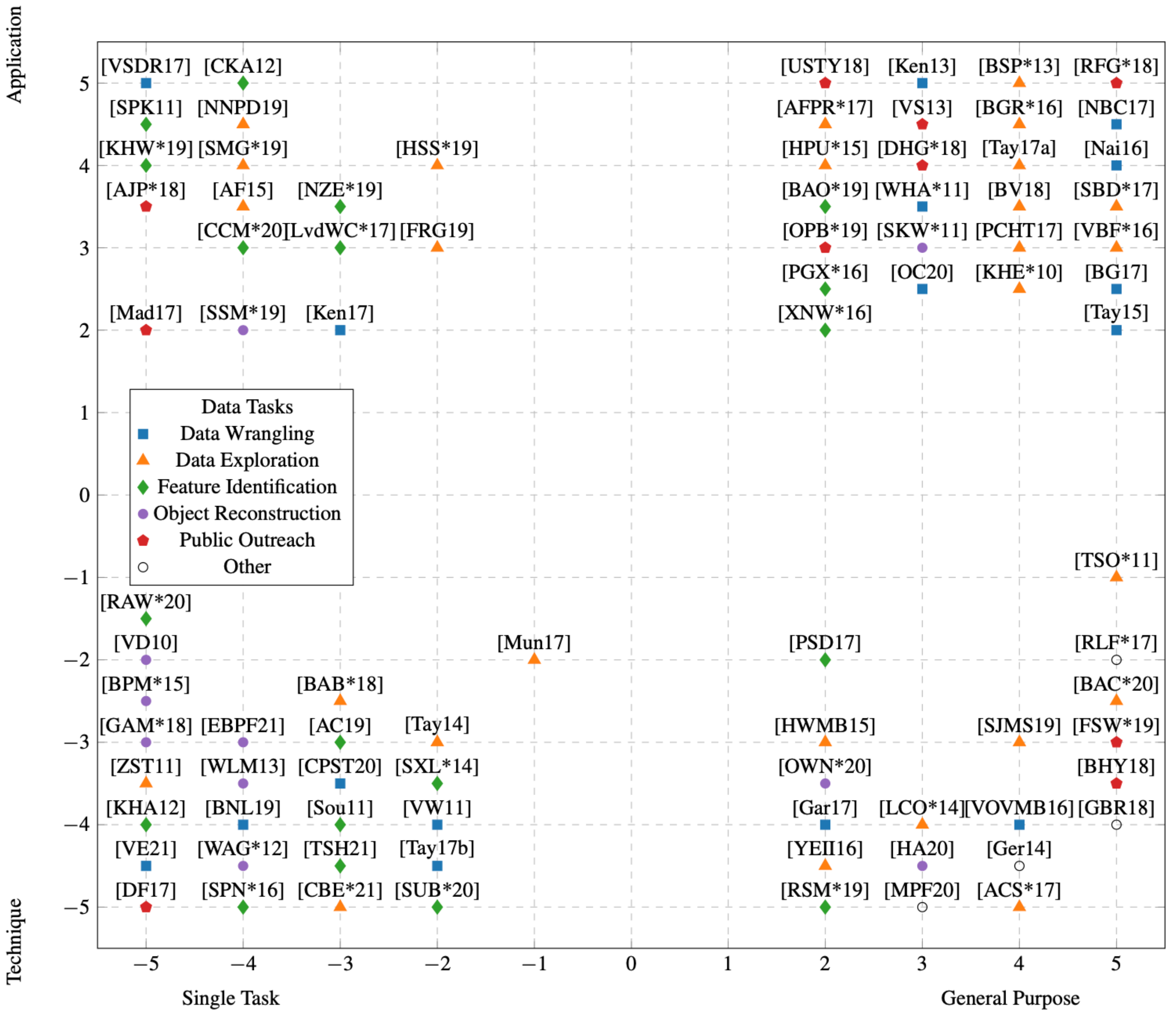}
     \caption{An alternative navigation tool that highlights the surveyed papers along two axes: single task vs. general purpose and technique vs application.}
     \label{table-nav-tool}
 \end{figure*}

Together with the classification and descriptions of the papers included in this survey, we complement our survey with a visual literature browser available at {\small{\url{https://tdavislab.github.io/astrovis-survis}}}. 
The visual browser follows the same classification scheme used in this report, where the users can use keyword searches to identify potential aspects in the field that are underserved. 
Additionally, we provide an alternative navigation tool within the visual browser (also illustrated in~\autoref{table-nav-tool}), where the surveyed papers are distributed along two axes. 
This tool provides a different viewpoint for the state-of-the-art survey. 

The first x-axis -- \textbf{single task vs. general purpose} -- specifies whether a specific paper addresses a singular challenge of visualization in astronomy (single task), or whether it describes a more general purpose system that can be applied to a wide array of potential applications (general purpose). A general purpose system also includes software systems that combine datasets of multiple modalities in a shared  contextualization.  
The second y-axis -- \textbf{technique vs. application} -- specifies whether a paper develops a specific visualization or analysis technique, or whether it  combines many different techniques to address a specific application.
The primary category -- data analysis tasks -- is double-encoded with colors and marker shapes. The "other" category represents relevant papers mentioned in the survey but that do not belong to any of the data analysis tasks. The coordinates of the papers in the navigation tool are based on our best estimation. We lay out the papers in their general areas in the figure to avoid overlap of labels.

%% file: sec-conclusion.tex
\section{Conclusions}
\label{sec:conclusions}

In this report, we provide an overview of the state of the art in astrophysics visualization.  
We have surveyed the literature and found that visualization in astrophysics can be categorized broadly into five categories based upon the primary user objectives: data wrangling, data exploration, feature identification, object reconstruction, and education and outreach.   

A major finding of this work is that there remains a significant gap between cutting-edge visualization techniques and astrophysical datasets.  
Among the 80+ papers surveyed, around 20 papers are from visualization venues. 
Given the scope of current and future datasets in astrophysics, as well as the advanced methodologies and capabilities in visualization research, the potential opportunity is great for scientific discovery in bridging this gap. However, this bridge will not build itself.

We therefore take this opportunity to issue a ``call to action'' for both the visualization and astrophysics communities to consider more robust and intentional ways of bridging the gap between visualization methods and astronomy data.  We make the specific recommendations below as concrete suggestions for improving this goal over the next decade.

We suggest the construction of a comprehensive \textbf{AstroVis Roadmap}  for bringing these disparate communities and stakeholders together at both the grassroots and institutional levels.  
In order to build community, we suggest regular annual joint meetings that will explicitly target this gap and bring together visualization and astrophysics domain expertise; the 2019 Dagstuhl Seminar on the topic of ``Astrographics: Interactive Data-Driven Journeys through Space'' is a good example~\cite{GoodmanHansenWeiskopf2019}. 
We specifically suggest yearly companion meetings to be held alternately at the Winter AAS or annual IAU meetings and the IEEE Visualization conferences.  Having explicit joint sponsorship of the professional society is an important step in growing this joint community.

We recognize and appreciate the ``grassroots'' efforts that bring together the visualization and astrophysics communities.  Indeed, this contribution is the direct result of a Carnegie-SCI workshop as well as the IEEE Vis2020 workshop ``Visualization in Astrophysics'' ({\small{\url{http://www.sci.utah.edu/~beiwang/visastro2020/}}}).  Other efforts include the \emph{RHytHM} (ResearcH using yt Highlights Meeting) at the Flatiron Institute in 2020, the application spotlight event ({\small{\url{https://sites.google.com/view/viscomsospotlight/}}}) at IEEE VIS 2020 discussing opportunities and challenges in cosmology visualization, and the annual \emph{glue-con} hackathon ({\small{\url{https://www.gluesolutions.io/glue-con}}}) that integrate astronomical software projects including \textsf{glue}, \textsf{yt}, \textsf{OpenSpace}, \textsf{WorldWide Telescope} into a centralized system. These meetings are critical, in addition to the larger meetings we suggest above.

To assist in the access to information and literature, we suggest that an ``\textbf{astrovis}'' \textbf{keyword} be added to papers -- published in either astrophysics journals or visualization publications -- to make interrogation of papers easy for the communities. 
Our visual literature browser that enables exploration of ``astrovis'' papers is a step in this direction. 
  
We further suggest starting \textbf{visualization/data challenges} within the large publicly available astrophysics data surveys. 
The ``solutions'' to these challenges should be made publicly available and thus applicable to other datasets. 
A few scientific visualization (SciVis) challenges have involved astronomy datasets at IEEE VIS conferences, notably the SciVis 2015 Contest using the Dark Sky Simulations ({\small{\url{https://darksky.slac.stanford.edu/}}}) and the SciVis 2019 Contest using the HACC cosmological simulation ({\small{\url{https://wordpress.cels.anl.gov/2019-scivis-contest/}}}). 
A recent example is the data challenge held at the IEEE VIS 2020 VisAstro workshop, where a visualization tool under development called \textsf{CosmoVis} was used by Burchett \etal~\cite{BurchettAbramovElek2020} to interactively analyze  cosmological simulations generated by \textsf{IllustrisTNG}.  
In terms of pedagogy, we recommend summer schools, hackathons, and workshops that help onboard members of each community engage in this joint effort.  The data challenges may serve as the seeding point for such workshops.  
Science not communicated is science not done.  Looking toward science education and public outreach, we highlight the important role played by museums and planetariums in transitioning scientific discovery to public education.  We suggest that these stakeholders be included organically in the discovery process and emphasize their key role in the scientific process.

We are encouraged by the progress that has been made in the last decade, and we look forward to the next decade of development, discovery, and education.

%% file: arXiv-astro-camera-ready.bbl
\newcommand{\etalchar}[1]{$^{#1}$}

%% file: sec-bio.tex
\section*{Short Biographies}

\paragraph*{Fangfei Lan} is a Ph.D. student in the Scientific Computing and Imaging (SCI) Institute, University of Utah.
She is interested in topological data analysis and visualization. She works at the intersection between theories in topological data analysis and their practical applications in science and engineering. 
Email: fangfei.lan@utah.edu. 

\paragraph*{Michael Young} is a Master's student in the Scientific Computing and Imaging (SCI) Institute, University of Utah. 
His research focuses on building interactive visualization systems and exploring visualization methods for astronomy datasets.  
Email: myoung@cs.utah.edu.

\paragraph*{Lauren Anderson} is a Carnegie Postdoctoral Fellow at the Carnegie Observatories. She obtained her Ph.D. in Astronomy from University of Washington. She works on applying computational data analysis techniques to interesting astronomy problems. Currently she is building a dust map of the Milky Way galaxy, and is in need of novel visualization techniques for this large, rich data set.
Email: landerson@carnegiescience.edu. 

\paragraph*{Anders Ynnerman} is a Professor of Visualization at Linköping University and the director of the Norrköping Visualization Center. He also heading the Media and Information Technology division at Linköping University. His research in visualization focuses on large scale data visualization in a range of application domains, and visualization for science communication is a central application theme.  He is the founder of the \emph{OpenSpace} project and a producer of planetarium shows. In 2018 he received the IEEE VGTC Technical Achievement Award. 
Email: anders.ynnerman@liu.se.

\paragraph*{Alexander Bock} is an Assistant Professor at Link\"oping University, Research Fellow at the University of Utah, and the Development Lead for the astrovisualization software \emph{OpenSpace}.  He uses visualization techniques to create applications that can make difficult data, for example CT/MRI scans or astronomical datasets, easier to understand for a specific target audience, such as domain experts or the general public.
Email: alexander.bock@liu.se. 

\paragraph*{Michelle A. Borkin} is an Assistant Professor in the Khoury College of Computer Sciences, Northeastern University.  Her research interests include data visualization, human-computer interaction, and application work across domains including astronomy, physics, and medical imaging.  She received a Ph.D. in 2014 and an M.S. in 2011, both in Applied Physics, and a B.A. in Astronomy and Astrophysics \& Physics from Harvard University. She is the visualization lead and original project member of the \textit{glue} multi-dimensional linked-data exploration software, originally developed for astronomy and designed to facilitate data analysis and visualization for NASA's James Webb Space Telescope.
Email: m.borkin@northeastern.edu. 

\paragraph*{Angus G. Forbes} is an Associate Professor in the Department of Computational Media at University of California, Santa Cruz, where he directs the UCSC Creative Coding Lab. He received his Ph.D. and M.S. degrees from University of California, Santa Barbara. His visualization research investigates novel techniques for representing and interacting with complex scientific information.  
Email: angus@ucsc.edu. 

\paragraph*{Juna A. Kollmeier} is a staff member at the Carnegie Observatories where she is the Founding Director of the Carnegie Theoretical Astrophysics Center. She is also the director of the fifth phase of the Sloan Digital Sky Survey (SDSS; which started in October 2020). SDSS has been one of the foundational datasets in astrophysics but also in data science more generally. SDSS pioneered the open-data model for large survey databases and is a major motivator of advanced visualization techniques in astrophysics. Her research interests are in theoretical astrophysics concerning the growth of cosmic structure on all scales. She works on cosmological hydrodynamics simulations as well as massively multiplexed spectroscopic datasets.   
Email: jak@carnegiescience.edu. 

\paragraph*{Bei Wang} is an assistant professor at the SCI Institute, School of Computing, University of Utah. She works on topological data analysis and visualization. She received her Ph.D. in Computer Science from Duke University. She is interested in the analysis and visualization of large and complex data. Her research interests include topological data analysis, data visualization, computational topology, computational geometry, machine learning and data mining. She is working on connecting data visualization with astronomy data, in particular, using topological techniques to separate signals from noise. 
Email: beiwang@sci.utah.edu.